\documentclass[prd,aps,epsfig,amsmath]{revtex4}

\usepackage{times}
\usepackage{hyperref}

\usepackage{graphicx}
\usepackage{dcolumn}
\usepackage{bm}

\begin{document}

\begin{flushright}
Dedicated to the memory of Professor Augusto Garc{\'\i}a, \\
teacher and friend
\end{flushright}

\title{Baryon magnetic moments in large-$N_c$ chiral perturbation theory}

\author{
Rub\'en Flores-Mendieta
}
\affiliation{
Instituto de F{\'\i}sica, Universidad Aut\'onoma de San Luis Potos{\'\i}, \'Alvaro Obreg\'on 64, Zona Centro, San Luis Potos{\'\i}, S.L.P.\ 78000, M\'exico
}

\date{\today}

\begin{abstract}
The baryon magnetic and transition magnetic moments are computed in heavy baryon chiral perturbation theory in the large-$N_c$ limit, where $N_c$ is the number of colors. One-loop nonanalytic corrections of orders $m_q^{1/2}$ and $m_q \ln m_q$ are incorporated into the analysis, where contributions of both intermediate octet and decuplet baryon states are explicitly included. Expressions are obtained in the limit of vanishing baryon mass differences and compared with the current experimental data. Furthermore, a comparison with conventional heavy baryon chiral perturbation theory is carried out for three light quarks flavors and at the physical value $N_c=3$.

\end{abstract}


\maketitle

\section{Introduction}\label{sec:intro}

From the theoretical point of view, the study of the magnetic moments of baryons presents an opportunity to shed light on an accurate test of QCD in the same way the magnetic moments of the electron and muon provided an accurate test of QED in the past. There are an important number of works focused on the analysis of baryon magnetic moments; the approaches include, among others, the quark model (and its variants) \cite{qm1,qm2,qm3,qm4,qm5,qm6,qm7}, QCD sum rules \cite{sum1,sum2,sum4,sum5}, the $1/N_c$ expansion, where $N_c$ is the number of colors \cite{jm94,dai,ji,lb}, chiral perturbation theory \cite{caldi,gasser,krause,jen92,milana,meiss,loyal,puglia,tib,geng,geng2}, and lattice gauge theory \cite{latt1}, to name but a few. The $1/N_c$ expansion and chiral perturbation theory, on general grounds, have had a major impact on the extraction of the low-energy consequences of QCD.

On the one hand, the generalization of QCD from $N_c=3$ to $N_c \gg 3$ colors, referred to as the large-$N_c$ limit, provides a framework for studying the nonperturbative QCD dynamics of hadrons. Specifically, in the large-$N_c$ limit, the baryon sector of QCD possesses an exact contracted SU($2N_f$) spin-flavor symmetry, where $N_f$ is the number of light quark flavors \cite{djm94,djm95}. The spin-flavor structure for baryons for finite $N_c$ is thus given by analyzing $1/N_c$ corrections to the large-$N_c$ limit \cite{djm94,djm95}. Apart from the calculation of baryon magnetic moments, the $1/N_c$ expansion has also been successfully used in the calculation of other static properties of baryons such as masses \cite{j315,djm94,djm95,jl} and vector and axial-vector couplings \cite{djm94,djm95,dai,rfm98,rfm04}. Most calculations include corrections of relative order $1/N_c^2$, and even of relative order $1/N_c^3$ in the case of baryon masses, for two and three light quark flavors.
The impact of the $1/N_c$ expansion in the calculation of baryon static properties can be assessed by comparing its predictions with experiment, which are in overall good agreement.

On the other hand, heavy baryon chiral perturbation theory \cite{jm255,jm259} is another formalism that has been implemented to systematically compute the properties of baryons. In this formalism, the expansion of the baryon chiral Lagrangian is in powers of $m_q$ and $1/M_B$, where $M_B$ is the baryon mass (for a recent review on chiral perturbation theory see for instance Ref.~\cite{vero}). The effective Lagrangian thus obtained can be used to compute chiral logarithms in the effective theory, yielding to a remarkable computational simplicity because there are no gamma matrices left. One of the earliest applications of heavy baryon chiral perturbation theory can be found in the computation of one-loop corrections to the leading axial-vector couplings in baryon semileptonic decays \cite{jm255,jm259}. An important result derived from this analysis was the observation of large cancellations in the loop corrections between graphs with intermediate spin-1/2 octet and spin-3/2 decuplet baryon states. These cancellations occur as a consequence of the spin-flavor symmetry which is present in the large-$N_c$ limit, and have already been proven both phenomenologically and analytically \cite{djm95,dai,fmhjm,rfm06}.

A further theoretical improvement has been achieved through a combined expansion in $1/N_c$ and chiral corrections \cite{lmr94,lmr95,jen96}, which can constrain the low-energy interactions of baryons with the meson nonet more effectively than either approach alone. In particular, in Ref.~\cite{jen96} a $1/N_c$ expansion of the chiral Lagrangian for baryons was proposed and applied to the calculation of the flavor-\textbf{27} nonanalytic meson-loop corrections to the baryon masses. A more recent application of this formalism can be found in the renormalization of the baryon axial-vector current \cite{rfm06}.

The earliest attempts of computing corrections to baryon magnetic moments beyond tree level in chiral perturbation theory can be traced back to the works of Caldi and Pagels \cite{caldi}, Gasser, Sainio and Svarc \cite{gasser} and Krause \cite{krause}. Relatively more recent analyses can be found in Refs.~\cite{jen92,meiss,loyal,puglia,tib,geng,geng2}.

The one-loop corrections to baryon magnetic moments have leading nonanalytic dependences on the quark masses and fall into two classes, namely, $m_q^{1/2}$ and $m_q \ln m_q$. Caldi and Pagels \cite{caldi} focused on calculating what were supposed to be the leading corrections, namely, those of order $m_s^{1/2}$, and concluded that, in general, the corrections turned out to be at least as large as the zeroth-order contribution so that the perturbation expansion would break down. In turn, Gasser \textit{et.\ al.} \cite{gasser} reanalyzed these corrections whereas Krause \cite{krause} dealt with the logarithmic term, namely, $m_s \ln m_s$. Afterwards, Jenkins \textit{et.\ al.} \cite{jen92} reexamined the problem in the framework of heavy baryon chiral perturbation theory, including explicitly contributions of both intermediate octet and decuplet baryon states. Contrary to expectations, it was found that the inclusion of the decuplet did not produce appreciably better agreement with the data than the case when only the octet was included. In other words, the large cancellations observed in the one-loop corrections to the baryon axial-vector current when both octet and decuplet intermediate baryon states are included would not occur in the magnetic moments. It was argued that there could be some evidence that chiral perturbation theory overestimated the size of the kaon loops so it was proposed compensating such an effect by using the one-loop corrected axial-vector couplings rather than the tree-level ones.

Meissner and Steininger \cite{meiss} also tackled the problem in the context of heavy baryon chiral perturbation theory but included all the terms up to order $\mathcal{O}(q^4)$. They thus took into account $1/M_B$ corrections and contributions of certain double derivative operators which occur to this order. They however did not include the decuplet explicitly in the analysis; rather, its effects were considered by computing its contributions to some low-energy constants. Durand and Ha \cite{loyal}, in turn, performed the calculation in the same context of Ref.~\cite{jen92}, emphasizing the role of the decuplet-octet mass difference; they found that there was not clear evidence of the convergence of the chiral series. Puglia and Ramsey-Musolf \cite{puglia} performed an analysis similar to the one of Ref.~\cite{meiss}, but included the decuplet explicitly and retained only nonanalytic one-loop corrections. Finally, Geng \textit{et.\ al.} \cite{geng,geng2}, using covariant perturbation theory, found small loop corrections leading to an improvement over the SU(3) description.

All in all, the analyses performed over the past decade and a half about baryon magnetic moments have yielded a number of interesting conclusions, some of them in contradiction with the others. We would like to highlight, however, the role the decuplet plays as an intermediate state in the one-loop graphs.

In the present paper, we will confine our attention to the calculation of baryon magnetic moments in a simultaneous expansion in $m_q$ and $1/N_c$ at one-loop order. The starting point will be the fact that, in the large-$N_c$ limit, the baryon magnetic moment and the baryon axial-vector current share the same kinematical properties so that they can be described in terms of the same operators. Thus, for the main aim of the analysis we will use the formalism introduced in Ref.~\cite{jen96} and follow a close parallelism with the analysis of Ref.~\cite{rfm06}. In this latter reference, the axial-vector form factor $g_A$, which enters into play in baryon semileptonic decays, was computed at one-loop order and compared, for $N_f=3$ and at the physical value $N_c=3$, with the corresponding one obtained in the framework of conventional heavy baryon chiral perturbation theory, \textit{i.e.}, the effective field theory with no $1/N_c$ expansion. The agreement observed, order by order, was remarkable.

We need to point out that, in Ref.~\cite{lmr95}, a similar analysis within the combined expansion has already been performed, based on the formulation implemented in Ref.~\cite{lmr94}, so we will pin down the similarities and/or differences of this approach with ours.

We have organized this paper as follows. In Sec.~\ref{sec:overview} we provide in broad terms an overview of the formalism on large-$N_c$ baryons in order to bring out the essential features of it. Our notation and conventions will be introduced accordingly. In Sec.~\ref{sec:tree} we construct the $1/N_c$ expansion of the baryon magnetic moment operator and then use it to obtain the tree-level values of the magnetic moments of the octet and decuplet baryons and the allowed octet-octet and decuplet-octet transitions. A total of 27 magnetic moments are obtained at this level. In Sec.~\ref{sec:oneloop} we turn to the computation of one-loop nonanalytic corrections of the types $m_q^{1/2}$ and $m_q\ln m_q$, to relative order $\mathcal{O}(1/N_c^3)$. In Sec.~\ref{sec:comparison} we compare the results obtained in the combined expansion with the ones obtained in the framework of heavy baryon chiral perturbation theory. The comparison is carried out by establishing the relations existing between the parameters of the $1/N_c$ expansion and the invariant couplings of the chiral expansion. We devote Sec.~\ref{sec:numerical} to performing a detailed numerical comparison of our expressions with the available experimental data \cite{part} by means of a least-squares fit. The best-fit parameters are then used to make some predictions of the unmeasured magnetic moments. Finally, in Sec.~\ref{sec:conclusion} we present a summary as well as the main conclusions of the study. We complement the paper with two appendices which contain the reduction of all the baryon operators that appear to relative order $\mathcal{O}(1/N_c^3)$, for $N_f$ and $N_c$ arbitrary.

\section{Overview on large-$N_c$ baryons\label{sec:overview}}

For a detailed outline of the formalism on large-$N_c$ baryons and all the mathematical groundwork, which contains a formidable amount of group theory, we refer to the original works \cite{djm94,djm95,jen96}, while we restrict us here to a short description of the method and to introduce our notation and conventions.

First of all, the static baryon matrix elements of a QCD operator have a $1/N_c$ expansion of the form \cite{djm95}
\begin{equation}
\mathcal{O}_{\mathrm{QCD}} = N_c \sum_n d_n \frac{1}{N_c^n} \mathcal{O}_n, \label{eq:qcdop}
\end{equation}
where the various $\mathcal{O}_n$'s are independent operators which transform according to the same spin$\otimes$flavor representations as $\mathcal{O}_{\mathrm{QCD}}$. For finite $N_c$, the sum on $n$ in Eq.~(\ref{eq:qcdop}) is over $0\leq n \leq N_c$ so that a given $\mathcal{O}_n$ is termed an $n$-body operator, which can be written as a monomial of degree $n$ in the baryon spin-flavor generators. On the other hand, each unknown operator coefficient $d_n(1/N_c)$ has an expansion in $1/N_c$ beginning at order unity. The factor $1/N_c^n$ is required since each spin-flavor generator in an $n$th-order operator product $\mathcal{O}_n$ comes along with a factor of $1/N_c$. Additionally, the overall factor of $N_c$ arises because a QCD 1-body operator has matrix elements that are at most of order $\mathcal{O}(N_c)$ when inserted on all quark lines on the baryon \cite{djm95}.

Specifically, for $N_f=3$, the large-$N_c$ spin-flavor symmetry for baryons is generated by the baryon spin, flavor and spin-flavor operators $J^k$, $T^c$ and $G^{kc}$, respectively, which can be written for large but finite $N_c$ as 1-body quark operators acting on the $N_c$-quark baryon states \cite{djm95} as
\begin{eqnarray}
J^k & = & \sum_\alpha q_\alpha^\dagger \left(\frac{\sigma^k}{2}\otimes \openone \right) q_\alpha, \nonumber \\
T^c & = & \sum_\alpha q_\alpha^\dagger \left(\openone \otimes \frac{\lambda^c}{2} \right) q_\alpha, \label{eq:su6gen}\\
G^{kc} & = & \sum_\alpha q_\alpha^\dagger \left(\frac{\sigma^k}{2}\otimes \frac{\lambda^c}{2} \right) q_\alpha, \nonumber
\end{eqnarray}
where $q_\alpha^\dagger$ and $q_\alpha$ are operators that create and annihilate states in the fundamental representation of SU(6) and the index $\alpha$ sums over the $N_c$ quarks. In addition, $\sigma^k$ and $\lambda^c$ are the Pauli spin and Gell-Mann flavor matrices, respectively, where the spin index $k$ runs from one to three and the flavor index $c$ runs from one to eight. Throughout this paper, unless explicitly noticed otherwise, the square of the spin operator is given by $J^2 \equiv J^rJ^r$, where the ordinary convention of summing over repeated indices is adopted. No confusion is expected to arise from this. Without loss of generality, the baryon matrix elements of the spin-flavor generators (\ref{eq:su6gen}) can be taken as the values in the nonrelativistic quark model, so this convention is usually referred to as the quark representation in the literature \cite{djm95}.

Analyzing the $N_c$ dependence of operator products appearing in the $1/N_c$ expansion (\ref{eq:qcdop}) is not an easy matter due to the fact that the operator matrix elements have a different $N_c$ dependence in different parts of the flavor weight diagrams \cite{djm95}. The $N_c$ dependence of the operator products is ultimately obtained by analyzing the $N_c$ dependence of the matrix elements of the baryon spin-flavor generators $J^k$, $T^c$, and $G^{kc}$ in the weight diagrams for the SU(3) flavor representations of the spin-1/2 and spin-3/2 baryons. In Ref.~\cite{fmhjm}, a naive $1/N_c$ counting rule was implemented. However, on a more detailed level, baryons with spins of order unity have matrix elements of the flavor generators $T^c$ that are $\mathcal{O}(1)$, $\mathcal{O}(\sqrt{N_c})$, and $\mathcal{O}(N_c)$ for $c=1,2,3$, $c=4,5,6,7$, and $c=8$, respectively, and matrix elements of the spin-flavor generators $G^{kc}$ that are $\mathcal{O}(1)$, $\mathcal{O}(\sqrt{N_c})$, and $\mathcal{O}(N_c)$ for $c=1,2,3$, $c=4,5,6,7$, and $c=8$, respectively. Thus, factors of $T^c/N_c$ and $G^{kc}/N_c$ are of order 1 somewhere in the weight diagram, whereas factors of $J^k/N_c$ are of order $1/N_c$ everywhere \cite{djm95}.

As an illustrative example of the $1/N_c$ expansion for a baryon operator, let us consider the baryon axial-vector current $A^{kc}$, which is a spin-1 object, an octet under SU(3), and odd under time reversal. The $1/N_c$ expansion of $A^{kc}$ can thus be written as \cite{djm95}
\begin{equation}
A^{kc} = a_1 G^{kc} + \sum_{n=2,3}^{N_c} b_n \frac{1}{N_c^{n-1}} \mathcal{D}_n^{kc} + \sum_{n=3,5}^{N_c} c_n
\frac{1}{N_c^{n-1}} \mathcal{O}_n^{kc}, \label{eq:akcfull}
\end{equation}
where $a_1$, $b_n$, and $c_n$ are unknown coefficients. The $\mathcal{D}_n^{kc}$ are diagonal operators with nonvanishing matrix elements only between states with the same spin, and the $\mathcal{O}_n^{kc}$ are purely off-diagonal operators with nonvanishing matrix elements only between states with different spin. The first few terms in expansion (\ref{eq:akcfull}) read
\begin{eqnarray}
\mathcal{D}_2^{kc} & = & J^kT^c, \label{eq:d2kc} \\
\mathcal{O}_2^{kc} & = & \epsilon^{ijk} \{J^i,G^{jc}\} = i [J^2,G^{kc}], \label{eq:o2kc} \\
\mathcal{D}_3^{kc} & = & \{J^k,\{J^r,G^{rc}\}\}, \label{eq:d3kc} \\
\mathcal{O}_3^{kc} & = & \{J^2,G^{kc}\} - \frac12 \{J^k,\{J^r,G^{rc}\}\}, \label{eq:o3kc}
\end{eqnarray}
whereas successive higher order operators are obtained as $\mathcal{D}_n^{kc}=\{J^2,\mathcal{D}_{n-2}^{kc}\}$ and $\mathcal{O}_n^{kc}=\{J^2,\mathcal{O}_{n-2}^{kc}\}$ for $n\geq 4$. The operators $\mathcal{O}_{m}^{kc}$, for $m$ even, are forbidden in expansion (\ref{eq:akcfull}) because they are even under time reversal. At the physical value $N_c=3$ the series can be truncated as
\begin{equation}
A^{kc} = a_1 G^{kc} + b_2 \frac{1}{N_c} \mathcal{D}_2^{kc} + b_3 \frac{1}{N_c^2} \mathcal{D}_3^{kc} + c_3 \frac{1}{N_c^2} \mathcal{O}_3^{kc}. \label{eq:akc}
\end{equation}

The matrix elements of the space components of $A^{kc}$ between SU(6) baryon symmetric states yield the actual values of the axial-vector couplings. For octet baryons, the axial-vector couplings are $g_A$ as usually defined in baryon semileptonic decay experiments. After renormalization, $g_A \approx 1.27$ \cite{rfm06} for neutron $\beta$-decay.

\section{Baryon magnetic moments at tree level\label{sec:tree}}

The static properties of baryons can be determined from their couplings to the weak and electromagnetic currents.
In particular, in this work, we will describe the magnetic moments in the context of large-$N_c$ chiral perturbation theory.

In the large-$N_c$ limit, the baryon magnetic moments possess the same kinematical properties as the baryon axial-vector couplings; as a result, the operators used to describe these quantities are practically identical \cite{dai}. The baryon magnetic moment operator, likewise the baryon axial-vector current operator $A^{kc}$, is a spin-1 object and an octet under SU(3). We will thus follow a close parallelism with the analysis of the renormalization of $A^{kc}$ performed in Ref.~\cite{rfm06} in order to achieve our goal.

For definiteness, in analogy with Eq.~(\ref{eq:akc}), we construct the $1/N_c$ expansion of the operator which yields the baryon magnetic moment as follows
\begin{equation}
M^{kc} = m_1 G^{kc} + m_2 \frac{1}{N_c} \mathcal{D}_2^{kc} + m_3 \frac{1}{N_c^2} \mathcal{D}_3^{kc} + m_4 \frac{1}{N_c^2} \mathcal{O}_3^{kc}, \label{eq:mmag}
\end{equation}
where we have truncated the series at the physical value $N_c=3$. If we assume SU(3) symmetry, the unknown coefficients  $m_i$ are independent of $k$ so they are unrelated to the ones of expansion (\ref{eq:akc}) in this limit. The magnetic moments are proportional to the quark charge matrix $\mathcal{Q}=\textrm{diag}(2/3,-1/3,-1/3)$, so they can be separated into isovector and isoscalar components, $M^{k3}$ and $M^{k8}$, respectively. Accordingly, from Eq.~(\ref{eq:mmag}), we define the baryon magnetic moment operator as
\begin{equation}
M^k = M^{kQ} \equiv M^{k3} + \frac{1}{\sqrt{3}} M^{k8}. \label{eq:mmsep}
\end{equation}
Hereafter, when computing matrix elements, the spin index $k$ of $M^k$ will be set to 3 whereas the flavor index $Q$ will stand for $Q=3+(1/\sqrt{3})8$ so any operator of the form $X^Q$ should be understood as $X^3+(1/\sqrt{3})X^8$.

In actual calculations, the one-body operators $T^c$ and $G^{ic}$, $c=3,8$, appear quite often; these operators can be rewritten in terms of quark number and spin operators as \cite{djm95}
\begin{eqnarray}
T^3 & = & \frac12 (N_u - N_d), \label{eq:t3} \\
T^8 & = & \frac{1}{2 \sqrt 3} (N_c - 3 N_s), \label{eq:t8} \\
G^{i3} & = & \frac12 (J_u^i - J_d^i), \label{eq:gi3} \\
G^{i8} & = & \frac{1}{2 \sqrt 3} (J^i - 3 J_s^i), \label{eq:gi8}
\end{eqnarray}
where $N_c=N_u+N_d+N_s$ and $J^i=J_u^i+J_d^i+J_s^i$. The $N_c$ dependence of the operators involved in relations (\ref{eq:t3})-(\ref{eq:gi8}) is now manifest.

Let us now turn to evaluate the matrix elements of $M^k$ for octet ($B$) and decuplet baryons ($T$) and the allowed octet-octet and decuplet-octet transition magnetic moments.

\subsection{Magnetic moments of octet baryons at tree level}

The magnetic moments at tree level of the octet baryons and the transition $\Sigma^0\Lambda$ can easily be obtained by computing the matrix elements of $M^k$ between SU(6) symmetric states.\footnote{Without further specification, all magnetic moments are given in units of nuclear magnetons, $\mu_N$.} We would like to remark that an analysis of baryon magnetic moments in the $1/N_c$ expansion alone is presented in Ref.~\cite{lb}. The operator basis used in this reference is somewhat different from ours since operators up to relative order $\mathcal{O}(1/N_c^2)$ are retained in the $1/N_c$ expansion. We thus have performed an independent computation of the matrix elements of our operator basis and crosschecked those in common with the ones of Ref.~\cite{lb} at $N_c=3$. All matrix elements agree, except for a change of sign of those corresponding to the transitions $\Sigma^0\Lambda$ and ${\Sigma^*}^0\Lambda$. This difference might come from a different choice of the phases of the baryon states and should not affect the calculations. At any rate, the matrix elements of the operators involved in Eq.~(\ref{eq:mmsep}) for octet baryons are listed in Table \ref{t:mm133B} for the sake of completeness.

\begin{table*}
\caption{\label{t:mm133B}Nontrivial matrix elements of the operators involved in the magnetic moments of octet baryons at tree level.}
\begin{ruledtabular}
\begin{tabular}{lccccccccc}
& $\displaystyle n$ & $\displaystyle p$ & $\displaystyle \Sigma^-$ & $\displaystyle \Sigma^0$ & $\displaystyle \Sigma^+$ & $\displaystyle \Xi^-$ & $\displaystyle \Xi^0$ & $\displaystyle \Lambda$ & $\displaystyle \Lambda\Sigma^0$ \\[2mm]
\hline
$\langle G^{33} \rangle$ & $\displaystyle -\frac{5}{12}$ & $\displaystyle \frac{5}{12}$ & $\displaystyle -\frac{1}{3}$ & $\displaystyle 0$ & $\displaystyle \frac{1}{3}$ & $\displaystyle \frac{1}{12}$ & $\displaystyle -\frac{1}{12}$ & $\displaystyle 0$ & $\displaystyle \frac{1}{2 \sqrt{3}}$ \\[2mm]
$\langle \mathcal{D}_2^{33} \rangle$ & $\displaystyle -\frac{1}{4}$ & $\displaystyle \frac{1}{4}$ & $\displaystyle -\frac{1}{2}$ & $\displaystyle 0$ & $\displaystyle \frac{1}{2}$ & $\displaystyle -\frac{1}{4}$ & $\displaystyle \frac{1}{4}$ & $\displaystyle 0$ & $\displaystyle 0$ \\[2mm]
$\langle \mathcal{D}_3^{33} \rangle$ & $\displaystyle -\frac{5}{4}$ & $\displaystyle \frac{5}{4}$ & $\displaystyle -1$ & $\displaystyle 0$ & $\displaystyle 1$ & $\displaystyle \frac{1}{4}$ & $\displaystyle -\frac{1}{4}$ & $\displaystyle 0$ & $\displaystyle \frac{\sqrt{3}}{2}$ \\[2mm]
$\langle \mathcal{D}_4^{33} \rangle$ & $\displaystyle -\frac{3}{8}$ & $\displaystyle \frac{3}{8}$ & $\displaystyle -\frac{3}{4}$ & $\displaystyle 0$ & $\displaystyle \frac{3}{4}$ & $\displaystyle -\frac{3}{8}$ & $\displaystyle \frac{3}{8}$ & $\displaystyle 0$ & $\displaystyle 0$ \\[2mm]
$\langle \mathcal{D}_5^{33} \rangle$ & $\displaystyle -\frac{15}{8}$ & $\displaystyle \frac{15}{8}$ & $\displaystyle -\frac{3}{2}$ & $\displaystyle 0$ & $\displaystyle \frac{3}{2}$ & $\displaystyle \frac{3}{8}$ & $\displaystyle -\frac{3}{8}$ & $\displaystyle 0$ & $\displaystyle \frac{3 \sqrt{3}}{4}$ \\[2mm]
\hline
$\langle G^{38} \rangle$ & $\displaystyle \frac{1}{4 \sqrt{3}}$ & $\displaystyle \frac{1}{4 \sqrt{3}}$ & $\displaystyle \frac{1}{2 \sqrt{3}}$ & $\displaystyle \frac{1}{2 \sqrt{3}}$ & $\displaystyle \frac{1}{2 \sqrt{3}}$ & $\displaystyle -\frac{\sqrt{3}}{4}$ & $\displaystyle -\frac{\sqrt{3}}{4}$ & $\displaystyle -\frac{1}{2 \sqrt{3}}$ & $\displaystyle 0$ \\[2mm]
$\langle \mathcal{D}_2^{38} \rangle$ & $\displaystyle \frac{\sqrt{3}}{4}$ & $\displaystyle \frac{\sqrt{3}}{4}$ & $\displaystyle 0$ & $\displaystyle 0$ & $\displaystyle 0$ & $\displaystyle -\frac{\sqrt{3}}{4}$ & $\displaystyle -\frac{\sqrt{3}}{4}$ & $\displaystyle 0$ & $\displaystyle 0$ \\[2mm]
$\langle \mathcal{D}_3^{38} \rangle$ & $\displaystyle \frac{\sqrt{3}}{4}$ & $\displaystyle \frac{\sqrt{3}}{4}$ & $\displaystyle \frac{\sqrt{3}}{2}$ & $\displaystyle \frac{\sqrt{3}}{2}$ & $\displaystyle \frac{\sqrt{3}}{2}$ & $\displaystyle -\frac{3 \sqrt{3}}{4}$ & $\displaystyle -\frac{3 \sqrt{3}}{4}$ & $\displaystyle -\frac{\sqrt{3}}{2}$ & $\displaystyle 0$ \\[2mm]
$\langle \mathcal{D}_4^{38} \rangle$ & $\displaystyle \frac{3 \sqrt{3}}{8}$ & $\displaystyle \frac{3 \sqrt{3}}{8}$ & $\displaystyle 0$ & $\displaystyle 0$ & $\displaystyle 0$ & $\displaystyle -\frac{3 \sqrt{3}}{8}$ & $\displaystyle -\frac{3 \sqrt{3}}{8}$ & $\displaystyle 0$ & $\displaystyle 0$ \\[2mm]
$\langle \mathcal{D}_5^{38} \rangle$ & $\displaystyle \frac{3 \sqrt{3}}{8}$ & $\displaystyle \frac{3 \sqrt{3}}{8}$ & $\displaystyle \frac{3 \sqrt{3}}{4}$ & $\displaystyle \frac{3 \sqrt{3}}{4}$ & $\displaystyle \frac{3 \sqrt{3}}{4}$ & $\displaystyle -\frac{9 \sqrt{3}}{8}$ & $\displaystyle -\frac{9 \sqrt{3}}{8}$ & $\displaystyle -\frac{3 \sqrt{3}}{4}$ & $\displaystyle 0$ \\[2mm]
\end{tabular}
\end{ruledtabular}
\end{table*}

The tree-level value of the magnetic moment of baryon $B$ is defined here as $\mu_B^{(0)} \equiv \langle B|M^3|B\rangle$, where the superscript attached to $\mu_B$ denotes the tree-level value and $M^3$ refers to the third component of $M^k$. For $N_f=N_c=3$, the various $\mu_B^{(0)}$ read
\begin{subequations}
\label{eq:tlo}
\begin{eqnarray}
&  & \mu_n^{(0)} = -\frac13 m_1 - \frac{1}{9} m_3, \\
&  & \mu_p^{(0)} = \frac12 m_1 + \frac{1}{6} m_2 + \frac{1}{6} m_3, \\
&  & \mu_\Lambda^{(0)} = -\frac16 m_1 - \frac{1}{18} m_3, \\
&  & \mu_{\Sigma^0}^{(0)} = \frac16 m_1 + \frac{1}{18} m_3, \\
&  & \mu_{\Sigma^+}^{(0)} = \frac12 m_1 + \frac{1}{6} m_2 + \frac{1}{6} m_3, \\
&  & \mu_{\Sigma^-}^{(0)} = -\frac16 m_1 - \frac{1}{6} m_2 - \frac{1}{18} m_3, \\
&  & \mu_{\Xi^0}^{(0)} = -\frac13 m_1 - \frac{1}{9} m_3, \\
&  & \mu_{\Xi^-}^{(0)} = -\frac16 m_1 - \frac{1}{6} m_2 - \frac{1}{18} m_3, \\
&  & \mu_{\Lambda\Sigma^0}^{(0)} = \frac{\sqrt{3}}{6} m_1 + \frac{\sqrt{3}}{18} m_3.
\end{eqnarray}
\end{subequations}

Let us observe that, for baryon octet states, the matrix elements of the operators $\mathcal{O}_n^{kQ}$ vanish whereas the matrix elements of the operators $\mathcal{D}_n^{kQ}$, with $n\geq 3$, are directly proportional to the matrix elements of $G^{kQ}$ and $\mathcal{D}_2^{kQ}$ for $n$ odd and even, respectively. Accordingly, we can define $m_1^\prime \equiv m_1 + m_3/3$ in Eqs.~(\ref{eq:tlo}) in such a way that we are left with only two parameters, $m_1^\prime$ and $m_2$, to parametrize the tree-level values, in complete agreement with the analysis performed in the framework of heavy baryon chiral perturbation theory \cite{jen92}, which does so in terms of the two couplings $\mu_D$ and $\mu_F$. We will deal with this issue in subsequent sections, but at any rate, we purposely keep for completeness the four parameters $m_i$ as they are introduced in Eq.~(\ref{eq:mmag}).

From Eqs.~(\ref{eq:tlo}), we readily verify that the Coleman-Glashow relations \cite{cg61}, valid in the SU(3) limit, are fulfilled, namely,
\begin{eqnarray}
\begin{array}{lcl}
\mu_{\Sigma^+}^{(0)} = \mu_p^{(0)}, & \qquad \qquad & \mu_{\Sigma^-}^{(0)} + \mu_n^{(0)} = -\mu_p^{(0)}, \\[4mm]
2\mu_\Lambda^{(0)} = \mu_n^{(0)}, & \qquad  \qquad & \mu_{\Xi^-}^{(0)} = \mu_{\Sigma^-}^{(0)}, \\[4mm]
\mu_{\Xi^0}^{(0)} = \mu_n^{(0)}, & \qquad  \qquad & 2\mu_{\Lambda\Sigma^0}^{(0)} = -\sqrt{3}\mu_n^{(0)}, \label{eq:cg}
\end{array}
\label{eq:treeval}
\end{eqnarray}
along with the isospin relation
\begin{equation}
\mu_{\Sigma^+}^{(0)} - 2 \mu_{\Sigma^0}^{(0)} + \mu_{\Sigma^-}^{(0)} = 0. \label{eq:isos}
\end{equation}

It is important to remark that relations (\ref{eq:treeval}) and (\ref{eq:isos}) are valid to \textit{all orders} in the $1/N_c$ expansion. Indeed, this must be the case since these relations were derived using SU(3) symmetry only.

Besides, let us also notice that the SU(6) prediction \cite{beg}, $3\mu_n^{\textrm{SU(6)}}+ 2\mu_p^{\textrm{SU(6)}}=0$, in our approach is written as
\begin{equation}
3\mu_n^{(0)} + 2 \mu_p^{(0)} = \frac{m_2}{N_c},
\end{equation}
\textit{i.e.}, it picks up a correction of relative order $\mathcal{O}(1/N_c)$, but stills holds at leading order in the $1/N_c$ expansion.\footnote{This should hold since the matrix elements of $J$, $T$ and $G$ contain both leading and subleading contributions in $N_c$. See the discussion on these issues in Sec.~\ref{sec:overview}.}

\subsection{Magnetic moments of decuplet baryons at tree level}

The magnetic moments of the decuplet baryons at tree level, $\mu_T^{(0)}=\langle T|M^3|T\rangle$, are given by using the matrix elements of the corresponding operators listed in Table \ref{t:mm133T}. For $N_f=N_c=3$ they read

\begin{table*}
\caption{\label{t:mm133T}Nontrivial matrix elements of the operators involved in the magnetic moments of decuplet baryons at tree level.}
\begin{ruledtabular}
\begin{tabular}{lcccccccccc}
& $\displaystyle \Delta^{++}$ & $\displaystyle \Delta^+$ & $\displaystyle \Delta^0$ & $\displaystyle \Delta^-$ & $\displaystyle {\Sigma^*}^+$ & $\displaystyle {\Sigma^*}^0$ & $\displaystyle {\Sigma^*}^-$ & $\displaystyle {\Xi^*}^0$ & $\displaystyle {\Xi^*}^-$ & $\displaystyle \Omega^-$ \\[2mm]
\hline
$\langle G^{33} \rangle$ & $\displaystyle \frac{3}{4}$ & $\displaystyle \frac{1}{4}$ & $\displaystyle -\frac{1}{4}$ & $\displaystyle -\frac{3}{4}$ & $\displaystyle \frac{1}{2}$ & $\displaystyle 0$ & $\displaystyle -\frac{1}{2}$ & $\displaystyle \frac{1}{4}$ & $\displaystyle -\frac{1}{4}$ & $\displaystyle 0$ \\[2mm]
$\langle \mathcal{D}_2^{33} \rangle$ & $\displaystyle \frac{9}{4}$ & $\displaystyle \frac{3}{4}$ & $\displaystyle -\frac{3}{4}$ & $\displaystyle -\frac{9}{4}$ & $\displaystyle \frac{3}{2}$ & $\displaystyle 0$ & $\displaystyle -\frac{3}{2}$ & $\displaystyle \frac{3}{4}$ & $\displaystyle -\frac{3}{4}$ & $\displaystyle 0$ \\[2mm]
$\langle \mathcal{D}_3^{33} \rangle$ & $\displaystyle \frac{45}{4}$ & $\displaystyle \frac{15}{4}$ & $\displaystyle -\frac{15}{4}$ & $\displaystyle -\frac{45}{4}$ & $\displaystyle \frac{15}{2}$ & $\displaystyle 0$ & $\displaystyle -\frac{15}{2}$ & $\displaystyle \frac{15}{4}$ & $\displaystyle -\frac{15}{4}$ & $\displaystyle 0$ \\[2mm]
$\langle \mathcal{D}_4^{33} \rangle$ & $\displaystyle \frac{135}{8}$ & $\displaystyle \frac{45}{8}$ & $\displaystyle -\frac{45}{8}$ & $\displaystyle -\frac{135}{8}$ & $\displaystyle \frac{45}{4}$ & $\displaystyle 0$ & $\displaystyle -\frac{45}{4}$ & $\displaystyle \frac{45}{8}$ & $\displaystyle -\frac{45}{8}$ & $\displaystyle 0$ \\[2mm]
$\langle \mathcal{D}_5^{33} \rangle$ & $\displaystyle \frac{675}{8}$ & $\displaystyle \frac{225}{8}$ & $\displaystyle -\frac{225}{8}$ & $\displaystyle -\frac{675}{8}$ & $\displaystyle \frac{225}{4}$ & $\displaystyle 0$ & $\displaystyle -\frac{225}{4}$ & $\displaystyle \frac{225}{8}$ & $\displaystyle -\frac{225}{8}$ & $\displaystyle 0$ \\[2mm]
\hline
$\langle G^{38} \rangle$ & $\displaystyle \frac{\sqrt{3}}{4}$ & $\displaystyle \frac{\sqrt{3}}{4}$ & $\displaystyle \frac{\sqrt{3}}{4}$ & $\displaystyle \frac{\sqrt{3}}{4}$ & $\displaystyle 0$ & $\displaystyle 0$ & $\displaystyle 0$ & $\displaystyle -\frac{\sqrt{3}}{4}$ & $\displaystyle -\frac{\sqrt{3}}{4}$ & $\displaystyle -\frac{\sqrt{3}}{2}$ \\[2mm]
$\langle \mathcal{D}_2^{38} \rangle$ & $\displaystyle \frac{3 \sqrt{3}}{4}$ & $\displaystyle \frac{3 \sqrt{3}}{4}$ & $\displaystyle \frac{3 \sqrt{3}}{4}$ & $\displaystyle \frac{3 \sqrt{3}}{4}$ & $\displaystyle 0$ & $\displaystyle 0$ & $\displaystyle 0$ & $\displaystyle -\frac{3 \sqrt{3}}{4}$ & $\displaystyle -\frac{3 \sqrt{3}}{4}$ & $\displaystyle -\frac{3 \sqrt{3}}{2}$ \\[2mm]
$\langle \mathcal{D}_3^{38} \rangle$ & $\displaystyle \frac{15 \sqrt{3}}{4}$ & $\displaystyle \frac{15 \sqrt{3}}{4}$ & $\displaystyle \frac{15 \sqrt{3}}{4}$ & $\displaystyle \frac{15 \sqrt{3}}{4}$ & $\displaystyle 0$ & $\displaystyle 0$ & $\displaystyle 0$ & $\displaystyle -\frac{15 \sqrt{3}}{4}$ & $\displaystyle -\frac{15 \sqrt{3}}{4}$ & $\displaystyle -\frac{15 \sqrt{3}}{2}$ \\[2mm]
$\langle \mathcal{D}_4^{38} \rangle$ & $\displaystyle \frac{45 \sqrt{3}}{8}$ & $\displaystyle \frac{45 \sqrt{3}}{8}$ & $\displaystyle \frac{45 \sqrt{3}}{8}$ & $\displaystyle \frac{45 \sqrt{3}}{8}$ & $\displaystyle 0$ & $\displaystyle 0$ & $\displaystyle 0$ & $\displaystyle -\frac{45 \sqrt{3}}{8}$ & $\displaystyle -\frac{45 \sqrt{3}}{8}$ & $\displaystyle -\frac{45 \sqrt{3}}{4}$ \\[2mm]
$\langle \mathcal{D}_5^{38} \rangle$ & $\displaystyle \frac{225 \sqrt{3}}{8}$ & $\displaystyle \frac{225 \sqrt{3}}{8}$ & $\displaystyle \frac{225 \sqrt{3}}{8}$ & $\displaystyle \frac{225 \sqrt{3}}{8}$ & $\displaystyle 0$ & $\displaystyle 0$ & $\displaystyle 0$ & $\displaystyle -\frac{225 \sqrt{3}}{8}$ & $\displaystyle -\frac{225 \sqrt{3}}{8}$ & $\displaystyle -\frac{225 \sqrt{3}}{4}$ \\[2mm]
\end{tabular}
\end{ruledtabular}
\end{table*}

\begin{subequations}
\label{eq:tlt}
\begin{eqnarray}
&  & \mu_{\Delta^{++}}^{(0)} = m_1 + m_2 + \frac53 m_3, \\
&  & \mu_{\Delta^{+}}^{(0)} = \frac12 m_1 + \frac12 m_2 + \frac56 m_3, \\
&  & \mu_{\Delta^{0}}^{(0)} = 0, \\
&  & \mu_{\Delta^{-}}^{(0)} = -\frac12 m_1 - \frac12 m_2 - \frac56 m_3, \\
&  & \mu_{{\Sigma^{*}}^+}^{(0)} = \frac12 m_1 + \frac12 m_2 + \frac56 m_3, \\
&  & \mu_{{\Sigma^{*}}^-}^{(0)} = -\frac12 m_1 - \frac12 m_2 - \frac56 m_3, \\
&  & \mu_{{\Sigma^{*}}^0}^{(0)} = 0, \\
&  & \mu_{{\Xi^{*}}^0}^{(0)} = 0, \\
&  & \mu_{{\Xi^{*}}^-}^{(0)} = -\frac12 m_1 - \frac12 m_2 - \frac56 m_3, \\
&  & \mu_{\Omega^-}^{(0)} = -\frac12 m_1 - \frac12 m_2 - \frac56 m_3.
\end{eqnarray}
\end{subequations}

A quick glance at Eqs.~(\ref{eq:tlt}) allows us to anticipate that the values listed are consistent with the ones obtained in the framework of heavy baryon chiral perturbation theory \cite{jen92}, where they are given in terms of a single invariant $\mu_C$, with a normalization such that the magnetic moment of the $i$th decuplet baryon of electric charge $q_i$ is $q_i\mu_C$ nuclear magnetons.

At this point we can verify that the isotensor combinations of magnetic moments with $I=2$ and $I=3$, introduced in Ref.~\cite{lb}, hold at tree level. For $I=2$ one has
\begin{equation}
\mu_{\Delta^{++}}^{(0)} - \mu_{\Delta^+}^{(0)} - \mu_{\Delta^0}^{(0)} + \mu_{\Delta^-}^{(0)} = 0, \label{eq:srd1}
\end{equation}
and
\begin{equation}
\mu_{{\Sigma^*}^+}^{(0)} -2 \mu_{{\Sigma^*}^0}^{(0)} + \mu_{{\Sigma^*}^-}^{(0)} = 0, \label{eq:srd2}
\end{equation}
whereas for $I=3$
\begin{equation}
\mu_{\Delta^{++}}^{(0)} - 3 \mu_{\Delta^+}^{(0)} + 3 \mu_{\Delta^0}^{(0)} - \mu_{\Delta^-}^{(0)} = 0. \label{eq:srd3}
\end{equation}

Finally, the SU(6) prediction $\mu_{\Omega^-}^{\textrm{SU(6)}} + \mu_p^{\textrm{SU(6)}} = 0$ \cite{beg} becomes
\begin{equation}
\mu_{\Omega^-}^{(0)} + \mu_p^{(0)} = -\frac{m_2}{N_c} - \frac{6m_3}{N_c^2},
\end{equation}
which remains valid to leading order but not to subleading orders in the $1/N_c$ expansion.

\subsection{Baryon decuplet to octet transition magnetic moments at tree level}

Expression (\ref{eq:mmsep}) can also be used to obtain the tree-level values of the baryon decuplet to octet transition magnetic moments by reading off the matrix elements of the pertinent operators displayed in Table \ref{t:mm133TO}. These values, denoted here by $\alpha_{TB}^{(0)} \equiv \langle T|M^3|B \rangle$, can be expressed as

\begin{table*}
\caption{\label{t:mm133TO}Nontrivial matrix elements of the operators involved in the decuplet to octet transition magnetic moments at tree level.}
\begin{ruledtabular}
\begin{tabular}{lcccccccc}
& $\displaystyle \Delta^+p$ & $\displaystyle \Delta^0n$ & $\displaystyle {\Sigma^*}^0\Lambda$ & $\displaystyle {\Sigma^*}^0\Sigma^0$ & $\displaystyle {\Sigma^*}^+\Sigma^+$ & $\displaystyle {\Sigma^*}^-\Sigma^-$ & $\displaystyle {\Xi^*}^0\Xi^0$ & $\displaystyle {\Xi^*}^-\Xi^-$ \\[2mm]
\hline
$\langle G^{33} \rangle$ & $\displaystyle \frac{\sqrt{2}}{3}$ & $\displaystyle \frac{\sqrt{2}}{3}$ & $\displaystyle -\frac{1}{\sqrt{6}}$ & $\displaystyle 0$ & $\displaystyle \frac{1}{3 \sqrt{2}}$ & $\displaystyle -\frac{1}{3 \sqrt{2}}$ & $\displaystyle \frac{1}{3 \sqrt{2}}$ & $\displaystyle -\frac{1}{3 \sqrt{2}}$ \\
$\langle \mathcal{O}_3^{33} \rangle$ & $\displaystyle \frac{3}{\sqrt{2}}$ & $\displaystyle \frac{3}{\sqrt{2}}$ & $\displaystyle -\frac32 \sqrt{\frac{3}{2}}$ & $\displaystyle 0$ & $\displaystyle \frac{3}{2
\sqrt{2}}$ & $\displaystyle -\frac{3}{2 \sqrt{2}}$ & $\displaystyle \frac{3}{2 \sqrt{2}}$ & $\displaystyle -\frac{3}{2 \sqrt{2}}$ \\
$\langle \mathcal{O}_5^{33} \rangle$ & $\displaystyle \frac{27}{2 \sqrt{2}}$ & $\displaystyle \frac{27}{2 \sqrt{2}}$ & $\displaystyle -\frac{27}{4} \sqrt{\frac{3}{2}}$ & $\displaystyle 0$ & $\displaystyle \frac{27}{4 \sqrt{2}}$ & $\displaystyle -\frac{27}{4 \sqrt{2}}$ & $\displaystyle \frac{27}{4 \sqrt{2}}$ & $\displaystyle -\frac{27}{4 \sqrt{2}}$ \\
\hline
$\langle G^{38} \rangle$ & $\displaystyle 0$ & $\displaystyle 0$ & $\displaystyle 0$ & $\displaystyle \frac{1}{\sqrt{6}}$ & $\displaystyle \frac{1}{\sqrt{6}}$ & $\displaystyle \frac{1}{\sqrt{6}}$ & $\displaystyle \frac{1}{\sqrt{6}}$ & $\displaystyle -\frac{1}{3 \sqrt{2}}$ \\
$\langle \mathcal{O}_3^{38} \rangle$ & $\displaystyle 0$ & $\displaystyle 0$ & $\displaystyle 0$ & $\displaystyle \frac32 \sqrt{\frac{3}{2}}$ & $\displaystyle \frac32 \sqrt{\frac{3}{2}}$ & $\displaystyle \frac32 \sqrt{\frac{3}{2}}$ & $\displaystyle \frac32 \sqrt{\frac{3}{2}}$ & $\displaystyle -\frac{3}{2 \sqrt{2}}$ \\
$\langle \mathcal{O}_5^{38} \rangle$ & $\displaystyle 0$ & $\displaystyle 0$ & $\displaystyle 0$ & $\displaystyle \frac{27}{4} \sqrt{\frac{3}{2}}$ & $\displaystyle \frac{27}{4} \sqrt{\frac{3}{2}}$ & $\displaystyle \frac{27}{4} \sqrt{\frac{3}{2}}$ & $\displaystyle \frac{27}{4} \sqrt{\frac{3}{2}}$ & $\displaystyle -\frac{27}{4 \sqrt{2}}$ \\
\end{tabular}
\end{ruledtabular}
\end{table*}

\begin{subequations}
\label{eq:tlto}
\begin{eqnarray}
&  & \mu_{\Delta^{+}p}^{(0)} = \frac{\sqrt{2}}{3}(m_1 + \frac12 m_4), \\
&  & \mu_{\Delta^{0}n}^{(0)} = \frac{\sqrt{2}}{3}(m_1 + \frac12 m_4), \\
&  & \mu_{{\Sigma^{*}}^0\Lambda}^{(0)} = -\frac{1}{\sqrt{6}}(m_1 + \frac12 m_4), \\
&  & \mu_{{\Sigma^{*}}^0\Sigma^0}^{(0)} = \frac{1}{3\sqrt{2}}(m_1 + \frac12 m_4), \\
&  & \mu_{{\Sigma^{*}}^+\Sigma^+}^{(0)} = \frac{\sqrt{2}}{3}(m_1 + \frac12 m_4), \\
&  & \mu_{{\Sigma^{*}}^-\Sigma^-}^{(0)} = 0, \\
&  & \mu_{{\Xi^{*}}^0\Xi^0}^{(0)} = \frac{\sqrt{2}}{3}(m_1 + \frac12 m_4), \\
&  & \mu_{{\Xi^{*}}^-\Xi^-}^{(0)} = 0.
\end{eqnarray}
\end{subequations}

Notice that Eqs.~(\ref{eq:tlto}) can be reexpressed in terms of a single invariant $\mu_T$ \cite{jen92}, in agreement with heavy baryon chiral perturbation theory results.

On the other hand, let us notice that $\mu_{{\Sigma^{*}}\Sigma}^{(0)} = \mu_{{\Xi^{*}}\Xi}^{(0)}$, which follows from $U$-spin symmetry at this order.

Following Ref.~\cite{lb}, the isotensor combinations of transition magnetic moments for $I=2$ read
\begin{equation}
\mu_{\Delta^{+}p}^{(0)} - \mu_{\Delta^{0}n}^{(0)} = 0, \label{eq:srdo1}
\end{equation}
and
\begin{equation}
\mu_{{\Sigma^{*}}^+\Sigma^+}^{(0)} - 2 \mu_{{\Sigma^{*}}^0\Sigma^0}^{(0)} + \mu_{{\Sigma^{*}}^-\Sigma^-}^{(0)} = 0. \label{eq:srdo2}
\end{equation}

\section{One-loop corrections to baryon magnetic moments\label{sec:oneloop}}

\begin{figure}[ht]
\scalebox{0.3}{\includegraphics{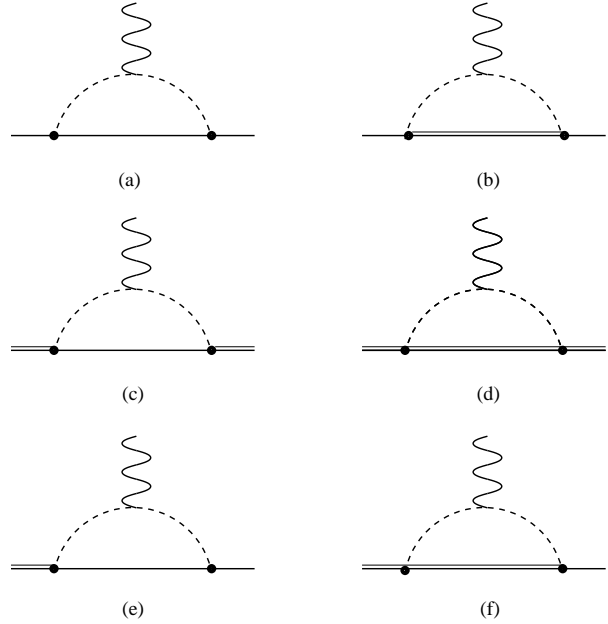}}
\caption{\label{fig:mmloop1}Feynman diagrams which yield nonanalytic $m_q^{1/2}$ corrections to baryon magnetic moments. Dashed lines denote mesons and single and double solid lines denote octet and decuplet baryons, respectively.}
\end{figure}

\begin{figure}[ht]
\scalebox{0.3}{\includegraphics{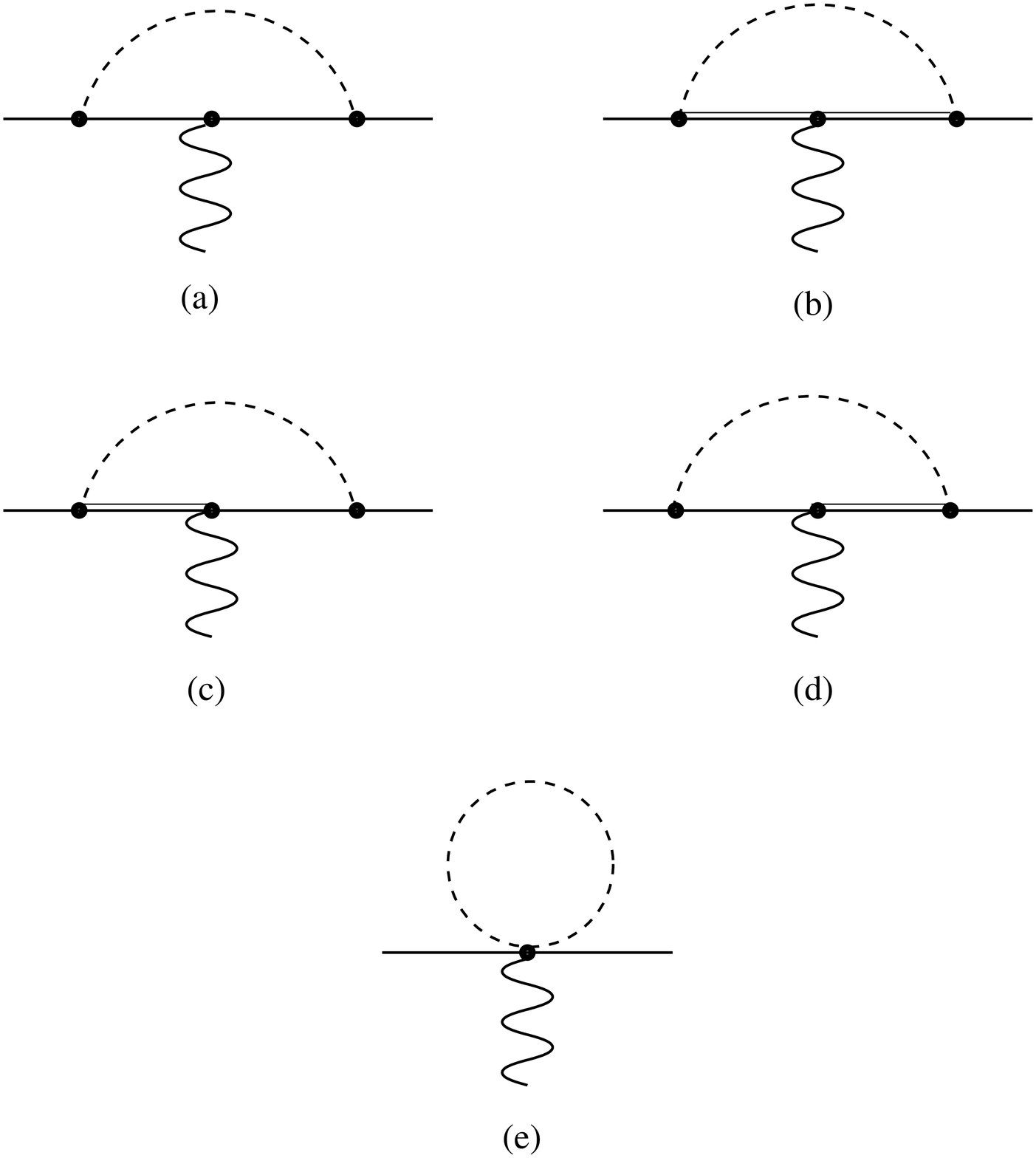}}
\caption{\label{fig:mmloop2}Feynman diagrams which yield nonanalytic $m_q \ln m_q$ corrections to the magnetic moments of octet baryons. Dashed lines denote mesons and single and double solid lines denote octet and decuplet baryons, respectively. Although the wavefunction renormalization graphs are omitted in the figure for simplicity, they are nevertheless taken into account in the analysis. For decuplet baryons and decuplet-octet transitions the diagrams are similar.}
\end{figure}

Having acquired the necessary physical motivation and mathematical tools, we are now ready to deal with baryon magnetic moments beyond tree level. The aim of this section is to provide a complete analysis of corrections to these observables at one-loop order in the framework of heavy baryon chiral perturbation theory in the large-$N_c$ limit. These corrections arise from the Feynman diagrams depicted in Figs.~\ref{fig:mmloop1} and \ref{fig:mmloop2}. The computation of the baryon operator structures involved in each diagram for finite $N_c$ (specifically for $N_c=3$) is presented in detail in this section. Computations at larger $N_c$ are less interesting physically and are by far more complicated to extrapolate to $N_c=3$ due to the participation of unphysical baryons as intermediate states in the loops \cite{weise}. We need, however, to emphasize that the approach we implement here has been simplified by working in the degeneracy limit $\Delta \equiv M_T - M_B\to 0$, where $M_T$ and $M_B$ are the SU(3) invariant masses of the decuplet and octet baryon multiplets, respectively. In the large-$N_c$ limit, although the mass of each baryon is order $\mathcal{O}(N_c)$, $\Delta$ is order $\mathcal{O}(1/N_c)$, so our assumption is a reasonable one. We now proceed to analyze each diagram separately.

\subsection{Nonanalytic corrections of order $m_q^{1/2}$}

The Feynman diagrams depicted in Fig.~\ref{fig:mmloop1} contribute to order $\mathcal{O}(m_q^{1/2})$ to baryon magnetic moments. Previous works \cite{lmr95,dai} have pointed out that this contribution should be the dominant source of SU(3) breaking. This diagram involves $\pi$ and $K$ emission and reabsorption only (the $\eta$ meson does not contribute). For degenerate heavy baryons interacting with mesons, the diagram depends on a function $I(m_\Pi)$ of the meson mass $m_\Pi$, which is obtained by performing the Feynman loop integration. Thus, in the $\Delta \to 0$ limit, this diagram can be written as\footnote{Hereafter, the label \lq\lq loop $n$" attached to a given quantity will identify the loop graph $n$ that originates it.}
\begin{equation}
M_{\textrm{loop 1}}^k = \epsilon^{ijk} A^{ia}A^{jb} \Gamma^{ab}, \label{eq:corrloop1}
\end{equation}
where we have used $A^{ia}$ and $A^{jb}$ of Eq.~(\ref{eq:akc}) at the meson-baryon vertices. Here $\Gamma^{ab}$ is an antisymmetric tensor which contains the integrals over the loops and has been discussed in detail in Ref.~\cite{dai}; therefore, we briefly give an account of its main mathematical properties.\footnote{For convenience, we denote such an antisymmetric tensor by $\Gamma^{ab}$ rather than $I^{ab}$, which is the one actually introduced in Eq.~(4.2) of Ref.~\cite{dai}.} $i\Gamma^{ab}$ can be represented by a Hermitian matrix which is diagonal in a basis corresponding to particles of definite quantum numbers. This matrix has eight eigenvalues: four of them are zero and correspond to the four neutral mesons, two of them are equal and opposite eigenvalues $\pm I(m_K)$ corresponding to $K^\pm$, respectively, and the remaining two are equal and opposite eigenvalues $\pm I(m_\pi)$ corresponding to $\pi^\pm$, respectively. Thus, $\Gamma^{ab}$ can be decomposed as \cite{dai}
\begin{equation}
\Gamma^{ab} = A_0 \Gamma_0^{ab} + A_1 \Gamma_1^{ab} + A_2 \Gamma_2^{ab},
\end{equation}
where the coefficients $A_i$ are linear combinations of the functions $I(m_\pi)$ and $I(m_K)$ and read
\begin{subequations}
\label{eq:ais}
\begin{eqnarray}
A_0 & = & \frac13 [ I(m_\pi)+2I(m_K) ], \\
A_1 & = & \frac13 [ I(m_\pi)-I(m_K) ], \\
A_2 & = & \frac{1}{\sqrt{3}}[ I(m_\pi)-I(m_K) ],
\end{eqnarray}
\end{subequations}
and the tensors $\Gamma_i^{ab}$ are written as
\begin{subequations}
\begin{eqnarray}
&  & \Gamma_0^{ab} = f^{abQ}, \\
&  & \Gamma_1^{ab} = f^{ab\overline{Q}}, \\
&  & \Gamma_2^{ab} = f^{aeQ}d^{be8} - f^{beQ}d^{ae8} - f^{abe}d^{eQ8}. \label{eq:tens}
\end{eqnarray}
\end{subequations}

Let us stress that, although $\Gamma_0^{ab}$ and $\Gamma_1^{ab}$ are both SU(3) octets, they have quite different physical interpretations. The former transforms as the electric charge whereas the latter also transforms as the electric charge but rotated by $\pi$ in isospin space \cite{dai}. This can be better seen by considering
\begin{eqnarray}
T^Q & = & T^3 + \frac{1}{\sqrt{3}}T^8 = \left( \begin{array}{ccc} 2/3 & 0 & 0 \\ 0 & -1/3 & 0 \\ 0 & 0 & -1/3 \end{array}  \right), \nonumber \\
T^{\overline{Q}} & = & T^3 - \frac{1}{\sqrt{3}}T^8 = \left( \begin{array}{ccc} 1/3 & 0 & 0 \\ 0 & -2/3 & 0 \\ 0 & 0 & 1/3 \end{array} \right).
\end{eqnarray}
In what follows any operator of the form $X^{\overline{Q}}$ should be understood as $X^3-(1/\sqrt{3})X^8$. One also should keep in mind than $X^Q$ and $X^{\overline{Q}}$ fall into different octet representations. On the other hand, $\Gamma_2^{ab}$ breaks SU(3) as $\mathbf{10}+\overline{\mathbf{10}}$ \cite{dai}.

In the degeneracy limit $\Delta \to 0$ and retaining only the nonanalytic pieces in $m_q$, the integral over the loop, which comprises the proper factors to give the correct dimensions, can be expressed as \cite{jen92}
\begin{equation}
I(m_\Pi) = \frac{1}{8\pi f^2} M_N m_\Pi, \label{eq:iloop}
\end{equation}
where $f\sim 93$ MeV is the pion decay constant and $M_N$ and $m_\Pi$ denote the nucleon and the meson masses, respectively. When $\Delta$ is not neglected, the resulting function can be found in Eq.~(28) of Ref.~\cite{jen92} and will not be repeated here.

Thus, the one-loop correction arising from Fig.~\ref{fig:mmloop1} can be decomposed into the pieces emerging from the $\mathbf{8}$ and $\mathbf{10}+\overline{\mathbf{10}}$ representations as follows,
\begin{equation}
M_{\textrm{loop 1}}^k = A_0 M_{\mathbf{8},\textrm{loop 1}}^{kQ} + A_1 M_{\mathbf{8},\textrm{loop 1}}^{k\overline{Q}} + A_2 M_{\mathbf{10}+\overline{\mathbf{10}},\textrm{loop 1}}^{kQ}, \label{eq:loop1}
\end{equation}
where the different contributions read
\begin{equation}
M_{\mathbf{8},\textrm{loop 1}}^{kc} = \epsilon^{ijk} f^{abc} A^{ia}A^{jb}, \label{eq:m8l1}
\end{equation}
and
\begin{equation}
M_{\mathbf{10}+\overline{\mathbf{10}},\textrm{loop 1}}^{kc} = \epsilon^{ijk}(f^{aec}d^{be8} - f^{bec}d^{ae8} - f^{abe}d^{ec8})A^{ia}A^{jb}. \label{eq:m10l1}
\end{equation}
For computational purposes, a free flavor index $c$ has been left in Eqs.~(\ref{eq:m8l1}) and (\ref{eq:m10l1}). This free index can be set to $Q=3+(1/\sqrt{3})8$ once the operator reductions on the right-hand sides of such equations have been performed.

Now, in the product operators such as $\epsilon^{ijk} f^{abc} A^{ia}A^{jb}$, $\epsilon^{ijk} f^{abe}d^{ec8} A^{ia}A^{jb}$ and so on found in Eqs.~(\ref{eq:m8l1}) and (\ref{eq:m10l1}), there will appear up to six-body operators if we truncate the $1/N_c$ expansion of $A^{kc}$ at the physical value $N_c=3$. The leading order in $1/N_c$ is contained in the product $\epsilon^{ijk} f^{abc} G^{ia}G^{jb}$ and similar terms with two $G$'s, which will be proportional to the square of $a_1$, which is the leading parameter introduced in Eq.~(\ref{eq:akc}). The analysis of Ref.~\cite{lmr95} is given to this order. In the present work we will proceed to compute not only leading but also subleading order terms. Because of the fact that the operator basis is complete \cite{djm95}, the reduction is always possible. We, however, consider pertinent to work out terms up to relative order $\mathcal{O}(1/N_c^3)$, which implies evaluating products up to five-body operators in Eqs.~(\ref{eq:m8l1}) and (\ref{eq:m10l1}). The contributions ignored will be proportional to $b_3^2$, $c_3^2$, and $b_3c_3$, which we consider small compared to the ones retained.

The reduction of the operator products contained in $M_{\mathbf{8}}^{kc}$ and $M_{\mathbf{10}+\overline{\mathbf{10}}}^{kc}$ to the order considered here are listed in Appendix \ref{app:bo1}. Gathering together partial results we find:

\begin{widetext}

(1) \textit{Flavor octet contribution}

\begin{eqnarray}
M_{\mathbf{8},\textrm{loop 1}}^{kc} & = & \left[ -\frac{N_c+3}{2} a_1^2 - \frac{3}{N_c} a_1 b_2 - \frac{2(N_c+3)}{N_c^2} a_1b_3 \right] G^{kc} + \frac12 \left[ a_1^2 - \frac{1}{N_c^2} ( 2a_1b_3 - 9a_1c_3 + 3b_2^2 ) \right] \mathcal{D}_2^{kc} \nonumber \\
&  & \mbox{} - \frac{1}{N_c^2} \left[ \frac{N_c+3}{2} a_1c_3 + \frac{3}{N_c} b_2b_3 \right] \mathcal{D}_3^{kc}
- \frac{1}{N_c^2} \left[N_c a_1b_2 + (N_c+3)a_1b_3 + \frac{N_c+3}{2} a_1c_3 + \frac{3}{N_c}b_2c_3 \right] \mathcal{O}_3^{kc} \nonumber \\
&  & \mbox{} + \frac{1}{N_c^2} a_1c_3 \mathcal{D}_4^{kc} - \frac{1}{N_c^3}b_2c_3 \mathcal{O}_5^{kc} + \mathcal{O}(\mathcal{D}_3\mathcal{D}_3). \label{eq:loop1Ared}
\end{eqnarray}

(2) \textit{Flavor $\mathbf{10}+\overline{\mathbf{10}}$ contribution}

\begin{eqnarray}
M_{\mathbf{10}+\overline{\mathbf{10}},\textrm{loop 1}}^{kc} & = & \left[ \frac12 a_1^2 + \frac{2}{N_c^2} a_1b_3 \right] \left( \{T^c,G^{k8}\} - \{G^{kc},T^8\} \right) - \frac{1}{N_c} a_1b_2 \left( \{G^{kc},\{J^r,G^{r8}\}\} - \{G^{k8},\{J^r,G^{rc}\}\} \right) \nonumber \\
&  & \mbox{} - \frac{1}{2N_c^2} \left( 2a_1b_3 - a_1c_3 \right) \left( \{\mathcal{D}_2^{kc},\{J^r,G^{r8}\}\} - \{\mathcal{D}_2^{k8},\{J^r,G^{rc}\}\} \right) - \frac{1}{2N_c^2} \left( 2a_1b_3 + a_1c_3 \right)  \nonumber \\
&  & \mbox{} \times \left( \{J^2,\{G^{kc},T^8\}\} - \{J^2,\{G^{k8},T^c\}\} \right) - \frac{3}{8N_c^3} b_2c_3 \left( \{J^2,[G^{kc},\{J^r,G^{r8}\}]\} - \{J^2,[G^{k8},\{J^r,G^{rc}\}]\} \right) \nonumber \\
&  & \mbox{} - \frac{3}{8N_c^3} b_2c_3 \left(\{[J^2,G^{kc}],\{J^r,G^{r8}\}\} - \{[J^2,G^{k8}],\{J^r,G^{rc}\}\} \right) + \frac{3}{8N_c^3} b_2c_3 \{J^k,[\{J^m,G^{mc}\},\{J^r,G^{r8}\}]\} \nonumber \\
&  & \mbox{} + \frac{1}{N_c^3} b_2c_3 \left(  \{J^2,\{G^{k8},\{J^r,G^{rc}\}\}\} - \{J^2,\{G^{kc},\{J^r,G^{r8}\}\}\} \right) + \mathcal{O}(\mathcal{D}_3\mathcal{D}_3). \label{eq:loop1Bred}
\end{eqnarray}

\end{widetext}
where the free flavor index $c$ will be set to $Q=3+(1/\sqrt{3})8$ or $\overline{Q}=3-(1/\sqrt{3})8$ as required in Eq.~(\ref{eq:loop1}). The symbol $\mathcal{O}(\mathcal{D}_3\mathcal{D}_3)$ in Eqs.~(\ref{eq:loop1Ared}) and (\ref{eq:loop1Bred}) means that, in the structures such as $\epsilon^{ijk}f^{abc}A^{ia}A^{jb}$, $\epsilon^{ijk} f^{aec}d^{be8}A^{ia}A^{jb}$ and so on we have included all terms up to five-body operators, such as $\mathcal{D}_2\mathcal{D}_3$, but have neglected contributions which are six-body operators -- like $\mathcal{D}_3\mathcal{D}_3$ -- or higher.

Equations (\ref{eq:loop1Ared}) and (\ref{eq:loop1Bred}) have been rearranged to exhibit explicitly leading and subleading terms in $1/N_c$. It is now evident that both expressions yield matrix elements at most of order $\mathcal{O}(N_c^2)$, according to the $N_c$ dependence of matrix elements of baryon operators discussed in Sec.~\ref{sec:overview}. In addition, $f$ and $M_N$, which appear in the loop-integral (\ref{eq:iloop}), are $\mathcal{O}(\sqrt{N_c})$ and $\mathcal{O}(N_c)$, respectively, so the one-loop contribution $M_{\textrm{loop 1}}^k$, Eq.~(\ref{eq:loop1}), is order $\mathcal{O}(N_c)$. In the limit of small $m_s$, the symmetry breaking part of $M_{\textrm{loop 1}}^k$ is $\mathcal{O}(m_s^{1/2})$ so the overall contribution of Eq.~(\ref{eq:loop1}) to baryon magnetic moments is $\mathcal{O}(m_s^{1/2}N_c)$.

In order to proceed further, we still need to evaluate the matrix elements of the operators in Eqs.~(\ref{eq:loop1Ared}) and (\ref{eq:loop1Bred}). To relative order $\mathcal{O}(1/N_c^3)$, we have identified 24 linearly independent spin-1 operators which fall into the $\mathbf{8}$ and $\mathbf{10}+\overline{\mathbf{10}}$ flavor representations. These basic operators are

\begin{widetext}

\begin{eqnarray}
\begin{tabular}{lll}
$Y_1^{kc} = d^{c8e} G^{ke}$, &
$Y_2^{kc} = \delta^{c8} J^k$, &
$Y_3^{kc} = d^{c8e} \mathcal{D}_2^{ke}$, \nonumber \\
$Y_4^{kc} = \{G^{kc},T^8\}$, &
$Y_5^{kc} = \{T^c,G^{k8}\}$, &
$Y_6^{kc} = d^{c8e} \mathcal{D}_3^{ke}$, \nonumber \\
$Y_7^{kc} = d^{c8e} \mathcal{O}_3^{ke}$, &
$Y_8^{kc} = \{G^{kc},\{J^r,G^{r8}\}\}$, &
$Y_9^{kc} = \{G^{k8},\{J^r,G^{rc}\}\}$, \nonumber \\
$Y_{10}^{kc} =\{J^k,\{T^c,T^8\}\}$, &
$Y_{11}^{kc} = \{J^k,\{G^{rc},G^{r8}\}\}$, &
$Y_{12}^{kc} = \delta^{c8} \{J^2,J^k\}$, \nonumber \\
$Y_{13}^{kc} = d^{c8e} \mathcal{D}_4^{ke}$, &
$Y_{14}^{kc} = \{\mathcal{D}_2^{kc},\{J^r,G^{r8}\}\}$, &
$Y_{15}^{kc} = \{J^2,\{G^{kc},T^8\}\}$, \nonumber \\
$Y_{16}^{kc} = \{J^2,\{T^c,G^{k8}\}\}$, &
$Y_{17}^{kc} = \{\mathcal{D}_2^{k8},\{J^r,G^{rc}\}\}$, &
$Y_{18}^{kc} = \{J^2,[G^{kc},\{J^r,G^{r8}\}]\}$, \nonumber \\
$Y_{19}^{kc} = \{J^2,[G^{k8},\{J^r,G^{rc}\}]\}$, &
$Y_{20}^{kc} = \{[J^2,G^{k8}],\{J^r,G^{rc}\}\}$, &
$Y_{21}^{kc} = \{[J^2,G^{kc}],\{J^r,G^{r8}\}\}$, \nonumber \\
$Y_{22}^{kc} = \{J^k,[\{J^m,G^{mc}\},\{J^r,G^{r8}\}]\}$, &
$Y_{23}^{kc} = \{J^2,\{G^{kc},\{J^r,G^{r8}\}\}\}$, &
$Y_{24}^{kc} = \{J^2,\{G^{k8},\{J^r,G^{rc}\}\}\}$,
\end{tabular}
\end{eqnarray}

\end{widetext}
whose matrix elements are displayed in Tables \ref{t:mm833O}-\ref{t:mm833TO} for the sake of completeness. In these tables we have kept only nontrivial contributions. For instance, $Y_7^{kc}=d^{c8e} \mathcal{O}_3^{ke}$ is an off-diagonal operator with nonzero matrix elements only for decuplet to octet transitions, and will vanish otherwise.

\begin{table*}
\caption{\label{t:mm833O}Nontrivial matrix elements of the operators involved in the magnetic moments of octet baryons: Flavor $\mathbf{8}$ and $\mathbf{10}+\overline{\mathbf{10}}$ representations. The entries correspond to $48\sqrt{3}\langle Y_{m}^{33}\rangle$ and $48\langle Y_{m}^{38} \rangle$.}
\begin{ruledtabular}
\begin{tabular}{lccccccccc}
& $\displaystyle n$ & $\displaystyle p$ & $\displaystyle \Sigma^-$ & $\displaystyle \Sigma^0$ & $\displaystyle \Sigma^+$ & $\displaystyle \Xi^-$ & $\displaystyle \Xi^0$ & $\displaystyle \Lambda$ & $\displaystyle \Lambda\Sigma^0$ \\[2mm]
\hline
$\langle Y_{1}^{33} \rangle$ & $-20$ & $20$ & $-16$ & $0$ & $16$ & $4$ & $-4$ & $0$ & $8 \sqrt{3}$ \\
$\langle Y_{2}^{33} \rangle$ & $0$ & $0$ & $0$ & $0$ & $0$ & $0$ & $0$ & $0$ & $0$ \\
$\langle Y_{3}^{33} \rangle$ & $-12$ & $12$ & $-24$ & $0$ & $24$ & $-12$ & $12$ & $0$ & $0$ \\
$\langle Y_{4}^{33} \rangle$ & $-60$ & $60$ & $0$ & $0$ & $0$ & $-12$ & $12$ & $0$ & $0$ \\
$\langle Y_{5}^{33} \rangle$ & $-12$ & $12$ & $-48$ & $0$ & $48$ & $36$ & $-36$ & $0$ & $0$ \\
$\langle Y_{6}^{33} \rangle$ & $-60$ & $60$ & $-48$ & $0$ & $48$ & $12$ & $-12$ & $0$ & $24 \sqrt{3}$ \\
$\langle Y_{8}^{33} \rangle$ & $-30$ & $30$ & $-48$ & $0$ & $48$ & $-18$ & $18$ & $0$ & $0$ \\
$\langle Y_{9}^{33} \rangle$ & $-30$ & $30$ & $-48$ & $0$ & $48$ & $-18$ & $18$ & $0$ & $0$ \\
$\langle Y_{10}^{33} \rangle$ & $-72$ & $72$ & $0$ & $0$ & $0$ & $72$ & $-72$ & $0$ & $0$ \\
$\langle Y_{11}^{33} \rangle$ & $-30$ & $30$ & $-96$ & $0$ & $96$ & $-66$ & $66$ & $0$ & $-24 \sqrt{3}$ \\
$\langle Y_{12}^{33} \rangle$ & $0$ & $0$ & $0$ & $0$ & $0$ & $0$ & $0$ & $0$ & $0$ \\
$\langle Y_{13}^{33} \rangle$ & $-18$ & $18$ & $-36$ & $0$ & $36$ & $-18$ & $18$ & $0$ & $0$ \\
$\langle Y_{14}^{33} \rangle$ & $-18$ & $18$ & $-72$ & $0$ & $72$ & $54$ & $-54$ & $0$ & $0$ \\
$\langle Y_{15}^{33} \rangle$ & $-90$ & $90$ & $0$ & $0$ & $0$ & $-18$ & $18$ & $0$ & $0$ \\
$\langle Y_{16}^{33} \rangle$ & $-18$ & $18$ & $-72$ & $0$ & $72$ & $54$ & $-54$ & $0$ & $0$ \\
$\langle Y_{17}^{33} \rangle$ & $-90$ & $90$ & $0$ & $0$ & $0$ & $-18$ & $18$ & $0$ & $0$ \\
$\langle Y_{18}^{33} \rangle$ & $0$ & $0$ & $0$ & $0$ & $0$ & $0$ & $0$ & $0$ & $36 \sqrt{3}$ \\
$\langle Y_{19}^{33} \rangle$ & $0$ & $0$ & $0$ & $0$ & $0$ & $0$ & $0$ & $0$ & $-36 \sqrt{3}$ \\
$\langle Y_{22}^{33} \rangle$ & $0$ & $0$ & $0$ & $0$ & $0$ & $0$ & $0$ & $0$ & $72 \sqrt{3}$ \\
$\langle Y_{23}^{33} \rangle$ & $-45$ & $45$ & $-72$ & $0$ & $72$ & $-27$ & $27$ & $0$ & $0$ \\
$\langle Y_{24}^{33} \rangle$ & $-45$ & $45$ & $-72$ & $0$ & $72$ & $-27$ & $27$ & $0$ & $0$ \\
\hline
$\langle Y_{1}^{38} \rangle$ & $-4$ & $-4$ & $-8$ & $-8$ & $-8$ & $12$ & $12$ & $8$ & $0$ \\
$\langle Y_{2}^{38} \rangle$ & $24$ & $24$ & $24$ & $24$ & $24$ & $24$ & $24$ & $24$ & $0$ \\
$\langle Y_{3}^{38} \rangle$ & $-12$ & $-12$ & $0$ & $0$ & $0$ & $12$ & $12$ & $0$ & $0$ \\
$\langle Y_{4}^{38} \rangle$ & $12$ & $12$ & $0$ & $0$ & $0$ & $36$ & $36$ & $0$ & $0$ \\
$\langle Y_{5}^{38} \rangle$ & $12$ & $12$ & $0$ & $0$ & $0$ & $36$ & $36$ & $0$ & $0$ \\
$\langle Y_{6}^{38} \rangle$ & $-12$ & $-12$ & $-24$ & $-24$ & $-24$ & $36$ & $36$ & $24$ & $0$ \\
$\langle Y_{8}^{38} \rangle$ & $6$ & $6$ & $24$ & $24$ & $24$ & $54$ & $54$ & $24$ & $0$ \\
$\langle Y_{9}^{38} \rangle$ & $6$ & $6$ & $24$ & $24$ & $24$ & $54$ & $54$ & $24$ & $0$ \\
$\langle Y_{10}^{38} \rangle$ & $72$ & $72$ & $0$ & $0$ & $0$ & $72$ & $72$ & $0$ & $0$ \\
$\langle Y_{11}^{38} \rangle$ & $6$ & $6$ & $72$ & $72$ & $72$ & $102$ & $102$ & $24$ & $0$ \\
$\langle Y_{12}^{38} \rangle$ & $36$ & $36$ & $36$ & $36$ & $36$ & $36$ & $36$ & $36$ & $0$ \\
$\langle Y_{13}^{38} \rangle$ & $-18$ & $-18$ & $0$ & $0$ & $0$ & $18$ & $18$ & $0$ & $0$ \\
$\langle Y_{14}^{38} \rangle$ & $18$ & $18$ & $0$ & $0$ & $0$ & $54$ & $54$ & $0$ & $0$ \\
$\langle Y_{15}^{38} \rangle$ & $18$ & $18$ & $0$ & $0$ & $0$ & $54$ & $54$ & $0$ & $0$ \\
$\langle Y_{16}^{38} \rangle$ & $18$ & $18$ & $0$ & $0$ & $0$ & $54$ & $54$ & $0$ & $0$ \\
$\langle Y_{17}^{38} \rangle$ & $18$ & $18$ & $0$ & $0$ & $0$ & $54$ & $54$ & $0$ & $0$ \\
$\langle Y_{18}^{38} \rangle$ & $0$ & $0$ & $0$ & $0$ & $0$ & $0$ & $0$ & $0$ & $0$ \\
$\langle Y_{19}^{38} \rangle$ & $0$ & $0$ & $0$ & $0$ & $0$ & $0$ & $0$ & $0$ & $0$ \\
$\langle Y_{22}^{38} \rangle$ & $0$ & $0$ & $0$ & $0$ & $0$ & $0$ & $0$ & $0$ & $0$ \\
$\langle Y_{23}^{38} \rangle$ & $9$ & $9$ & $36$ & $36$ & $36$ & $81$ & $81$ & $36$ & $0$ \\
$\langle Y_{24}^{38} \rangle$ & $9$ & $9$ & $36$ & $36$ & $36$ & $81$ & $81$ & $36$ & $0$ \\
\end{tabular}
\end{ruledtabular}
\end{table*}

\begin{table*}
\caption{\label{t:mm833T}Nontrivial matrix elements of the operators involved in the magnetic moments of decuplet baryons: Flavor $\mathbf{8}$ and $\mathbf{10}+\overline{\mathbf{10}}$ representations. The entries correspond to $16\sqrt{3}\langle Y_{m}^{33}\rangle$ and $16\langle Y_{m}^{38} \rangle$.}
\begin{ruledtabular}
\begin{tabular}{lcccccccccc}
& $\displaystyle \Delta^{++}$ & $\displaystyle \Delta^+$ & $\displaystyle \Delta^0$ & $\displaystyle \Delta^-$ & $\displaystyle {\Sigma^*}^+$ & $\displaystyle {\Sigma^*}^0$ & $\displaystyle {\Sigma^*}^-$ & $\displaystyle {\Xi^*}^0$ & $\displaystyle {\Xi^*}^-$ & $\displaystyle \Omega^-$ \\[2mm]
\hline
$\langle Y_{1}^{33} \rangle$ & $12$ & $4$ & $-4$ & $-12$ & $8$ & $0$ & $-8$ & $4$ & $-4$ & $0$ \\
$\langle Y_{2}^{33} \rangle$ & $0$ & $0$ & $0$ & $0$ & $0$ & $0$ & $0$ & $0$ & $0$ & $0$ \\
$\langle Y_{3}^{33} \rangle$ & $36$ & $12$ & $-12$ & $-36$ & $24$ & $0$ & $-24$ & $12$ & $-12$ & $0$ \\
$\langle Y_{4}^{33} \rangle$ & $36$ & $12$ & $-12$ & $-36$ & $0$ & $0$ & $0$ & $-12$ & $12$ & $0$ \\
$\langle Y_{5}^{33} \rangle$ & $36$ & $12$ & $-12$ & $-36$ & $0$ & $0$ & $0$ & $-12$ & $12$ & $0$ \\
$\langle Y_{6}^{33} \rangle$ & $180$ & $60$ & $-60$ & $-180$ & $120$ & $0$ & $-120$ & $60$ & $-60$ & $0$ \\
$\langle Y_{8}^{33} \rangle$ & $90$ & $30$ & $-30$ & $-90$ & $0$ & $0$ & $0$ & $-30$ & $30$ & $0$ \\
$\langle Y_{9}^{33} \rangle$ & $90$ & $30$ & $-30$ & $-90$ & $0$ & $0$ & $0$ & $-30$ & $30$ & $0$ \\
$\langle Y_{10}^{33} \rangle$ & $216$ & $72$ & $-72$ & $-216$ & $0$ & $0$ & $0$ & $-72$ & $72$ & $0$ \\
$\langle Y_{11}^{33} \rangle$ & $90$ & $30$ & $-30$ & $-90$ & $24$ & $0$ & $-24$ & $-6$ & $6$ & $0$ \\
$\langle Y_{12}^{33} \rangle$ & $0$ & $0$ & $0$ & $0$ & $0$ & $0$ & $0$ & $0$ & $0$ & $0$ \\
$\langle Y_{13}^{33} \rangle$ & $270$ & $90$ & $-90$ & $-270$ & $180$ & $0$ & $-180$ & $90$ & $-90$ & $0$ \\
$\langle Y_{14}^{33} \rangle$ & $270$ & $90$ & $-90$ & $-270$ & $0$ & $0$ & $0$ & $-90$ & $90$ & $0$ \\
$\langle Y_{15}^{33} \rangle$ & $270$ & $90$ & $-90$ & $-270$ & $0$ & $0$ & $0$ & $-90$ & $90$ & $0$ \\
$\langle Y_{16}^{33} \rangle$ & $270$ & $90$ & $-90$ & $-270$ & $0$ & $0$ & $0$ & $-90$ & $90$ & $0$ \\
$\langle Y_{17}^{33} \rangle$ & $270$ & $90$ & $-90$ & $-270$ & $0$ & $0$ & $0$ & $-90$ & $90$ & $0$ \\
$\langle Y_{23}^{33} \rangle$ & $675$ & $225$ & $-225$ & $-675$ & $0$ & $0$ & $0$ & $-225$ & $225$ & $0$ \\
$\langle Y_{24}^{33} \rangle$ & $675$ & $225$ & $-225$ & $-675$ & $0$ & $0$ & $0$ & $-225$ & $225$ & $0$ \\
\hline
$\langle Y_{1}^{38} \rangle$ & $-4$ & $-4$ & $-4$ & $-4$ & $0$ & $0$ & $0$ & $4$ & $4$ & $8$ \\
$\langle Y_{2}^{38} \rangle$ & $24$ & $24$ & $24$ & $24$ & $24$ & $24$ & $24$ & $24$ & $24$ & $24$ \\
$\langle Y_{3}^{38} \rangle$ & $-12$ & $-12$ & $-12$ & $-12$ & $0$ & $0$ & $0$ & $12$ & $12$ & $24$ \\
$\langle Y_{4}^{38} \rangle$ & $12$ & $12$ & $12$ & $12$ & $0$ & $0$ & $0$ & $12$ & $12$ & $48$ \\
$\langle Y_{5}^{38} \rangle$ & $12$ & $12$ & $12$ & $12$ & $0$ & $0$ & $0$ & $12$ & $12$ & $48$ \\
$\langle Y_{6}^{38} \rangle$ & $-60$ & $-60$ & $-60$ & $-60$ & $0$ & $0$ & $0$ & $60$ & $60$ & $120$ \\
$\langle Y_{8}^{38} \rangle$ & $30$ & $30$ & $30$ & $30$ & $0$ & $0$ & $0$ & $30$ & $30$ & $120$ \\
$\langle Y_{9}^{38} \rangle$ & $30$ & $30$ & $30$ & $30$ & $0$ & $0$ & $0$ & $30$ & $30$ & $120$ \\
$\langle Y_{10}^{38} \rangle$ & $72$ & $72$ & $72$ & $72$ & $0$ & $0$ & $0$ & $72$ & $72$ & $288$ \\
$\langle Y_{11}^{38} \rangle$ & $30$ & $30$ & $30$ & $30$ & $24$ & $24$ & $24$ & $54$ & $54$ & $120$ \\
$\langle Y_{12}^{38} \rangle$ & $180$ & $180$ & $180$ & $180$ & $180$ & $180$ & $180$ & $180$ & $180$ & $180$ \\
$\langle Y_{13}^{38} \rangle$ & $-90$ & $-90$ & $-90$ & $-90$ & $0$ & $0$ & $0$ & $90$ & $90$ & $180$ \\
$\langle Y_{14}^{38} \rangle$ & $90$ & $90$ & $90$ & $90$ & $0$ & $0$ & $0$ & $90$ & $90$ & $360$ \\
$\langle Y_{15}^{38} \rangle$ & $90$ & $90$ & $90$ & $90$ & $0$ & $0$ & $0$ & $90$ & $90$ & $360$ \\
$\langle Y_{16}^{38} \rangle$ & $90$ & $90$ & $90$ & $90$ & $0$ & $0$ & $0$ & $90$ & $90$ & $360$ \\
$\langle Y_{17}^{38} \rangle$ & $90$ & $90$ & $90$ & $90$ & $0$ & $0$ & $0$ & $90$ & $90$ & $360$ \\
$\langle Y_{23}^{38} \rangle$ & $225$ & $225$ & $225$ & $225$ & $0$ & $0$ & $0$ & $225$ & $225$ & $900$ \\
$\langle Y_{24}^{38} \rangle$ & $225$ & $225$ & $225$ & $225$ & $0$ & $0$ & $0$ & $225$ & $225$ & $900$ \\
\end{tabular}
\end{ruledtabular}
\end{table*}

\begin{table*}
\caption{\label{t:mm833TO}Nontrivial matrix elements of the operators involved in the decuplet to octet transition magnetic moments: Flavor $\mathbf{8}$ and $\mathbf{10}+\overline{\mathbf{10}}$ representations. The entries correspond to $12\sqrt{6}\langle Y_{m}^{33}\rangle$ and $12\sqrt{2}\langle Y_{m}^{38} \rangle$.}
\begin{ruledtabular}
\begin{tabular}{lcccccccc}
& $\displaystyle \Delta^+p$ & $\displaystyle \Delta^0n$ & $\displaystyle {\Sigma^*}^0\Lambda$ & $\displaystyle {\Sigma^*}^0\Sigma^0$ & $\displaystyle {\Sigma^*}^+\Sigma^+$ & $\displaystyle {\Sigma^*}^-\Sigma^-$ & $\displaystyle {\Xi^*}^0\Xi^0$ & $\displaystyle {\Xi^*}^-\Xi^-$ \\[2mm]
\hline
$\langle Y_{1}^{33} \rangle$ & $8$ & $8$ & $-4 \sqrt{3}$ & $0$ & $4$ & $-4$ & $4$ & $-4$ \\
$\langle Y_{4}^{33} \rangle$ & $24$ & $24$ & $0$ & $0$ & $0$ & $0$ & $-12$ & $12$ \\
$\langle Y_{5}^{33} \rangle$ & $0$ & $0$ & $0$ & $0$ & $24$ & $-24$ & $12$ & $-12$ \\
$\langle Y_{7}^{33} \rangle$ & $36$ & $36$ & $-18 \sqrt{3}$ & $0$ & $18$ & $-18$ & $18$ & $-18$ \\
$\langle Y_{8}^{33} \rangle$ & $36$ & $36$ & $6 \sqrt{3}$ & $0$ & $6$ & $-6$ & $-24$ & $24$ \\
$\langle Y_{9}^{33} \rangle$ & $0$ & $0$ & $6 \sqrt{3}$ & $0$ & $42$ & $-42$ & $12$ & $-12$ \\
$\langle Y_{15}^{33} \rangle$ & $108$ & $108$ & $0$ & $0$ & $0$ & $0$ & $-54$ & $54$ \\
$\langle Y_{16}^{33} \rangle$ & $0$ & $0$ & $0$ & $0$ & $108$ & $-108$ & $54$ & $-54$ \\
$\langle Y_{18}^{33} \rangle$ & $-108$ & $-108$ & $0$ & $0$ & $27$ & $-27$ & $27$ & $-27$ \\
$\langle Y_{19}^{33} \rangle$ & $0$ & $0$ & $0$ & $0$ & $-81$ & $81$ & $-81$ & $81$ \\
$\langle Y_{20}^{33} \rangle$ & $0$ & $0$ & $0$ & $0$ & $126$ & $-126$ & $36$ & $-36$ \\
$\langle Y_{21}^{33} \rangle$ & $108$ & $108$ & $0$ & $0$ & $18$ & $-18$ & $-72$ & $72$ \\
$\langle Y_{23}^{33} \rangle$ & $162$ & $162$ & $27 \sqrt{3}$ & $0$ & $27$ & $-27$ & $-108$ & $108$ \\
$\langle Y_{24}^{33} \rangle$ & $0$ & $0$ & $27 \sqrt{3}$ & $0$ & $189$ & $-189$ & $54$ & $-54$ \\
\hline
$\langle Y_{1}^{38} \rangle$ & $0$ & $0$ & $0$ & $-4$ & $-4$ & $-4$ & $-4$ & $-4$ \\
$\langle Y_{4}^{38} \rangle$ & $0$ & $0$ & $0$ & $0$ & $0$ & $0$ & $-12$ & $-12$ \\
$\langle Y_{5}^{38} \rangle$ & $0$ & $0$ & $0$ & $0$ & $0$ & $0$ & $-12$ & $-12$ \\
$\langle Y_{7}^{38} \rangle$ & $0$ & $0$ & $0$ & $-18$ & $-18$ & $-18$ & $-18$ & $-18$ \\
$\langle Y_{8}^{38} \rangle$ & $0$ & $0$ & $0$ & $6$ & $6$ & $6$ & $-24$ & $-24$ \\
$\langle Y_{9}^{38} \rangle$ & $0$ & $0$ & $0$ & $6$ & $6$ & $6$ & $-24$ & $-24$ \\
$\langle Y_{15}^{38} \rangle$ & $0$ & $0$ & $0$ & $0$ & $0$ & $0$ & $-54$ & $-54$ \\
$\langle Y_{16}^{38} \rangle$ & $0$ & $0$ & $0$ & $0$ & $0$ & $0$ & $-54$ & $-54$ \\
$\langle Y_{18}^{38} \rangle$ & $0$ & $0$ & $0$ & $27$ & $27$ & $27$ & $27$ & $27$ \\
$\langle Y_{19}^{38} \rangle$ & $0$ & $0$ & $0$ & $27$ & $27$ & $27$ & $27$ & $27$ \\
$\langle Y_{20}^{38} \rangle$ & $0$ & $0$ & $0$ & $18$ & $18$ & $18$ & $-72$ & $-72$ \\
$\langle Y_{21}^{38} \rangle$ & $0$ & $0$ & $0$ & $18$ & $18$ & $18$ & $-72$ & $-72$ \\
$\langle Y_{23}^{38} \rangle$ & $0$ & $0$ & $0$ & $27$ & $27$ & $27$ & $-108$ & $-108$ \\
$\langle Y_{24}^{38} \rangle$ & $0$ & $0$ & $0$ & $27$ & $27$ & $27$ & $-108$ & $-108$ \\
\end{tabular}
\end{ruledtabular}
\end{table*}

At this point we are able to compute the one-loop contribution of Fig.~\ref{fig:mmloop1} and provide analytical expressions. Such a contribution is given by
\begin{equation}
\mu_B^{(\textrm{loop 1})} = \langle B|M_{\textrm{loop 1}}^3 |B \rangle, \label{eq:mmloop1gr}
\end{equation}
where $B$ stands for either an octet or a decuplet baryon. In the former case the correction arises from Fig.~\ref{fig:mmloop1}(a,b) whereas for the latter comes from Fig.~\ref{fig:mmloop1}(c,d). Furthermore, the contribution to the decuplet to octet transition magnetic moment, Fig.~\ref{fig:mmloop1}(e,f), can be obtained as
\begin{equation}
\mu_{TB}^{(\textrm{loop 1})} = \langle T|M_{\textrm{loop 1}}^3 |B \rangle. \label{eq:mmloop1trans}
\end{equation}

Analytical expressions can be readily found by reading off the matrix elements of the pertinent operators from Table \ref{t:mm133B}. We just show the one corresponding to the neutron as a case example because the others can be obtained analogously. We thus have,
\begin{widetext}
\begin{eqnarray}
\mu_n^{\textrm{(loop 1)}} & = & \left[ a_1^2 + \frac13 a_1b_2 + \frac49 a_1b_3 + \frac19 b_2b_3 + \frac13 a_1c_3\right] A_0 + \left[ \frac54a_1^2 + \frac12 a_1b_2 + \frac{1}{12} b_2^2 + \frac{13}{18} a_1b_3 + \frac16 b_2b_3 + \frac16 a_1c_3\right] A_1 \nonumber \\
&  & \mbox{} + \frac{1}{\sqrt{3}} \left[ \frac12 a_1^2 + \frac29 a_1b_3 + \frac16 a_1c_3\right] A_2,
\end{eqnarray}

\end{widetext}
where the different coefficients $A_i$, which contain the integrals over the loops, are given in Eq.~(\ref{eq:ais}).

Corrections of order $\mathcal{O}(m_q^{1/2}N_c)$ have some important effects on the relations among the magnetic moments referred to in Sec.~\ref{sec:tree}. First, the term that comes along with $A_0$, $M_{\mathbf{8},\textrm{loop 1}}^{kQ}$ in Eq.~(\ref{eq:loop1}), yields baryon magnetic moments that satisfy the Coleman-Glashow relations (\ref{eq:treeval}) whereas violations to them are due to the terms that accompany to $A_1$ and $A_2$, which are $M_{\mathbf{8},\textrm{loop 1}}^{k\overline{Q}}$ and $M_{\mathbf{10}+\overline{\mathbf{10}},\textrm{loop 1}}^{kQ}$. Hence, for the Coleman-Glashow relations one gets the generic expressions
\begin{equation}
\mu_{L}^{\textrm{(loop 1)}} - \mu_{R}^{\textrm{(loop 1)}} = D_{(L,R)} [I(m_K) - I(m_\Pi)],
\end{equation}
where $\mu_{L}^{\textrm{(loop 1)}}$ $[\mu_{R}^{\textrm{(loop 1)}}]$ represents the left-hand [right-hand] side of the corresponding relation in Eq.~(\ref{eq:treeval}) and $D_{(L,R)}$ is a quadratic function of the unknown coefficients of the $1/N_c$ expansion of $A^{kc}$. For instance, for the first relation one has
\begin{equation}
\mu_{\Sigma^+}^{\textrm{(loop 1)}} - \mu_p^{\textrm{(loop 1)}} = D_{(\Sigma^+,p)} [I(m_K) - I(m_\pi)],
\end{equation}
where, for $N_c=3$,
\begin{eqnarray}
&  & D_{(\Sigma^+,p)} = \nonumber \\
&  & \mbox{} -\frac{1}{180}(63 a_1^2 + 6a_1b_2 + 22 a_1b_3 + 30 a_1c_3 - 3b_2^2 + 2b_2b_3). \nonumber \\
\end{eqnarray}
Analogous results are obtained for the remaining relations and will not be listed here.

Caldi and Pagels \cite{caldi} first evaluated these corrections and found a dependence on the meson mass difference $m_K-m_\pi$. This dependence is already contained in the expression $I(m_K)-I(m_\pi)$ in our results. These authors also derived some sum rules, which in this approach we can check they are fulfilled, \textit{i.e.},
\begin{equation}
\mu_{\Sigma^+}^{\textrm{(loop 1)}} + 2\mu_\Lambda^{\textrm{(loop 1)}} + \mu_{\Sigma^-}^{\textrm{(loop 1)}} = 0, \label{eq:cp1}
\end{equation}
\begin{equation}
\mu_{\Xi^0}^{\textrm{(loop 1)}} + \mu_{\Xi^-}^{\textrm{(loop 1)}} + \mu_n^{\textrm{(loop 1)}} - 2\mu_\Lambda^{\textrm{(loop 1)}} + 2\mu_p^{\textrm{(loop 1)}} = 0, \label{eq:cp2}
\end{equation}
and
\begin{equation}
\mu_\Lambda^{\textrm{(loop 1)}} - \sqrt{3} \mu_{\Lambda\Sigma^0}^{\textrm{(loop 1)}} - \mu_{\Xi^0}^{\textrm{(loop 1)}} -
\mu_n^{\textrm{(loop 1)}} = 0. \label{eq:cp3}
\end{equation}

In turn, the relation
\begin{equation}
\mu_{\Sigma^+}^{\textrm{(loop 1)}} - 2\mu_{\Sigma^0}^{\textrm{(loop 1)}} + \mu_{\Sigma^-}^{\textrm{(loop 1)}} = 0, \label{eq:isosloop1}
\end{equation}
also holds to this order.

On the other hand, for decuplet baryons, the analogs of Eqs.~(\ref{eq:srd1})-(\ref{eq:srd3}) are
\begin{equation}
\mu_{\Delta^{++}}^{\textrm{(loop 1)}} - \mu_{\Delta^+}^{\textrm{(loop 1)}} - \mu_{\Delta^0}^{\textrm{(loop 1)}} + \mu_{\Delta^-}^{\textrm{(loop 1)}} = 0,
\end{equation}
\begin{equation}
\mu_{{\Sigma^*}^+}^{\textrm{(loop 1)}} - 2 \mu_{{\Sigma^*}^0}^{\textrm{(loop 1)}} + \mu_{{\Sigma^*}^-}^{\textrm{(loop 1)}} = 0,
\end{equation}
and
\begin{equation}
\mu_{\Delta^{++}}^{\textrm{(loop 1)}} - 3 \mu_{\Delta^+}^{\textrm{(loop 1)}} + 3 \mu_{\Delta^0}^{\textrm{(loop 1)}} - \mu_{\Delta^-}^{\textrm{(loop 1)}} = 0.
\end{equation}
whereas for transition magnetic moments, the analogs of Eqs.~(\ref{eq:srdo1})-(\ref{eq:srdo2}) read
\begin{equation}
\mu_{\Delta^{+}p}^{\textrm{(loop 1)}} - \mu_{\Delta^{0}n}^{\textrm{(loop 1)}} = 0,
\end{equation}
and
\begin{equation}
\mu_{{\Sigma^{*}}^+\Sigma^+}^{\textrm{(loop 1)}} - 2 \mu_{{\Sigma^{*}}^0\Sigma^0}^{\textrm{(loop 1)}} + \mu_{{\Sigma^{*}}^-\Sigma^-}^{\textrm{(loop 1)}} = 0.
\end{equation}

In Ref.~\cite{milana}, Eq.~(36), some other relations among magnetic moments of the decuplet baryons are presented which are satisfied at this order. We have explicitly checked that these relations are also satisfied within our approach.

\subsection{Nonanalytic corrections of order $m_q \ln m_q$}

The one-loop corrections to baryon magnetic moments arising from the Feynman diagrams of Fig.~\ref{fig:mmloop2} have a nonanalytic dependence on the quark mass of the form $m_q \ln m_q$. Compared to the case discussed previously, the computation of these diagrams requires a rather formidable effort when performing the algebraic reduction of the operator products involved. A great deal of computational ease is gained by noticing that this contribution has the same operator structure as the one found in the renormalized baryon axial-vector current $A^{kc}+\delta A^{kc}$ computed in Ref.~\cite{rfm06} so that some modifications and/or new partial computations are required. Algebraic manipulations of the equivalent diagrams \ref{fig:mmloop2}(a-d) and \ref{fig:mmloop2}(e) for the axial-vector current show that they can be combined in the double commutator structures given in Eqs.~(31) and (40), respectively, of Ref.~\cite{fmhjm}. Explicit computations of these double commutators are given in Ref.~\cite{rfm06} to relative order $\mathcal{O}(1/N_c^3)$.

In our case, we then follow a close parallelism to the analyses referred to above. Accordingly, we will discuss separately diagrams \ref{fig:mmloop2}(a-d) and \ref{fig:mmloop2}(e), as they involve rather different computational complication.

\subsubsection{Contribution of diagrams \ref{fig:mmloop2}(a-d)}

The first set of diagrams, Fig.~\ref{fig:mmloop2}(a-d), taking into account the wavefunction renormalization graphs, contribute to the baryon magnetic moment operator as
\begin{equation}
M_{\textrm{loop 2(a-d)}}^k = \left[A^{ja},\left[A^{jb},M^k \right] \right] \Pi^{ab}. \label{eq:corrloop2}
\end{equation}
Here, $\Pi^{ab}$ is a symmetric tensor which contains meson-loop integrals with the exchange of a single meson: A meson of flavor $a$ is emitted and a meson of flavor $b$ is reabsorbed. $\Pi^{ab}$ decomposes into flavor singlet, flavor $\mathbf{8}$, and flavor $\mathbf{27}$ representations as \cite{jen96}
\begin{equation}
\Pi^{ab} = F_\mathbf{1} \delta^{ab} + F_\mathbf{8} d^{ab8} + F_\mathbf{27} \left[ \delta^{a8} \delta^{b8} - \frac18 \delta^{ab} - \frac35 d^{ab8} d^{888}\right], \label{eq:pisym}
\end{equation}
where
\begin{equation}
F_\mathbf{1} = \frac18 \left[3F(m_\pi,\mu) + 4F(m_K,\mu) + F(m_\eta,\mu) \right], \label{eq:F1}
\end{equation}
\begin{equation}
F_\mathbf{8} = \frac{2\sqrt 3}{5} \left[\frac32 F(m_\pi,\mu) - F(m_K,\mu) - \frac12 F(m_\eta,\mu) \right], \label{eq:F8}
\end{equation}
and
\begin{equation}
F_\mathbf{27} = \frac13 F(m_\pi,\mu) - \frac43 F(m_K,\mu) + F(m_\eta,\mu). \label{eq:F27}
\end{equation}
Equations (\ref{eq:F1})-(\ref{eq:F27}) are linear combinations of $F(m_\pi,\mu)$, $F(m_K,\mu)$, and $F(m_\eta,\mu)$, where $F(m_\Pi,\mu)$ contains the result of performing the loop integral. In the degeneracy limit $\Delta\to 0$, this function reduces to \cite{jen92}
\begin{equation}
F(m_\Pi,\mu) = \frac{m_\Pi^2}{32\pi^2f^2} \ln{\frac{m_\Pi^2}{\mu^2}}, \label{eq:fprime}
\end{equation}
where $\mu$ is the scale of dimensional regularization and only nonanalytic terms in $m_q$ have been kept.

Now, in the operator reduction of the structure (\ref{eq:corrloop2}) some subtleties arise. The appearance of
the new parameters $m_i$ makes unfeasible the direct application of Eqs.~(30)-(32) of Ref.~\cite{rfm06} to obtain the corresponding loop contribution $M_{\textrm{loop 2(a-d)}}^k$. Indeed, new terms need be calculated. We remark that, because the operator basis is complete, the reduction is doable. In Appendix \ref{app:bo2} we present the individual contributions of the double commutator in (\ref{eq:corrloop2}) to the order implemented here. After a long but otherwise standard calculation, we can gather together partial results to get
\begin{equation}
M_{\textrm{loop 2(a-d)}}^k = F_\mathbf{1} M_{\mathbf{1},\textrm{loop 2(a-d)}}^{kQ} + F_\mathbf{8} M_{\mathbf{8},\textrm{loop 2(a-d)}}^{kQ} + F_\mathbf{27} M_{\mathbf{27},\textrm{loop 2(a-d)}}^{kQ}, \label{eq:loop2(a-d)}
\end{equation}
where the group structures of the double commutator read as follows:
\begin{widetext}

(1) \textit{Flavor singlet contribution}

\begin{eqnarray}
M_{\textbf{1},\textrm{loop 2(a-d)}}^{kc} & = & \left[ \frac{23}{12}a_1^2m_1 - \frac{N_c+3}{3N_c} ( -a_1b_2m_1 + 3 a_1^2m_2 ) - \frac{N_c^2+6N_c+4}{N_c^2} a_1^2m_3 - \frac{N_c^2+6N_c-3}{N_c^2}a_1^2m_4 \right. \nonumber \\
&  & \left. + \frac{N_c^2+6N_c-18}{6N_c^2}b_2^2m_1 + \frac{2}{N_c^2} ( a_1b_3m_1 - 3a_1b_2m_2 ) - \frac{4(N_c+3)}{N_c^3} ( b_2b_3m_1 + a_1b_3m_2 + a_1b_2m_3 ) \right] G^{kc} \nonumber \\
&  & \mbox{} + \frac{1}{N_c} \left[ \frac52 a_1b_2m_1 + \frac{71}{12} a_1^2m_2 + \frac{N_c+3}{6N_c} ( 6a_1b_3m_1 - 9a_1c_3m_1 + 8a_1b_2m_2 - 24a_1^2m_3 + 6a_1^2m_4 ) \right. \nonumber \\
&  & \mbox{} + \frac{N_c^2+6N_c+6}{N_c^2} a_1b_3m_2 - \frac{3(N_c^2+6N_c-12)}{2N_c^2} a_1c_3m_2 + \frac{N_c^2+6N_c-18}{6N_c^2} b_2^2m_2 + \frac{1}{N_c^2} ( -2b_2b_3m_1 \nonumber \\
&  & \mbox{} + 9b_2c_3m_1 - 2a_1b_2m_3 + 9a_1b_2m_4 ) \left] \mathcal{D}_2^{kc} \right. + \frac{1}{N_c^2} \left[ \frac{11}{6} a_1b_3m_1 + 2a_1c_3m_1 + \frac54 b_2^2m_1 + \frac32 a_1b_2m_2 \right. \nonumber \\
&  & \mbox{} \left. + \frac{131}{12} a_1^2m_3  + \frac{N_c+3}{6N_c} ( 14 b_2b_3m_1 - 6b_2c_3m_1 + 6a_1b_3m_2 - 15a_1c_3m_2 + 14a_1b_2m_3 - 6a_1b_2m_4 ) \right] \mathcal{D}_3^{kc} \nonumber \\
&  & \mbox{} + \frac{1}{N_c^2} \left[ \frac73 a_1b_3m_1 + 3a_1c_3m_1 + \frac72 b_2^2m_1 - 2 a_1b_2m_2 + \frac{131}{12} a_1^2m_4 + \frac{N_c+3}{3N_c} \left(22 b_2b_3m_1 - 3 b_2c_3m_1 \right. \right. \nonumber \\
&  & \mbox{} \left. \left. - 6 a_1b_3m_2 - 3 a_1c_3m_2 - 6 a_1b_2m_3 + 4 a_1b_2m_4 \right) \right] \mathcal{O}_3^{kc} \nonumber \\
&  & \mbox{} + \frac{1}{6N_c^3} \left( 6b_2b_3m_1 + 12b_2c_3m_1 + 10 a_1b_3m_2 + 90 a_1c_3m_2 + 15 b_2^2m_2 + 6a_1b_2m_3 + 12a_1b_2m_4 \right) \mathcal{D}_4^{kc} \nonumber \\
&  & + \mathcal{O}(G\mathcal{D}_3\mathcal{D}_3). \label{eq:loop2Sing}
\end{eqnarray}
\end{widetext}

\begin{widetext}
(2) \textit{Flavor octet Contribution}

\begin{eqnarray}
M_{\textbf{8},\textrm{loop 2(a-d)}}^{kc} & = & \left[ \frac{11}{24} a_1^2m_1 - \frac{N_c+3}{6N_c} \left(a_1b_2m_1 + 3a_1^2m_2 \right) - \frac{1}{2N_c^2} \left( 3b_2^2m_1 + 2a_1b_3m_1 + 6a_1b_2m_2 + 8a_1^2m_3 - 3a_1^2m_4 \right) \right. \nonumber \\
&  & \left. - \frac{2(N_c+3)}{N_c^3} \left( b_2b_3m_1 +a_1b_3m_2 + a_1b_2m_3 \right) \right] d^{c8e} G^{ke} + \left[ \frac{5}{18} a_1^2m_1 + \frac{N_c+3}{9N_c} a_1b_2m_1 - \frac{N_c+6}{12N_c} \left( 3a_1c_3m_1 \right. \right. \nonumber \\
&  & \mbox{} \left. \left. - 2a_1^2m_4 \right) + \frac{N_c^2+6N_c+4}{6N_c^2} a_1b_3m_1 - \frac{2(N_c^2+6N_c-1)}{3N_c^2} a_1^2m_3 \right] \delta^{c8}J^k + \frac{1}{N_c} \left[ \frac54 a_1b_2m_1 + \frac{13}{8} a_1^2m_2 \right. \nonumber \\
&  & \mbox{} + \frac{N_c+3}{4N_c} \left( 2a_1b_3m_1 - 3a_1c_3m_1 - 8a_1^2m_3 + 2a_1^2m_4 \right) + \frac{1}{2N_c^2} \left( -2b_2b_3m_1 + 9b_2c_3m_1 - 3b_2^2m_2 - 2a_1b_3m_2 \right. \nonumber \\
&  & \mbox{} \left. + 18a_1c_3m_2 - 2a_1b_2m_3 + 9a_1b_2m_4 \right) \left] d^{c8e} \mathcal{D}_2^{ke} + \frac{1}{N_c} \left[ \frac12 a_1b_2m_1 + \frac43 a_1^2m_2 + \frac{N_c+3}{3N_c} a_1b_2m_2 \right. \right. \nonumber \\
&  & \mbox{} \left. + \frac{2}{N_c^2} \left(b_2b_3m_1 + 2a_1b_3m_2 + a_1b_2m_3 \right) \right] \{T^c,G^{k8}\} - \frac{1}{N_c} \left[ \frac16 a_1b_2m_1 + \frac12 a_1^2m_2 + \frac{N_c+3}{6N_c} \left( b_2^2m_1 + 6a_1^2m_3 \right. \right. \nonumber \\
&  & \mbox{} \left. + 6a_1^2m_4 \right) \left. + \frac{2}{N_c^2} \left( b_2b_3m_1 + a_1b_3m_2 + a_1b_2m_3 \right) \right] \{G^{kc},T^8\} + \frac{1}{N_c^2} \left[ \frac38 b_2^2m_1 + \frac{7}{12} a_1b_3m_1 + \frac23 a_1c_3m_1 \right. \nonumber \\
&  & \mbox{} + \frac34 a_1b_2m_2 + \frac{17}{8} a_1^2m_3 - \frac13 a_1^2m_4 + \frac{N_c+3}{12N_c} ( 6b_2b_3m_1 - 4b_2c_3m_1 + 6a_1b_3m_2 - 15a_1c_3m_2 + 6a_1b_2m_3 \nonumber \\
&  & \mbox{} - 4a_1b_2m_4 ) \left] d^{c8e} \mathcal{D}_3^{ke} + \frac{1}{N_c^2} \right[ - \frac56 a_1b_3m_1 + \frac76 a_1c_3m_1 + \frac54 b_2^2m_1 - a_1b_2m_2 - \frac23 a_1^2m_3 + \frac{43}{24} a_1^2 m_4 \nonumber \\
&  & \mbox{} \left. + \frac{N_c+3}{6N_c} ( 8b_2b_3m_1 - 2b_2c_3m_1 - 6a_1b_3m_2 - 3a_1c_3m_2 - 4a_1b_2m_3 + a_1b_2m_4) \right] d^{c8e} \mathcal{O}_3^{ke} \nonumber \\
&  & \mbox{} + \frac{1}{N_c^2} \left[ \frac14 a_1b_3m_1  - \frac38 a_1c_3m_1 + \frac16 a_1b_2m_2 - a_1^2m_3 + \frac14 a_1^2m_4 - \frac{N_c+3}{12N_c} ( - 6a_1b_3m_2 + 9a_1c_3m_2 + b_2^2m_2 ) \right] \nonumber \\
&  & \mbox{} \times \{J^k,\{T^c,T^8\}\} + \frac{1}{N_c^2} \left[ -a_1b_3m_1 + \frac76 a_1c_3m_1 + 5a_1^2m_3 - \frac{11}{6} a_1^2m_4 - \frac{N_c+3}{3N_c} ( b_2c_3m_1 + a_1b_2m_4 ) \right] \nonumber \\
&  & \mbox{} \times \{J^k,\{G^{rc},G^{r8}\}\} + \frac{1}{N_c^2} \left[ - a_1b_3m_1 + \frac13 a_1c_3m_1 + \frac32 b_2^2m_1 - a_1b_2m_2 + \frac23 a_1^2m_3  + \frac{14}{3} a_1^2m_4 \right. \nonumber \\
&  & \mbox{} \left. - \frac{N_c+3}{6N_c} ( -6b_2b_3m_1 + b_2c_3m_1 + 2a_1b_2m_3 - a_1b_2m_4 ) \right] \{G^{kc},\{J^r,G^{r8}\}\} \nonumber \\
&  & \mbox{} + \frac{1}{N_c^2} \left[ \frac53 a_1b_3m_1 - \frac13 a_1c_3m_1 - \frac12 b_2^2m_1 + a_1b_2m_2 + 3a_1^2m_3 -
\frac{13}{6} a_1^2m_4 + \frac{N_c+3}{6N_c} ( -2b_2b_3m_1 + b_2c_3m_1 \right. \nonumber \\
&  & \mbox{} \left. + 6a_1b_2m_3 - a_1b_2m_4 \right) \left] \{G^{k8},\{J^r,G^{rc}\}\} + \frac{1}{N_c^2} \left[ -\frac19 a_1b_3m_1 + \frac{17}{18} a_1c_3m_1 + \frac53 a_1^2m_3 - \frac29 a_1^2m_4 \right. \right. \nonumber \\
&  & \mbox{} \left. + \frac{N_c+3}{9N_c} \left(b_2c_3m_1 + a_1b_2m_4 \right) \right] \delta^{c8} \{J^2,J^k\} + \frac{1}{N_c^3} \left[ \frac12 b_2b_3m_1 + b_2c_3m_1 + \frac12 a_1b_3m_2 + 4a_1c_3m_2 + \frac34 b_2^2m_2 \right. \nonumber \\
&  & \mbox{} \left. + \frac12 a_1b_2m_3 + a_1b_2m_4 \right] d^{c8e} \mathcal{D}_4^{ke} + \frac{1}{N_c^3} \left[ -b_2b_3m_1 + \frac12 b_2c_3m_1 + \frac{10}{3} a_1b_3m_2 + \frac43 a_1c_3m_2 + a_1b_2m_3 \right] \nonumber \\
&  & \mbox{} \times \{J^2,\{T^c,G^{k8}\}\} + \frac{1}{N_c^3} \left[ \frac73 b_2b_3m_1 - \frac12 b_2c_3m_1 - a_1b_3m_2 - \frac12 a_1c_3m_2 - a_1b_2m_3 + \frac13 a_1b_2m_4 \right] \nonumber \\
&  & \mbox{} \times \{J^2,\{G^{kc},T^8\}\} + \frac{1}{N_c^3} \left[ \frac12 b_2c_3m_1 - 5a_1b_3m_2 + \frac{31}{6} a_1c_3m_2  + b_2^2m_2 - 2a_1b_2m_3 + a_1b_2m_4 \right] \nonumber \\
&  & \mbox{} \times \{\mathcal{D}_2^{kc},\{J^r,G^{r8}\}\}  + \frac{1}{N_c^3} \left[ -\frac23 b_2b_3m_1 - \frac12 b_2c_3m_1 + 2a_1b_3m_2 - 2a_1c_3m_2 + \frac83 a_1b_2m_3 - \frac43 a_1b_2m_4 \right] \nonumber \\
&  & \mbox{} \times\{\mathcal{D}_2^{k8},\{J^r,G^{rc}\}\} + \frac{1}{N_c^3} \left[ \frac{5}{32} b_2b_3m_1 - \frac{13}{64} b_2c_3m_1 - \frac{15}{64} a_1b_2m_3 - \frac{45}{128} a_1b_2m_4 \right] \nonumber \\
&  & \mbox{} \times \left( \{J^2,[G^{kc},\{J^r,G^{r8}\}]\} - \{J^2,[G^{k8},\{J^r,G^{rc}\}]\} - \{J^k,[\{J^m,G^{mc}\},\{J^r,G^{r8}\}]\} \right) + \mathcal{O}(G\mathcal{D}_3\mathcal{D}_3). \label{eq:loop2Oct}
\end{eqnarray}
\end{widetext}

\begin{widetext}

(3) \textit{Flavor \textbf{27} contribution}

\begin{eqnarray}
M_{\textbf{27},\textrm{loop 2(a-d)}}^{kc} & = & \frac16 a_1^2m_1 ( 3 d^{c8e} d^{8eg} G^{kg} + 2 \delta^{c8} G^{k8} + d^{c88} J^k) + \frac{2}{3N_c} a_1b_2m_1 \delta^{c8} \mathcal{D}_2^{k8} + \frac{1}{3N_c} a_1^2 m_2 \delta^{88} \mathcal{D}_2^{kc} \nonumber \\
&  & \mbox{} + \frac{1}{N_c} a_1b_2m_1 d^{c8e} \{G^{ke},T^8\} + \frac{1}{2N_c} a_1^2m_2 d^{88e} \{G^{ke},T^c\} + \frac{1}{2N_c^2} (a_1c_3m_1 + a_1^2m_3) d^{c8e} d^{8eg} \mathcal{D}_3^{kg} \nonumber \\
&  & \mbox{} + \frac{1}{2N_c^2} (2a_1b_3m_1 + a_1c_3m_1 + a_1^2m_4) d^{c8e} d^{8eg} \mathcal{O}_3^{kg} + \frac{1}{3N_c^2}
(2a_1b_3m_1 - a_1^2m_3 + a_1^2m_4) \delta^{c8} \mathcal{D}_3^{k8} \nonumber \\
&  & \mbox{} + \frac{1}{3N_c^2} ( 2a_1c_3m_1 + 3a_1^2m_4 ) \delta^{c8} \mathcal{O}_3^{k8} + \frac{2}{3N_c^2} a_1^2m_3 \delta^{88} \mathcal{D}_3^{kc} + \frac{2}{3N_c^2} a_1^2m_4 \delta^{88} \mathcal{O}_3^{kc}
\nonumber \\
&  & \mbox{} + \frac{1}{3N_c^2} (a_1c_3m_1 + a_1^2m_3) d^{c88} \{J^2,J^k\} + \frac{1}{N_c^2} a_1b_2m_2 \{G^{k8},\{T^c,T^8\}\} + \frac{1}{2N_c^2} b_2^2m_1 \{G^{kc},\{T^8,T^8\}\} \nonumber \\
&  & \mbox{} - \frac{1}{N_c^2} (2 a_1^2m_3 + a_1^2m_4) \{G^{kc},\{G^{r8},G^{r8}\}\} + \frac{1}{N_c^2} (2 a_1^2m_3 - a_1^2m_4) \{G^{k8},\{G^{rc},G^{r8}\}\} \nonumber \\
&  & \mbox{} - \frac{1}{2N_c^2} (2a_1c_3m_1 + 6a_1^2m_3 - a_1^2m_4) d^{c8e} \{J^k,\{G^{re},G^{r8}\}\} + \frac{1}{2N_c^2} (2a_1^2m_3 - a_1^2m_4) d^{88e} \{J^k,\{G^{rc},G^{re}\}\} \nonumber \\
&  & \mbox{} + \frac{1}{N_c^2} a_1^2m_4 d^{88e} \{G^{kc},\{J^r,G^{re}\}\} + \frac{1}{2N_c^2} (6a_1b_3m_1 - a_1c_3m_1 + 2a_1^2m_3 - a_1^2m_4) d^{c8e} \{G^{ke},\{J^r,G^{r8}\}\} \nonumber \\
&  & \mbox{} + \frac{1}{2N_c^2} (2a_1^2m_3 - a_1^2m_4) d^{88e} \{G^{ke},\{J^r,G^{rc}\}\} + \frac{1}{2N_c^2} (-2a_1b_3m_1 + a_1c_3m_1 + 2a_1^2m_4) d^{c8e} \{G^{k8},\{J^r,G^{re}\}\} \nonumber \\
&  & \mbox{} + \frac{1}{2N_c^3} (2b_2b_3m_1 + b_2c_3m_1 + 2a_1b_2m_3 + a_1b_2m_4) d^{c8e} \{J^2,\{G^{ke},T^8\}\} \nonumber \\
&  & \mbox{} + \frac{1}{2N_c^3} (2 a_1b_3m_2 + a_1c_3m_2) d^{88e} \{J^2,\{G^{ke},T^c\}\} + \frac{2}{3N_c^3} (b_2c_3m_1 + a_1b_2m_4) \delta^{c8} \mathcal{D}_4^{k8} + \frac{2}{3N_c^3} a_1c_3m_2 \delta^{88} \mathcal{D}_4^{kc} \nonumber \\
&  & \mbox{} + \frac{1}{2N_c^3} b_2^2m_2  \{\mathcal{D}_2^{kc},\{T^8,T^8\}\} - \frac{2}{N_c^3} a_1c_3m_2 \{\mathcal{D}_2^{kc},\{G^{r8},G^{r8}\}\} \nonumber \\
&  & \mbox{} - \frac{1}{2N_c^3} (2 a_1b_3m_2 - a_1c_3m_2) d^{88e} \{\mathcal{D}_2^{kc},\{J^r,G^{re}\}\} - \frac{2}{N_c^3} (b_2c_3m_1 + a_1b_2m_4) \{\mathcal{D}_2^{k8},\{G^{rc},G^{r8}\}\} \nonumber \\
&  & \mbox{} + \frac{1}{2N_c^3} (-2 b_2b_3m_1 + b_2c_3 m_1 - 2a_1b_2m_3 + a_1b_2m_4) d^{c8e} \{\mathcal{D}_2^{k8},\{J^r,G^{re}\}\} \nonumber \\
&  & \mbox{}  + \frac{1}{2N_c^3} (-2 b_2b_3m_1 + b_2c_3m_1 + 6a_1b_2m_3 - a_1b_2m_4) \{\{J^r,G^{rc}\},\{G^{k8},T^8\}\} \nonumber \\
&  & \mbox{} + \frac{1}{2N_c^3} (6 b_2b_3m_1 - b_2c_3m_1 - 2 a_1b_2m_3 + a_1b_2m_4) \{\{J^r,G^{r8}\},\{G^{kc},T^8\}\} \nonumber \\
&  & \mbox{} + \frac{2}{N_c^3} a_1b_3m_2 \{\{J^r,G^{r8}\},\{G^{k8},T^c\}\} + \mathcal{O}(G\mathcal{D}_3\mathcal{D}_3). \label{eq:loop2Twen}
\end{eqnarray}
\end{widetext}
Notice that, in order for $M_{\textbf{27},\textrm{loop 2(a-d)}}^{kc}$ to be a truly $\mathbf{27}$ contribution, singlet and octet pieces must be subtracted off.

In Eqs.~(\ref{eq:loop2Sing})-(\ref{eq:loop2Twen}) the symbol $\mathcal{O}(G\mathcal{D}_3\mathcal{D}_3)$ refers to the fact that in the double commutator structure $[A^{ja},[A^{jb},M^{kc}]]$ we have included all the terms up to six-body operators, such as $[G^{ja},[\mathcal{D}_2^{jb},\mathcal{O}_3^{kc}]]$, but have neglected contributions which are seven-body operators or higher. We have done this because the commutator of an $m$-body operator with an $n$-body operator is an $(m+n-1)$-operator. On the other hand, we have also rearranged these expressions to display leading and subleading terms in $1/N_c$ explicitly. The resulting formulas are rather lengthy but also illuminating. We can check that large-$N_c$ cancellations occur (as expected) in the double commutator structure in such a way that it is at most of order $\mathcal{O}(N_c)$. Besides, the factor $f^2$ in the denominator of the loop integral introduces an extra suppression of $1/N_c$ in such a way that the net contribution of $M_{\textrm{loop 2(a-d)}}^k$ to the magnetic moments is $\mathcal{O}(1)$, or equivalently, $1/N_c$ times the tree-level value which is order $\mathcal{O}(N_c)$. Thus, in principle, the dominant source of SU(3) symmetry breaking should come from the contribution $M_{\textrm{loop 1}}^k$ rather than $M_{\textrm{loop 2(a-d)}}^k$. We think that this statement is pointless unless we perform a numerical comparison of the theoretical expressions with the available experimental data. This analysis is postponed to Sec.~\ref{sec:numerical}. We now turn to evaluate the diagram corresponding to Fig.~\ref{fig:mmloop2}(e).

\subsubsection{Contribution of diagram \ref{fig:mmloop2}(e)}

The contribution to the magnetic moments of the Feynman diagram displayed in Fig.~\ref{fig:mmloop2}(e) possesses the structure
\begin{equation}
M_{\textrm{loop 2e}}^k = \left[T^a,\left[T^b,M^k \right] \right] \Pi^{ab}, \label{eq:corrloop2e}
\end{equation}
where $\Pi^{ab}$ is the symmetric tensor introduced in Eq.~(\ref{eq:pisym}). In a similar way to Eq.~(\ref{eq:loop2(a-d)}), Eq.~(\ref{eq:corrloop2e}) can be separated into flavor singlet, flavor octet, and flavor $\mathbf{27}$ pieces as
\begin{equation}
M_{\textrm{loop 2e}}^k = F_\mathbf{1} M_{\mathbf{1},\textrm{loop 2e}}^{kQ} + F_\mathbf{8} M_{\mathbf{8},\textrm{loop 2e}}^{kQ} + F_\mathbf{27} M_{\mathbf{27},\textrm{loop 2e}}^{kQ}, \label{eq:loop2sum2}
\end{equation}
where this time the group structures of the double commutator read as follows:

(1) \textit{Flavor singlet contribution}
\begin{eqnarray}
M_{\mathbf{1},\textrm{loop 2e}}^{kc} & = & [T^a,[T^a,M^{kc}]] \nonumber \\
& = & 3 M^{kc}. \label{eq:sind}
\end{eqnarray}

(2) \textit{ Flavor octet contribution}
\begin{eqnarray}
M_{\mathbf{8},\textrm{loop 2e}}^{kc} & = & d^{ab8} [T^a,[T^b,M^{kc}]] \nonumber \\
& = & \frac32 d^{c8e} M^{ke}. \label{eq:octd}
\end{eqnarray}

(3) \textit{Flavor $\mathbf{27}$ contribution}
\begin{eqnarray}
M_{\mathbf{27},\textrm{loop 2e}}^{kc} & = & [T^8,[T^8,M^{kc}]] \nonumber \\
& = & f^{c8e} f^{8eg} M^{kg}. \label{eq:27d}
\end{eqnarray}

Let us notice that, in order for $M_{\textbf{27},\textrm{loop 2e}}^{kc}$ to be a truly $\mathbf{27}$ contribution, singlet and octet pieces must be subtracted off. In the above equations, the free flavor index $c$ will be set to $Q=3+(1/\sqrt{3})8$. By doing this, expression (\ref{eq:27d}) as it stands, will vanish.

As before, in order to proceed further, we need to compute the matrix elements of operators that fall into the flavor $\mathbf{27}$ representation. To relative order $\mathcal{O}(1/N_c^3)$, this time we have identified 36 spin-1 operators in such a representaction. They read

\begin{widetext}
\begin{eqnarray}
\begin{tabular}{lll}
$Z_{1}^{kc}=d^{c8e} d^{8eg} G^{kg}$, &
$Z_{2}^{kc}=\delta^{c8} G^{k8}$, &
$Z_{3}^{kc}=d^{c88} J^k$, \nonumber \\
$Z_{4}^{kc}=\delta^{c8} \mathcal{D}_2^{k8}$, &
$Z_{5}^{kc}=\delta^{88} \mathcal{D}_2^{kc}$, &
$Z_{6}^{kc}=d^{c8e} \{G^{ke},T^8\}$, \nonumber \\
$Z_{7}^{kc}=d^{88e} \{G^{ke},T^c\}$, &
$Z_{8}^{kc}=d^{c8e} d^{8eg} \mathcal{D}_3^{kg}$, &
$Z_{9}^{kc}=d^{c8e} d^{8eg} \mathcal{O}_3^{kg}$, \\
$Z_{10}^{kc}=\delta^{c8} \mathcal{D}_3^{k8}$, &
$Z_{11}^{kc}=\delta^{c8} \mathcal{O}_3^{k8}$, &
$Z_{12}^{kc}=\delta^{88} \mathcal{D}_3^{kc}$, \\
$Z_{13}^{kc}=\delta^{88} \mathcal{O}_3^{kc}$, &
$Z_{14}^{kc}=d^{c88} \{J^2,J^k\}$, &
$Z_{15}^{kc}=\{G^{kc},\{T^8,T^8\}\}$, \nonumber \\
$Z_{16}^{kc}=\{G^{k8},\{T^c,T^8\}\}$, &
$Z_{17}^{kc}=\{G^{kc},\{G^{r8},G^{r8}\}\}$, &
$Z_{18}^{kc}=\{G^{k8},\{G^{rc},G^{r8}\}\}$, \\
$Z_{19}^{kc}=d^{c8e} \{J^k,\{G^{re},G^{r8}\}\}$, &
$Z_{20}^{kc}=d^{88e} \{J^k,\{G^{rc},G^{re}\}\}$, &
$Z_{21}^{kc}=d^{c8e} \{G^{k8},\{J^r,G^{re}\}\}$, \nonumber \\
$Z_{22}^{kc}=d^{88e} \{G^{kc},\{J^r,G^{re}\}\}$, &
$Z_{23}^{kc}=d^{c8e} \{G^{ke},\{J^r,G^{r8}\}\}$, &
$Z_{24}^{kc}=d^{88e} \{G^{ke},\{J^r,G^{rc}\}\}$, \\
$Z_{25}^{kc}=\delta^{c8} \mathcal{D}_4^{k8}$, &
$Z_{26}^{kc}=\delta^{88} \mathcal{D}_4^{kc}$, &
$Z_{27}^{kc}=\{\mathcal{D}_2^{kc},\{T^8,T^8\}\}$, \\
$Z_{28}^{kc}=\{\mathcal{D}_2^{kc},\{G^{r8},G^{r8}\}\}$, &
$Z_{29}^{kc}=\{\mathcal{D}_2^{k8},\{G^{rc},G^{r8}\}\}$, &
$Z_{30}^{kc}=d^{c8e}\{\mathcal{D}_2^{k8},\{J^r,G^{re}\}\}$, \nonumber \\
$Z_{31}^{kc}=d^{88e}\{\mathcal{D}_2^{kc},\{J^r,G^{re}\}\}$, &
$Z_{32}^{kc}=d^{c8e}\{J^2,\{G^{ke},T^8\}\}$, &
$Z_{33}^{kc}=d^{88e}\{J^2,\{G^{ke},T^c\}\}$, \\
$Z_{34}^{kc}=\{\{J^r,G^{rc}\},\{G^{k8},T^8\}\}$, &
$Z_{35}^{kc}=\{\{J^r,G^{r8}\},\{G^{kc},T^8\}\}$, &
$Z_{36}^{kc}=\{\{J^r,G^{r8}\},\{G^{k8},T^c\}\}$. \nonumber \\
\end{tabular}
\end{eqnarray}
\end{widetext}
Their corresponding  matrix elements are listed in Tables \ref{t:mm2733O}-\ref{t:mm2733TO} for baryon octet, decuplet, and decuplet-octet transitions.

\begin{table*}
\caption{\label{t:mm2733O}Nontrivial matrix elements of the operators involved in the magnetic moments of octet baryons: Flavor $\mathbf{27}$ representation. The entries correspond to $144\langle Z_{m}^{33}\rangle$.}
\begin{ruledtabular}
\begin{tabular}{lccccccccc}
& $\displaystyle n$ & $\displaystyle p$ & $\displaystyle \Sigma^-$ & $\displaystyle \Sigma^0$ & $\displaystyle \Sigma^+$ & $\displaystyle \Xi^-$ & $\displaystyle \Xi^0$ & $\displaystyle \Lambda$ & $\displaystyle \Lambda\Sigma^0$ \\[2mm]
\hline
$\langle Z_{1}^{33} \rangle$ & $-20$ & $20$ & $-16$ & $0$ & $16$ & $4$ & $-4$ & $0$ & $8 \sqrt{3}$ \\
$\langle Z_{2}^{33} \rangle$ & $0$ & $0$ & $0$ & $0$ & $0$ & $0$ & $0$ & $0$ & $0$ \\
$\langle Z_{3}^{33} \rangle$ & $0$ & $0$ & $0$ & $0$ & $0$ & $0$ & $0$ & $0$ & $0$ \\
$\langle Z_{4}^{33} \rangle$ & $0$ & $0$ & $0$ & $0$ & $0$ & $0$ & $0$ & $0$ & $0$ \\
$\langle Z_{5}^{33} \rangle$ & $-36$ & $36$ & $-72$ & $0$ & $72$ & $-36$ & $36$ & $0$ & $0$ \\
$\langle Z_{6}^{33} \rangle$ & $-60$ & $60$ & $0$ & $0$ & $0$ & $-12$ & $12$ & $0$ & $0$ \\
$\langle Z_{7}^{33} \rangle$ & $12$ & $-12$ & $48$ & $0$ & $-48$ & $-36$ & $36$ & $0$ & $0$ \\
$\langle Z_{8}^{33} \rangle$ & $-60$ & $60$ & $-48$ & $0$ & $48$ & $12$ & $-12$ & $0$ & $24 \sqrt{3}$ \\
$\langle Z_{10}^{33} \rangle$ & $0$ & $0$ & $0$ & $0$ & $0$ & $0$ & $0$ & $0$ & $0$ \\
$\langle Z_{12}^{33} \rangle$ & $-180$ & $180$ & $-144$ & $0$ & $144$ & $36$ & $-36$ & $0$ & $72 \sqrt{3}$ \\
$\langle Z_{14}^{33} \rangle$ & $0$ & $0$ & $0$ & $0$ & $0$ & $0$ & $0$ & $0$ & $0$ \\
$\langle Z_{15}^{33} \rangle$ & $-180$ & $180$ & $0$ & $0$ & $0$ & $36$ & $-36$ & $0$ & $0$ \\
$\langle Z_{16}^{33} \rangle$ & $-36$ & $36$ & $0$ & $0$ & $0$ & $-108$ & $108$ & $0$ & $0$ \\
$\langle Z_{17}^{33} \rangle$ & $-15$ & $15$ & $-144$ & $0$ & $144$ & $51$ & $-51$ & $0$ & $48 \sqrt{3}$ \\
$\langle Z_{18}^{33} \rangle$ & $-15$ & $15$ & $-96$ & $0$ & $96$ & $99$ & $-99$ & $0$ & $0$ \\
$\langle Z_{19}^{33} \rangle$ & $-30$ & $30$ & $-96$ & $0$ & $96$ & $-66$ & $66$ & $0$ & $-24 \sqrt{3}$ \\
$\langle Z_{20}^{33} \rangle$ & $30$ & $-30$ & $96$ & $0$ & $-96$ & $66$ & $-66$ & $0$ & $24 \sqrt{3}$ \\
$\langle Z_{21}^{33} \rangle$ & $-30$ & $30$ & $-48$ & $0$ & $48$ & $-18$ & $18$ & $0$ & $0$ \\
$\langle Z_{22}^{33} \rangle$ & $30$ & $-30$ & $48$ & $0$ & $-48$ & $18$ & $-18$ & $0$ & $0$ \\
$\langle Z_{23}^{33} \rangle$ & $-30$ & $30$ & $-48$ & $0$ & $48$ & $-18$ & $18$ & $0$ & $0$ \\
$\langle Z_{24}^{33} \rangle$ & $30$ & $-30$ & $48$ & $0$ & $-48$ & $18$ & $-18$ & $0$ & $0$ \\
$\langle Z_{25}^{33} \rangle$ & $0$ & $0$ & $0$ & $0$ & $0$ & $0$ & $0$ & $0$ & $0$ \\
$\langle Z_{26}^{33} \rangle$ & $-54$ & $54$ & $-108$ & $0$ & $108$ & $-54$ & $54$ & $0$ & $0$ \\
$\langle Z_{27}^{33} \rangle$ & $-108$ & $108$ & $0$ & $0$ & $0$ & $-108$ & $108$ & $0$ & $0$ \\
$\langle Z_{28}^{33} \rangle$ & $-9$ & $9$ & $-216$ & $0$ & $216$ & $-153$ & $153$ & $0$ & $0$ \\
$\langle Z_{29}^{33} \rangle$ & $-45$ & $45$ & $0$ & $0$ & $0$ & $99$ & $-99$ & $0$ & $0$ \\
$\langle Z_{30}^{33} \rangle$ & $-90$ & $90$ & $0$ & $0$ & $0$ & $-18$ & $18$ & $0$ & $0$ \\
$\langle Z_{31}^{33} \rangle$ & $18$ & $-18$ & $72$ & $0$ & $-72$ & $-54$ & $54$ & $0$ & $0$ \\
$\langle Z_{32}^{33} \rangle$ & $-90$ & $90$ & $0$ & $0$ & $0$ & $-18$ & $18$ & $0$ & $0$ \\
$\langle Z_{33}^{33} \rangle$ & $18$ & $-18$ & $72$ & $0$ & $-72$ & $-54$ & $54$ & $0$ & $0$ \\
$\langle Z_{34}^{33} \rangle$ & $-90$ & $90$ & $0$ & $0$ & $0$ & $54$ & $-54$ & $0$ & $0$ \\
$\langle Z_{35}^{33} \rangle$ & $-90$ & $90$ & $0$ & $0$ & $0$ & $54$ & $-54$ & $0$ & $0$ \\
$\langle Z_{36}^{33} \rangle$ & $-18$ & $18$ & $-144$ & $0$ & $144$ & $-162$ & $162$ & $0$ & $0$ \\
\end{tabular}
\end{ruledtabular}
\end{table*}

\begin{table*}
\caption{\label{t:mm2738O} Nontrivial matrix elements of the operators involved in the magnetic moments of octet baryons: Flavor $\mathbf{27}$ representation. The entries correspond to $144\sqrt{3}\langle Z_{m}^{38} \rangle$.}
\begin{ruledtabular}
\begin{tabular}{lccccccccc}
& $\displaystyle n$ & $\displaystyle p$ & $\displaystyle \Sigma^-$ & $\displaystyle \Sigma^0$ & $\displaystyle \Sigma^+$ & $\displaystyle \Xi^-$ & $\displaystyle \Xi^0$ & $\displaystyle \Lambda$ & $\displaystyle \Lambda\Sigma^0$ \\[2mm]
\hline
$\langle Z_{1}^{38} \rangle$ & $12$ & $12$ & $24$ & $24$ & $24$ & $-36$ & $-36$ & $-24$ & $0$ \\
$\langle Z_{2}^{38} \rangle$ & $36$ & $36$ & $72$ & $72$ & $72$ & $-108$ & $-108$ & $-72$ & $0$ \\
$\langle Z_{3}^{38} \rangle$ & $-72$ & $-72$ & $-72$ & $-72$ & $-72$ & $-72$ & $-72$ & $-72$ & $0$ \\
$\langle Z_{4}^{38} \rangle$ & $108$ & $108$ & $0$ & $0$ & $0$ & $-108$ & $-108$ & $0$ & $0$ \\
$\langle Z_{5}^{38} \rangle$ & $108$ & $108$ & $0$ & $0$ & $0$ & $-108$ & $-108$ & $0$ & $0$ \\
$\langle Z_{6}^{38} \rangle$ & $-36$ & $-36$ & $0$ & $0$ & $0$ & $-108$ & $-108$ & $0$ & $0$ \\
$\langle Z_{7}^{38} \rangle$ & $-36$ & $-36$ & $0$ & $0$ & $0$ & $-108$ & $-108$ & $0$ & $0$ \\
$\langle Z_{8}^{38} \rangle$ & $36$ & $36$ & $72$ & $72$ & $72$ & $-108$ & $-108$ & $-72$ & $0$ \\
$\langle Z_{10}^{38} \rangle$ & $108$ & $108$ & $216$ & $216$ & $216$ & $-324$ & $-324$ & $-216$ & $0$ \\
$\langle Z_{12}^{38} \rangle$ & $108$ & $108$ & $216$ & $216$ & $216$ & $-324$ & $-324$ & $-216$ & $0$ \\
$\langle Z_{14}^{38} \rangle$ & $-108$ & $-108$ & $-108$ & $-108$ & $-108$ & $-108$ & $-108$ & $-108$ & $0$ \\
$\langle Z_{15}^{38} \rangle$ & $108$ & $108$ & $0$ & $0$ & $0$ & $-324$ & $-324$ & $0$ & $0$ \\
$\langle Z_{16}^{38} \rangle$ & $108$ & $108$ & $0$ & $0$ & $0$ & $-324$ & $-324$ & $0$ & $0$ \\
$\langle Z_{17}^{38} \rangle$ & $9$ & $9$ & $216$ & $216$ & $216$ & $-459$ & $-459$ & $-72$ & $0$ \\
$\langle Z_{18}^{38} \rangle$ & $9$ & $9$ & $216$ & $216$ & $216$ & $-459$ & $-459$ & $-72$ & $0$ \\
$\langle Z_{19}^{38} \rangle$ & $-18$ & $-18$ & $-216$ & $-216$ & $-216$ & $-306$ & $-306$ & $-72$ & $0$ \\
$\langle Z_{20}^{38} \rangle$ & $-18$ & $-18$ & $-216$ & $-216$ & $-216$ & $-306$ & $-306$ & $-72$ & $0$ \\
$\langle Z_{21}^{38} \rangle$ & $-18$ & $-18$ & $-72$ & $-72$ & $-72$ & $-162$ & $-162$ & $-72$ & $0$ \\
$\langle Z_{22}^{38} \rangle$ & $-18$ & $-18$ & $-72$ & $-72$ & $-72$ & $-162$ & $-162$ & $-72$ & $0$ \\
$\langle Z_{23}^{38} \rangle$ & $-18$ & $-18$ & $-72$ & $-72$ & $-72$ & $-162$ & $-162$ & $-72$ & $0$ \\
$\langle Z_{24}^{38} \rangle$ & $-18$ & $-18$ & $-72$ & $-72$ & $-72$ & $-162$ & $-162$ & $-72$ & $0$ \\
$\langle Z_{25}^{38} \rangle$ & $162$ & $162$ & $0$ & $0$ & $0$ & $-162$ & $-162$ & $0$ & $0$ \\
$\langle Z_{26}^{38} \rangle$ & $162$ & $162$ & $0$ & $0$ & $0$ & $-162$ & $-162$ & $0$ & $0$ \\
$\langle Z_{27}^{38} \rangle$ & $324$ & $324$ & $0$ & $0$ & $0$ & $-324$ & $-324$ & $0$ & $0$ \\
$\langle Z_{28}^{38} \rangle$ & $27$ & $27$ & $0$ & $0$ & $0$ & $-459$ & $-459$ & $0$ & $0$ \\
$\langle Z_{29}^{38} \rangle$ & $27$ & $27$ & $0$ & $0$ & $0$ & $-459$ & $-459$ & $0$ & $0$ \\
$\langle Z_{30}^{38} \rangle$ & $-54$ & $-54$ & $0$ & $0$ & $0$ & $-162$ & $-162$ & $0$ & $0$ \\
$\langle Z_{31}^{38} \rangle$ & $-54$ & $-54$ & $0$ & $0$ & $0$ & $-162$ & $-162$ & $0$ & $0$ \\
$\langle Z_{32}^{38} \rangle$ & $-54$ & $-54$ & $0$ & $0$ & $0$ & $-162$ & $-162$ & $0$ & $0$ \\
$\langle Z_{33}^{38} \rangle$ & $-54$ & $-54$ & $0$ & $0$ & $0$ & $-162$ & $-162$ & $0$ & $0$ \\
$\langle Z_{34}^{38} \rangle$ & $54$ & $54$ & $0$ & $0$ & $0$ & $-486$ & $-486$ & $0$ & $0$ \\
$\langle Z_{35}^{38} \rangle$ & $54$ & $54$ & $0$ & $0$ & $0$ & $-486$ & $-486$ & $0$ & $0$ \\
$\langle Z_{36}^{38} \rangle$ & $54$ & $54$ & $0$ & $0$ & $0$ & $-486$ & $-486$ & $0$ & $0$ \\
\end{tabular}
\end{ruledtabular}
\end{table*}

\begin{table*}
\caption{\label{t:mm2733T}Nontrivial matrix elements of the operators involved in the magnetic moments of decuplet baryons: Flavor $\mathbf{27}$ representation. The entries correspond to $48\langle Z_{m}^{33}\rangle$.}
\begin{ruledtabular}
\begin{tabular}{lcccccccccc}
& $\displaystyle \Delta^{++}$ & $\displaystyle \Delta^+$ & $\displaystyle \Delta^0$ & $\displaystyle \Delta^-$ & $\displaystyle {\Sigma^*}^+$ & $\displaystyle {\Sigma^*}^0$ & $\displaystyle {\Sigma^*}^-$ & $\displaystyle {\Xi^*}^0$ & $\displaystyle {\Xi^*}^-$ & $\displaystyle \Omega^-$ \\[2mm]
\hline
$\langle Z_{1}^{33} \rangle$ & $12$ & $4$ & $-4$ & $-12$ & $8$ & $0$ & $-8$ & $4$ & $-4$ & $0$ \\
$\langle Z_{2}^{33} \rangle$ & $0$ & $0$ & $0$ & $0$ & $0$ & $0$ & $0$ & $0$ & $0$ & $0$ \\
$\langle Z_{3}^{33} \rangle$ & $0$ & $0$ & $0$ & $0$ & $0$ & $0$ & $0$ & $0$ & $0$ & $0$ \\
$\langle Z_{4}^{33} \rangle$ & $0$ & $0$ & $0$ & $0$ & $0$ & $0$ & $0$ & $0$ & $0$ & $0$ \\
$\langle Z_{5}^{33} \rangle$ & $108$ & $36$ & $-36$ & $-108$ & $72$ & $0$ & $-72$ & $36$ & $-36$ & $0$ \\
$\langle Z_{6}^{33} \rangle$ & $36$ & $12$ & $-12$ & $-36$ & $0$ & $0$ & $0$ & $-12$ & $12$ & $0$ \\
$\langle Z_{7}^{33} \rangle$ & $-36$ & $-12$ & $12$ & $36$ & $0$ & $0$ & $0$ & $12$ & $-12$ & $0$ \\
$\langle Z_{8}^{33} \rangle$ & $180$ & $60$ & $-60$ & $-180$ & $120$ & $0$ & $-120$ & $60$ & $-60$ & $0$ \\
$\langle Z_{10}^{33} \rangle$ & $0$ & $0$ & $0$ & $0$ & $0$ & $0$ & $0$ & $0$ & $0$ & $0$ \\
$\langle Z_{12}^{33} \rangle$ & $540$ & $180$ & $-180$ & $-540$ & $360$ & $0$ & $-360$ & $180$ & $-180$ & $0$ \\
$\langle Z_{14}^{33} \rangle$ & $0$ & $0$ & $0$ & $0$ & $0$ & $0$ & $0$ & $0$ & $0$ & $0$ \\
$\langle Z_{15}^{33} \rangle$ & $108$ & $36$ & $-36$ & $-108$ & $0$ & $0$ & $0$ & $36$ & $-36$ & $0$ \\
$\langle Z_{16}^{33} \rangle$ & $108$ & $36$ & $-36$ & $-108$ & $0$ & $0$ & $0$ & $36$ & $-36$ & $0$ \\
$\langle Z_{17}^{33} \rangle$ & $45$ & $15$ & $-15$ & $-45$ & $24$ & $0$ & $-24$ & $27$ & $-27$ & $0$ \\
$\langle Z_{18}^{33} \rangle$ & $45$ & $15$ & $-15$ & $-45$ & $0$ & $0$ & $0$ & $3$ & $-3$ & $0$ \\
$\langle Z_{19}^{33} \rangle$ & $90$ & $30$ & $-30$ & $-90$ & $24$ & $0$ & $-24$ & $-6$ & $6$ & $0$ \\
$\langle Z_{20}^{33} \rangle$ & $-90$ & $-30$ & $30$ & $90$ & $-24$ & $0$ & $24$ & $6$ & $-6$ & $0$ \\
$\langle Z_{21}^{33} \rangle$ & $90$ & $30$ & $-30$ & $-90$ & $0$ & $0$ & $0$ & $-30$ & $30$ & $0$ \\
$\langle Z_{22}^{33} \rangle$ & $-90$ & $-30$ & $30$ & $90$ & $0$ & $0$ & $0$ & $30$ & $-30$ & $0$ \\
$\langle Z_{23}^{33} \rangle$ & $90$ & $30$ & $-30$ & $-90$ & $0$ & $0$ & $0$ & $-30$ & $30$ & $0$ \\
$\langle Z_{24}^{33} \rangle$ & $-90$ & $-30$ & $30$ & $90$ & $0$ & $0$ & $0$ & $30$ & $-30$ & $0$ \\
$\langle Z_{25}^{33} \rangle$ & $0$ & $0$ & $0$ & $0$ & $0$ & $0$ & $0$ & $0$ & $0$ & $0$ \\
$\langle Z_{26}^{33} \rangle$ & $810$ & $270$ & $-270$ & $-810$ & $540$ & $0$ & $-540$ & $270$ & $-270$ & $0$ \\
$\langle Z_{27}^{33} \rangle$ & $324$ & $108$ & $-108$ & $-324$ & $0$ & $0$ & $0$ & $108$ & $-108$ & $0$ \\
$\langle Z_{28}^{33} \rangle$ & $135$ & $45$ & $-45$ & $-135$ & $72$ & $0$ & $-72$ & $81$ & $-81$ & $0$ \\
$\langle Z_{29}^{33} \rangle$ & $135$ & $45$ & $-45$ & $-135$ & $0$ & $0$ & $0$ & $9$ & $-9$ & $0$ \\
$\langle Z_{30}^{33} \rangle$ & $270$ & $90$ & $-90$ & $-270$ & $0$ & $0$ & $0$ & $-90$ & $90$ & $0$ \\
$\langle Z_{31}^{33} \rangle$ & $-270$ & $-90$ & $90$ & $270$ & $0$ & $0$ & $0$ & $90$ & $-90$ & $0$ \\
$\langle Z_{32}^{33} \rangle$ & $270$ & $90$ & $-90$ & $-270$ & $0$ & $0$ & $0$ & $-90$ & $90$ & $0$ \\
$\langle Z_{33}^{33} \rangle$ & $-270$ & $-90$ & $90$ & $270$ & $0$ & $0$ & $0$ & $90$ & $-90$ & $0$ \\
$\langle Z_{34}^{33} \rangle$ & $270$ & $90$ & $-90$ & $-270$ & $0$ & $0$ & $0$ & $90$ & $-90$ & $0$ \\
$\langle Z_{35}^{33} \rangle$ & $270$ & $90$ & $-90$ & $-270$ & $0$ & $0$ & $0$ & $90$ & $-90$ & $0$ \\
$\langle Z_{36}^{33} \rangle$ & $270$ & $90$ & $-90$ & $-270$ & $0$ & $0$ & $0$ & $90$ & $-90$ & $0$ \\
\end{tabular}
\end{ruledtabular}
\end{table*}

\begin{table*}
\caption{\label{t:mm2738T}Nontrivial matrix elements of the operators involved in the magnetic moments of decuplet baryons: Flavor $\mathbf{27}$ representation. The entries correspond to $48\sqrt{3}\langle Z_{m}^{38} \rangle$.}
\begin{ruledtabular}
\begin{tabular}{lcccccccccc}
& $\displaystyle \Delta^{++}$ & $\displaystyle \Delta^+$ & $\displaystyle \Delta^0$ & $\displaystyle \Delta^-$ & $\displaystyle {\Sigma^*}^+$ & $\displaystyle {\Sigma^*}^0$ & $\displaystyle {\Sigma^*}^-$ & $\displaystyle {\Xi^*}^0$ & $\displaystyle {\Xi^*}^-$ & $\displaystyle \Omega^-$ \\[2mm]
\hline
$\langle Z_{1}^{38} \rangle$ & $12$ & $12$ & $12$ & $12$ & $0$ & $0$ & $0$ & $-12$ & $-12$ & $-24$ \\
$\langle Z_{2}^{38} \rangle$ & $36$ & $36$ & $36$ & $36$ & $0$ & $0$ & $0$ & $-36$ & $-36$ & $-72$ \\
$\langle Z_{3}^{38} \rangle$ & $-72$ & $-72$ & $-72$ & $-72$ & $-72$ & $-72$ & $-72$ & $-72$ & $-72$ & $-72$ \\
$\langle Z_{4}^{38} \rangle$ & $108$ & $108$ & $108$ & $108$ & $0$ & $0$ & $0$ & $-108$ & $-108$ & $-216$ \\
$\langle Z_{5}^{38} \rangle$ & $108$ & $108$ & $108$ & $108$ & $0$ & $0$ & $0$ & $-108$ & $-108$ & $-216$ \\
$\langle Z_{6}^{38} \rangle$ & $-36$ & $-36$ & $-36$ & $-36$ & $0$ & $0$ & $0$ & $-36$ & $-36$ & $-144$ \\
$\langle Z_{7}^{38} \rangle$ & $-36$ & $-36$ & $-36$ & $-36$ & $0$ & $0$ & $0$ & $-36$ & $-36$ & $-144$ \\
$\langle Z_{8}^{38} \rangle$ & $180$ & $180$ & $180$ & $180$ & $0$ & $0$ & $0$ & $-180$ & $-180$ & $-360$ \\
$\langle Z_{10}^{38} \rangle$ & $540$ & $540$ & $540$ & $540$ & $0$ & $0$ & $0$ & $-540$ & $-540$ & $-1080$ \\
$\langle Z_{12}^{38} \rangle$ & $540$ & $540$ & $540$ & $540$ & $0$ & $0$ & $0$ & $-540$ & $-540$ & $-1080$ \\
$\langle Z_{14}^{38} \rangle$ & $-540$ & $-540$ & $-540$ & $-540$ & $-540$ & $-540$ & $-540$ & $-540$ & $-540$ & $-540$ \\
$\langle Z_{15}^{38} \rangle$ & $108$ & $108$ & $108$ & $108$ & $0$ & $0$ & $0$ & $-108$ & $-108$ & $-864$ \\
$\langle Z_{16}^{38} \rangle$ & $108$ & $108$ & $108$ & $108$ & $0$ & $0$ & $0$ & $-108$ & $-108$ & $-864$ \\
$\langle Z_{17}^{38} \rangle$ & $45$ & $45$ & $45$ & $45$ & $0$ & $0$ & $0$ & $-81$ & $-81$ & $-360$ \\
$\langle Z_{18}^{38} \rangle$ & $45$ & $45$ & $45$ & $45$ & $0$ & $0$ & $0$ & $-81$ & $-81$ & $-360$ \\
$\langle Z_{19}^{38} \rangle$ & $-90$ & $-90$ & $-90$ & $-90$ & $-72$ & $-72$ & $-72$ & $-162$ & $-162$ & $-360$ \\
$\langle Z_{20}^{38} \rangle$ & $-90$ & $-90$ & $-90$ & $-90$ & $-72$ & $-72$ & $-72$ & $-162$ & $-162$ & $-360$ \\
$\langle Z_{21}^{38} \rangle$ & $-90$ & $-90$ & $-90$ & $-90$ & $0$ & $0$ & $0$ & $-90$ & $-90$ & $-360$ \\
$\langle Z_{22}^{38} \rangle$ & $-90$ & $-90$ & $-90$ & $-90$ & $0$ & $0$ & $0$ & $-90$ & $-90$ & $-360$ \\
$\langle Z_{23}^{38} \rangle$ & $-90$ & $-90$ & $-90$ & $-90$ & $0$ & $0$ & $0$ & $-90$ & $-90$ & $-360$ \\
$\langle Z_{24}^{38} \rangle$ & $-90$ & $-90$ & $-90$ & $-90$ & $0$ & $0$ & $0$ & $-90$ & $-90$ & $-360$ \\
$\langle Z_{25}^{38} \rangle$ & $810$ & $810$ & $810$ & $810$ & $0$ & $0$ & $0$ & $-810$ & $-810$ & $-1620$ \\
$\langle Z_{26}^{38} \rangle$ & $810$ & $810$ & $810$ & $810$ & $0$ & $0$ & $0$ & $-810$ & $-810$ & $-1620$ \\
$\langle Z_{27}^{38} \rangle$ & $324$ & $324$ & $324$ & $324$ & $0$ & $0$ & $0$ & $-324$ & $-324$ & $-2592$ \\
$\langle Z_{28}^{38} \rangle$ & $135$ & $135$ & $135$ & $135$ & $0$ & $0$ & $0$ & $-243$ & $-243$ & $-1080$ \\
$\langle Z_{29}^{38} \rangle$ & $135$ & $135$ & $135$ & $135$ & $0$ & $0$ & $0$ & $-243$ & $-243$ & $-1080$ \\
$\langle Z_{30}^{38} \rangle$ & $-270$ & $-270$ & $-270$ & $-270$ & $0$ & $0$ & $0$ & $-270$ & $-270$ & $-1080$ \\
$\langle Z_{31}^{38} \rangle$ & $-270$ & $-270$ & $-270$ & $-270$ & $0$ & $0$ & $0$ & $-270$ & $-270$ & $-1080$ \\
$\langle Z_{32}^{38} \rangle$ & $-270$ & $-270$ & $-270$ & $-270$ & $0$ & $0$ & $0$ & $-270$ & $-270$ & $-1080$ \\
$\langle Z_{33}^{38} \rangle$ & $-270$ & $-270$ & $-270$ & $-270$ & $0$ & $0$ & $0$ & $-270$ & $-270$ & $-1080$ \\
$\langle Z_{34}^{38} \rangle$ & $270$ & $270$ & $270$ & $270$ & $0$ & $0$ & $0$ & $-270$ & $-270$ & $-2160$ \\
$\langle Z_{35}^{38} \rangle$ & $270$ & $270$ & $270$ & $270$ & $0$ & $0$ & $0$ & $-270$ & $-270$ & $-2160$ \\
$\langle Z_{36}^{38} \rangle$ & $270$ & $270$ & $270$ & $270$ & $0$ & $0$ & $0$ & $-270$ & $-270$ & $-2160$ \\
\end{tabular}
\end{ruledtabular}
\end{table*}

\begin{table*}
\caption{\label{t:mm2733TO}Nontrivial matrix elements of the operators involved in the decuplet to octet transition magnetic moments: Flavor $\mathbf{27}$ representation. The entries correspond to $36\sqrt{2}\langle Z_{m}^{33}\rangle$ and $36\sqrt{6}\langle Z_{m}^{38} \rangle$.}
\begin{ruledtabular}
\begin{tabular}{lcccccccc}
& $\displaystyle \Delta^+p$ & $\displaystyle \Delta^0n$ & $\displaystyle {\Sigma^*}^0\Lambda$ & $\displaystyle {\Sigma^*}^0\Sigma^0$ & $\displaystyle {\Sigma^*}^+\Sigma^+$ & $\displaystyle {\Sigma^*}^-\Sigma^-$ & $\displaystyle {\Xi^*}^0\Xi^0$ & $\displaystyle {\Xi^*}^-\Xi^-$ \\[2mm]
\hline
$\langle Z_{1}^{33} \rangle$ & $8$ & $8$ & $-4 \sqrt{3}$ & $0$ & $4$ & $-4$ & $4$ & $-4$ \\
$\langle Z_{2}^{33} \rangle$ & $0$ & $0$ & $0$ & $0$ & $0$ & $0$ & $0$ & $0$ \\
$\langle Z_{6}^{33} \rangle$ & $24$ & $24$ & $0$ & $0$ & $0$ & $0$ & $-12$ & $12$ \\
$\langle Z_{7}^{33} \rangle$ & $0$ & $0$ & $0$ & $0$ & $-24$ & $24$ & $-12$ & $12$ \\
$\langle Z_{9}^{33} \rangle$ & $36$ & $36$ & $-18 \sqrt{3}$ & $0$ & $18$ & $-18$ & $18$ & $-18$ \\
$\langle Z_{11}^{33} \rangle$ & $0$ & $0$ & $0$ & $0$ & $0$ & $0$ & $0$ & $0$ \\
$\langle Z_{13}^{33} \rangle$ & $108$ & $108$ & $-54 \sqrt{3}$ & $0$ & $54$ & $-54$ & $54$ & $-54$ \\
$\langle Z_{15}^{33} \rangle$ & $72$ & $72$ & $0$ & $0$ & $0$ & $0$ & $36$ & $-36$ \\
$\langle Z_{16}^{33} \rangle$ & $0$ & $0$ & $0$ & $0$ & $0$ & $0$ & $-36$ & $36$ \\
$\langle Z_{17}^{33} \rangle$ & $18$ & $18$ & $-12 \sqrt{3}$ & $0$ & $24$ & $-24$ & $39$ & $-39$ \\
$\langle Z_{18}^{33} \rangle$ & $0$ & $0$ & $0$ & $0$ & $30$ & $-30$ & $15$ & $-15$ \\
$\langle Z_{21}^{33} \rangle$ & $0$ & $0$ & $6 \sqrt{3}$ & $0$ & $42$ & $-42$ & $12$ & $-12$ \\
$\langle Z_{22}^{33} \rangle$ & $-36$ & $-36$ & $-6 \sqrt{3}$ & $0$ & $-6$ & $6$ & $24$ & $-24$ \\
$\langle Z_{23}^{33} \rangle$ & $36$ & $36$ & $6 \sqrt{3}$ & $0$ & $6$ & $-6$ & $-24$ & $24$ \\
$\langle Z_{24}^{33} \rangle$ & $0$ & $0$ & $-6 \sqrt{3}$ & $0$ & $-42$ & $42$ & $-12$ & $12$ \\
$\langle Z_{32}^{33} \rangle$ & $108$ & $108$ & $0$ & $0$ & $0$ & $0$ & $-54$ & $54$ \\
$\langle Z_{33}^{33} \rangle$ & $0$ & $0$ & $0$ & $0$ & $-108$ & $108$ & $-54$ & $54$ \\
$\langle Z_{34}^{33} \rangle$ & $0$ & $0$ & $0$ & $0$ & $0$ & $0$ & $-36$ & $36$ \\
$\langle Z_{35}^{33} \rangle$ & $108$ & $108$ & $0$ & $0$ & $0$ & $0$ & $72$ & $-72$ \\
$\langle Z_{36}^{33} \rangle$ & $0$ & $0$ & $0$ & $0$ & $36$ & $-36$ & $-72$ & $72$ \\
\hline
$\langle Z_{1}^{38} \rangle$ & $0$ & $0$ & $0$ & $12$ & $12$ & $12$ & $12$ & $12$ \\
$\langle Z_{2}^{38} \rangle$ & $0$ & $0$ & $0$ & $36$ & $36$ & $36$ & $36$ & $36$ \\
$\langle Z_{6}^{38} \rangle$ & $0$ & $0$ & $0$ & $0$ & $0$ & $0$ & $36$ & $36$ \\
$\langle Z_{7}^{38} \rangle$ & $0$ & $0$ & $0$ & $0$ & $0$ & $0$ & $36$ & $36$ \\
$\langle Z_{9}^{38} \rangle$ & $0$ & $0$ & $0$ & $54$ & $54$ & $54$ & $54$ & $54$ \\
$\langle Z_{11}^{38} \rangle$ & $0$ & $0$ & $0$ & $162$ & $162$ & $162$ & $162$ & $162$ \\
$\langle Z_{13}^{38} \rangle$ & $0$ & $0$ & $0$ & $162$ & $162$ & $162$ & $162$ & $162$ \\
$\langle Z_{15}^{38} \rangle$ & $0$ & $0$ & $0$ & $0$ & $0$ & $0$ & $108$ & $108$ \\
$\langle Z_{16}^{38} \rangle$ & $0$ & $0$ & $0$ & $0$ & $0$ & $0$ & $108$ & $108$ \\
$\langle Z_{17}^{38} \rangle$ & $0$ & $0$ & $0$ & $72$ & $72$ & $72$ & $117$ & $117$ \\
$\langle Z_{18}^{38} \rangle$ & $0$ & $0$ & $0$ & $72$ & $72$ & $72$ & $117$ & $117$ \\
$\langle Z_{21}^{38} \rangle$ & $0$ & $0$ & $0$ & $-18$ & $-18$ & $-18$ & $72$ & $72$ \\
$\langle Z_{22}^{38} \rangle$ & $0$ & $0$ & $0$ & $-18$ & $-18$ & $-18$ & $72$ & $72$ \\
$\langle Z_{23}^{38} \rangle$ & $0$ & $0$ & $0$ & $-18$ & $-18$ & $-18$ & $72$ & $72$ \\
$\langle Z_{24}^{38} \rangle$ & $0$ & $0$ & $0$ & $-18$ & $-18$ & $-18$ & $72$ & $72$ \\
$\langle Z_{32}^{38} \rangle$ & $0$ & $0$ & $0$ & $0$ & $0$ & $0$ & $162$ & $162$ \\
$\langle Z_{33}^{38} \rangle$ & $0$ & $0$ & $0$ & $0$ & $0$ & $0$ & $162$ & $162$ \\
$\langle Z_{34}^{38} \rangle$ & $0$ & $0$ & $0$ & $0$ & $0$ & $0$ & $216$ & $216$ \\
$\langle Z_{35}^{38} \rangle$ & $0$ & $0$ & $0$ & $0$ & $0$ & $0$ & $216$ & $216$ \\
$\langle Z_{36}^{38} \rangle$ & $0$ & $0$ & $0$ & $0$ & $0$ & $0$ & $216$ & $216$ \\
\end{tabular}
\end{ruledtabular}
\end{table*}

\subsubsection{Total contribution of Fig.~\ref{fig:mmloop2}}

The total correction arising from Fig.~\ref{fig:mmloop2} is then given by
\begin{equation}
M_{\textrm{loop 2}}^k = M_{\textrm{loop 2(a-d)}}^k + M_{\textrm{loop 2e}}^k,
\end{equation}
where the first and second summands on the right-hand side of the above expression are given by Eqs.~(\ref{eq:loop2(a-d)}) and (\ref{eq:loop2sum2}), respectively.

Corrections to the baryon magnetic moments are then obtained by computing the matrix elements of operator $M_{\textrm{loop 2}}^k$ between SU(6) baryon states, namely,
\begin{equation}
\mu_B^{(\textrm{loop 2})} = \langle B|M_{\textrm{loop 2}}^3 |B \rangle, \label{eq:mmloop2gr}
\end{equation}
where $B$ stands for either an octet or a decuplet baryon. For decuplet to octet transition magnetic moments, we also have
\begin{equation}
\mu_{TB}^{(\textrm{loop 2})} = \langle T|M_{\textrm{loop 2}}^3 |B \rangle. \label{eq:mmloop2trans}
\end{equation}

The singlet piece of $M_{\textrm{loop 2}}^k$ yields magnetic moments that satisfy the Coleman-Glashow relations
(\ref{eq:treeval}) whereas violations to them are due to the $\mathbf{8}$ and $\mathbf{27}$ pieces. Their effects can be better seen in the sum rules (\ref{eq:cp1})-(\ref{eq:cp3}), which are no longer satisfied at this order. We shall not write down the resulting expressions because they can be obtained without trouble by reading off the matrix elements of the corresponding operators displayed in Tables \ref{t:mm133B}-\ref{t:mm2733TO}.

We also notice that, for decuplet baryons, the analogs of Eqs.~(\ref{eq:srd1})-(\ref{eq:srd3}) read
\begin{equation}
\mu_{\Delta^{++}}^{\textrm{(loop 2)}} - \mu_{\Delta^+}^{\textrm{(loop 2)}} - \mu_{\Delta^0}^{\textrm{(loop 2)}} + \mu_{\Delta^-}^{\textrm{(loop 2)}} = 0,
\end{equation}
\begin{equation}
\mu_{{\Sigma^*}^+}^{\textrm{(loop 2)}} - 2 \mu_{{\Sigma^*}^0}^{\textrm{(loop 2)}} + \mu_{{\Sigma^*}^-}^{\textrm{(loop 2)}} = 0,
\end{equation}
and
\begin{equation}
\mu_{\Delta^{++}}^{\textrm{(loop 2)}} - 3 \mu_{\Delta^+}^{\textrm{(loop 2)}} + 3 \mu_{\Delta^0}^{\textrm{(loop 2)}} - \mu_{\Delta^-}^{\textrm{(loop 2)}} = 0,
\end{equation}
whereas for transition magnetic moments, the analogs of Eqs.~(\ref{eq:srdo1})-(\ref{eq:srdo2}) are
\begin{equation}
\mu_{\Delta^{+}p}^{\textrm{(loop 2)}} - \mu_{\Delta^{0}n}^{\textrm{(loop 2)}} = 0,
\end{equation}
and
\begin{equation}
\mu_{{\Sigma^{*}}^+\Sigma^+}^{\textrm{(loop 2)}} - 2 \mu_{{\Sigma^{*}}^0\Sigma^0}^{\textrm{(loop 2)}} + \mu_{{\Sigma^{*}}^-\Sigma^-}^{\textrm{(loop 2)}} = 0.
\end{equation}

On the other hand, to this order, sum rules (36) of Ref.~\cite{milana} are no longer satisfied.

In passing, let us mention that the flavor $\mathbf{27}$ piece of $\langle B|M_{\textrm{loop 2(a-d)}}^3|B\rangle$ is responsible for the small difference observed in the relation amongst magnetic moments of octet baryons given in Eq.~(21) of Ref.~\cite{jen92}, namely,
\begin{eqnarray}
&  & (6\mu_{\Lambda} + \mu_{\Sigma^-} - 4\sqrt{3}\mu_{\Lambda\Sigma^0}) - (4\mu_n-\mu_{\Sigma^+} + 4\mu_{\Xi^0}) \nonumber \\
&  & \mbox{} = F_{\mathbf{27}} f(a_1,\ldots,c_3,m_1,\ldots,m_4),
\end{eqnarray}
where $F_{\mathbf{27}}$ is the flavor $\mathbf{27}$ combination of the integrals over the loops given in Eq.~(\ref{eq:F27}) and $f(a_1,\ldots,c_3,m_1,\ldots,m_4)$ is a function which depends quadratically on $a_1,\ldots,c_3$ but linearly on $m_1,\ldots,m_4$. The function $F_{\mathbf{27}}$ is highly suppressed with respect to the flavor singlet and octet combinations, which explains such a small discrepancy.

\subsection{Total one-loop corrections to baryon magnetic moments}

At this point, we can summarize our findings and provide analytic results. Thus, the final expression of the magnetic moment of baryon $B$ up to one-loop order can be organized succinctly as
\begin{equation}
\mu_B = \mu_B^{(0)} + \mu_B^{(\textrm{loop 1})} + \mu_B^{(\textrm{loop 2})}. \label{eq:fullexp}
\end{equation}
Applications of this expression will be given in subsequent sections.

\subsubsection{Neutron magnetic moment as a case example}

Here we present the full expression at one-loop order for the magnetic moment of neutron just as an example. Although the form of the operators which originate it might look breathtaking, the final result gets simplified to a great extent. Thus one has
\begin{widetext}

\begin{eqnarray}
\mu_n & = & -\frac{m_1}{3} - \frac{m_3}{9} + I(m_\pi) \left[ \frac{13}{12} a_1^2 + \frac13 a_1 b_2 + \frac{1}{36} b_2^2 + \frac{29}{54} a_1 b_3 + \frac19 b_2 b_3 + \frac{5}{18} a_1 c_3 \right] \nonumber \\
&  & \mbox{} + I(m_K) \left[ \frac{5}{12} a_1^2 + \frac16 a_1 b_2 - \frac{1}{36} b_2^2 + \frac{7}{54} a_1 b_3 + \frac{1}{18} b_2b_3 + \frac{2}{9} a_1 c_3 \right] \nonumber \\
&  & \mbox{} + F(m_\pi) \left[ -\frac{5}{6} m_1 - \frac{m_2}{6} - \frac{5}{18} m_3 - \frac{7}{24} a_1^2 m_1-\frac{1}{36} a_1 b_2 m_1-\frac{7}{216} b_2^2 m_1-\frac{5}{108} a_1 b_3 m_1-\frac{35}{324} b_2
b_3 m_1 \right. \nonumber \\
&  & \mbox{\hglue2.0truecm} - \frac{2}{9} a_1 c_3 m_1 + \frac{4}{27} b_2 c_3 m_1 + \frac{35}{72} a_1^2 m_2 + \frac{5}{108} a_1b_2m_2 + \frac{1}{216} b_2^2m_2 + \frac{25}{324} a_1 b_3 m_2 \nonumber \\
&  & \mbox{\hglue2.0truecm} \left. + \frac{10}{27} a_1 c_3 m_2 - \frac{7}{72} a_1^2 m_3 - \frac{35}{324} a_1 b_2 m_3 + \frac{20}{27} a_1^2 m_4+\frac{4}{27} a_1 b_2 m_4 \right] \nonumber \\
&  & \mbox{} + F(m_K) \left[ -\frac{m_1}{6} +\frac{m_2}{6} -\frac{m_3}{18} - \frac{7}{24} a_1^2 m_1-\frac{1}{12} a_1 b_2 m_1-\frac{23}{216} b_2^2 m_1-\frac{7}{36} a_1 b_3 m_1 - \frac{25}{324} b_2 b_3
m_1 \right. \nonumber \\
&  & \mbox{\hglue2.0truecm} + \frac{2}{27} b_2 c_3 m_1 + \frac{13}{72} a_1^2 m_2+\frac{1}{108} a_1 b_2 m_2-\frac{1}{216} b_2^2 m_2 - \frac{1}{324} a_1 b_3 m_2 + \frac{5}{27} a_1 c_3 m_2 \nonumber \\
&  & \mbox{\hglue2.0truecm} \left. + \frac{11}{216} a_1^2 m_3-\frac{25}{324} a_1 b_2 m_3+\frac{4}{27} a_1^2 m_4+\frac{2}{27} a_1 b_2 m_4 \right] \nonumber \\
&  & \mbox{} + F(m_\eta) \left[ -\frac{1}{18} a_1^2 m_1-\frac{1}{9} a_1 b_2 m_1-\frac{1}{18} b_2^2 m_1-\frac{1}{27} a_1 b_3 m_1-\frac{1}{27} b_2 b_3 m_1-\frac{1}{54}
a_1^2 m_3-\frac{1}{27} a_1 b_2 m_3 \right]. \label{eq:neutronmm}
\end{eqnarray}
\end{widetext}

Expression (\ref{eq:neutronmm}) along with the additional 26 remaining are the ones actually used in the comparison with other analytic results and the experiment. Let us carry on with the analysis and perform such comparisons for the sake of completeness.

\section{Comparison with conventional heavy baryon chiral perturbation theory}\label{sec:comparison}

It is instructive to compare our computation of baryon magnetic moments at the physical value $N_c=3$ with the one obtained in the framework of conventional heavy baryon chiral perturbation theory, \textit{i.e.}, the effective field theory with no $1/N_c$ expansion. In Ref.~\cite{jen96} it has been shown that there is a one-to-one correspondence between the parameters of the $1/N_c$ baryon chiral Lagrangian at $N_c=3$ and the octet and decuplet chiral Lagrangian. The baryon-pion couplings are related to the coefficients of the $1/N_c$ expansion of $A^{ia}$, Eq.~(\ref{eq:akc}), at $N_c=3$ by \cite{jen96}
\begin{subequations}
\label{eq:rel1}
\begin{eqnarray}
&  & D = \frac12 a_1 + \frac16 b_3, \\
&  & F = \frac13 a_1 + \frac16 b_2 + \frac19 b_3, \\
&  & \mathcal{C} = - a_1 - \frac12 c_3, \\
&  & \mathcal{H} = - \frac32 a_1 - \frac32 b_2 - \frac52 b_3.
\end{eqnarray}
\end{subequations}

On the other hand, the magnetic moments in conventional heavy baryon chiral perturbation theory are parametrized by four SU(3) invariants $\mu_D$, $\mu_F$, $\mu_C$ and $\mu_T$ \cite{jen92} while in the present analysis they are parametrized in terms of $m_i$, with $i=1,\ldots,4$, introduced in Eq.~(\ref{eq:mmag}). We recall that Eqs.~(\ref{eq:tlo}), (\ref{eq:tlt}), and (\ref{eq:tlto}) suggest a close connection between these two sets of parameters. This is indeed the case and, at $N_c=3$, they are related by
\begin{subequations}
\label{eq:rel2}
\begin{eqnarray}
&  & \mu_D = \frac12 m_1 + \frac16 m_3, \\
&  & \mu_F = \frac13 m_1 + \frac16 m_2 + \frac19 m_3, \\
&  & \mu_C = \frac12 m_1 + \frac12 m_2 + \frac56 m_3, \\
&  & \mu_T = -2m_1 - m_4.
\end{eqnarray}
\end{subequations}

In the literature, there are some analyses of baryon magnetic moments within heavy baryon chiral perturbation theory which allow us to carry out a comparison of our respective results in the limit $\Delta \to 0$, where $\Delta$ is the decuplet-octet mass difference. The work by Jenkins \textit{et.\ al.} \cite{jen92} about octet baryons allows a full comparison between one-loop corrections whereas the papers by Geng \textit{et.\ al.} \cite{geng2} for decuplet baryons, and Arndt and Tiburzi \cite{tib} for decuplet-octet transitions only allow partial comparisons of contributions emerging from loop graphs of Fig.~\ref{fig:mmloop1}.

For octet baryons, we obtain a remarkable agreement with the theoretical expressions displayed in Eq.~(16) of Ref.~\cite{jen92} when using relations (\ref{eq:rel1}) and (\ref{eq:rel2}), except for the global factor $-5/2$ missing in the loop contributions of Fig.~2(c) of this reference (which corresponds to Fig.~\ref{fig:mmloop2}(b) in the present paper). When fixing this omission, the agreement is achieved for all nine observables.

As for decuplet baryons, starting from Eq.~(17) of Ref.~\cite{geng2} and working in the limit $\Delta\to 0$, we find a perfect agreement between their results and ours, once their conventions about the couplings $\mathcal{C}$ and $\mathcal{H}$ are taking into account.

Finally, Arndt and Tiburzi \cite{tib} present the calculation of baryon decuplet to octet electromagnetic transition form factors in quenched and partially quenched chiral perturbation theory, and provide the corresponding SU(3) coefficients emerging from these schemes. They also present the counterparts of such coefficients that appear in chiral perturbation theory. These are precisely the coefficients we need to compare with. We find that, except for transitions ${\Sigma^*}^+ \Sigma^+$ and ${\Xi^*}^0\Xi^0$, the theoretical expressions differ by a global sign.

We would like to close this section by stating that, to the order of approximation implemented here, both approaches lead to the same results. This fact causes no surprise. Previous works have shown this matching for baryon masses \cite{jen96,rfmzac} and axial-vector couplings \cite{rfm06} in a systematic way.

We now examine another aspect of the suggested comparisons, this time with experimental data.

\section{Numerical Analysis}\label{sec:numerical}

At this point we are able to perform a detailed numerical comparison of the expressions obtained in our analysis with the available experimental data \cite{part} through a least-squares fit. The data consists of seven out of the eight possible magnetic moments of the baryon octet ($\mu_{\Sigma^0}$ has not been measured yet), together with $\mu_{\Omega^-}$, $\mu_{\Lambda\Sigma^0}$ and $\mu_{\Delta^+p}$. Another piece of information which can also be incorporated is
$\mu_{\Delta^{++}}=6.14 \pm 0.51 \mu_N$, value obtained through a study of radiative $\pi^+p$ scattering within a dynamical model in Ref.~\cite{lopez}. All this information is summarized in the second column (from left to right) of Table \ref{t:numerical}. All in all, we have 11 observables at our disposal to perform the fit.

The analytic expressions used are written in terms of two sets of parameters: the first one is constituted by $a_1$, $b_2$, $b_3$ and $c_3$ arising from the $1/N_c$ expansion of $A^{kc}$, Eq.~(\ref{eq:akc}), and the latter is formed by $m_1,\ldots,m_4$ arising from the $1/N_c$ expansion of $M^{kc}$, Eq.~(\ref{eq:mmag}). According to the naive large-$N_c$ counting, these parameters should be of order $\mathcal{O}(N_c^0)$. Previous works \cite{dai,fmhjm,rfm06} have found that this is indeed the case for the first set of parameters. However, in order to ensure that this also occurs for $m_i$, we can follow Ref.~\cite{lb} and introduce an appropriate scale $\alpha_0=2\mu_p^{\textrm{exp}}$ which multiplies $m_i$. The reasoning for doing so is that $\mu_p$, being the best measured magnetic moment, to leading order in the $1/N_c$ expansion reduces to $m_1/2$ [see Eq.~(\ref{eq:tlo}b)]. Thus, the natural choice for such a scale is the one pointed out above. Notice that we should exercise some caution at this point because the parameters $m_i$ enter linearly at tree level and one-loop order only in contributions of Fig.~\ref{fig:mmloop2}, whereas contributions of Fig.~\ref{fig:mmloop1} do not depend on them at all. Therefore, the actual theoretical expressions we use in the fit take on the form
\begin{equation}
\mu_B = \alpha_0 (\mu_B^{\textrm{(0)}} + \mu_B^{\textrm{(loop 2)}} )  + \mu_B^{\textrm{(loop 1)}}.
\end{equation}

We can proceed with the numerical comparison in several ways. For instance, the first set of parameters can be borrowed from the analyses on baryon semileptonic decays in the form of either the invariant couplings $D$, $F$, $\mathcal{C}$ and $\mathcal{H}$ \cite{jm255,jm259} or the parameters of the $1/N_c$ expansion themselves \cite{rfm06}, both at tree-level and one-loop corrected values. We however found that none of these options lead to a reasonable fit because the corresponding $\chi^2$ was so high that the expansion would break down.

We shall follow a more pragmatic approach by allowing all eight unknown variables to enter into the fit as free parameters. The available experimental data and the total number of free parameters allow this issue. Now, in order to get a meaningful $\chi^2$, we add a roughly estimated theoretical error of $0.05 \, \mu_N$ to each magnetic moment, guessing that the higher order effects in symmetry breaking are at a few percent level. This procedure will also avoid a bias towards the best measured quantities.

Thus, after a standard procedure, we find that the best-fit parameters are
\begin{eqnarray}
\begin{array}{lcl}
a_1 =  1.06 \pm 0.12, & \quad & m_1 =  1.29 \pm 0.04, \\[3mm]
b_2 = -1.05 \pm 0.19, &       & m_2 =  0.34 \pm 0.16, \\[3mm]
b_3 = -1.11 \pm 0.19, &       & m_3 = -0.14 \pm 0.10, \\[3mm]
c_3 = -0.91 \pm 0.16, &       & m_4 =  0.07 \pm 0.24, \\[3mm]
\end{array}
\end{eqnarray}
with $\chi^2=7.53$ for 3 degrees of freedom and the quoted errors come from the fit only. The higher contributions to $\chi^2$ come from $\mu_n$ $(\Delta \chi^2=1.55)$, $\mu_\Lambda$ $(\Delta \chi^2=1.15)$ and $\mu_{\Delta^{++}}$ $(\Delta \chi^2=2.14)$. We should remark that this relatively high $\chi^2$ is a consequence of our working assumptions. From relations (\ref{eq:rel1}), we find numerically that $F=0.05$, $D=0.34$, $\mathcal{C}=0.60$ and $\mathcal{H}=-2.76$, values that differ from their counterparts extracted from baryon semileptonic decays \cite{jm255,jm259}. Thus, the extraction of these parameters from baryon magnetic moments will have to await for both new and better measurements. In a similar fashion, we also find $\mu_F = 2.63$, $\mu_D = 3.47$, $\mu_C = 3.88$ and $\mu_H = -14.83$, which also differ from other determinations \cite{jen92}.

Nevertheless, the best-fit parameters obtained are quite interesting and agree with expectations. We notice that the first set of parameters are order $\mathcal{O}(1)$, as expected. As for the second set, with the introduction of the scale $\alpha_0$, the values are in quite good agreement with the $1/N_c$ predictions: The leading order parameter $m_1$ is order $\mathcal{O}(1)$, whereas $m_2$, and $m_3$ and $m_4$ are roughly suppressed by $1/N_c$ and $1/N_c^2$, respectively, relative to the leading order parameter.

\begin{table*}
\caption{\label{t:numerical}Numerical values of baryon magnetic moments found in this work and comparison with the available experimental data. The entries are given in nuclear magnetons.}
\begin{ruledtabular}
\begin{tabular}{lcccccc}
\textrm{Baryon} & \textrm{Experimental data} & \textrm{Total} & \textrm{Tree level} & \textrm{Loop 1} & \textrm{Loop 2(a-d)} & \textrm{Loop 2(e)} \\
\hline
$n$                    & $-1.913 \pm 0.000$ & $-1.975$ & $-2.315$ & $-0.075$ & $ 0.136$ & $ 0.278$ \\
$p$                    & $ 2.793 \pm 0.000$ & $ 2.759$ & $ 3.785$ & $-0.106$ & $-0.086$ & $-0.833$ \\
$\Sigma^-$             & $-1.160 \pm 0.025$ & $-1.179$ & $-1.470$ & $ 0.146$ & $ 0.104$ & $ 0.041$ \\
$\Sigma^0$             & $                $ & $ 0.625$ & $ 1.157$ & $-0.037$ & $-0.058$ & $-0.437$ \\
$\Sigma^+$             & $ 2.458 \pm 0.010$ & $ 2.428$ & $ 3.785$ & $-0.220$ & $-0.220$ & $-0.916$ \\
$\Xi^-$                & $-0.651 \pm 0.003$ & $-0.691$ & $-1.470$ & $ 0.085$ & $ 0.056$ & $ 0.638$ \\
$\Xi^0$                & $-1.250 \pm 0.014$ & $-1.301$ & $-2.315$ & $ 0.169$ & $ 0.053$ & $ 0.792$ \\
$\Lambda$              & $-0.613 \pm 0.004$ & $-0.559$ & $-1.157$ & $ 0.037$ & $ 0.124$ & $ 0.437$ \\
$\Lambda\Sigma^0$      & $ 1.61 \pm 0.08$ & $ 1.594$ & $ 2.005$ & $-0.033$ & $-0.013$ & $-0.365$ \\
$\Delta^{++}$          & $ 6.14  \pm 0.51\footnote{Value reported in Ref.~\cite{lopez}.} $ & $ 5.390$ & $ 7.752$ & $-0.185$ & $-0.386$ & $-1.791$ \\
$\Delta^+$             & $                $ & $ 2.383$ & $ 3.876$ & $-0.110$ & $-0.299$ & $-1.085$ \\
$\Delta^0$             & $                $ & $-0.625$ & $ 0.000$ & $-0.034$ & $-0.211$ & $-0.379$ \\
$\Delta^-$             & $                $ & $-3.632$ & $-3.876$ & $ 0.041$ & $-0.123$ & $ 0.327$ \\
${\Sigma^*}^+$         & $                $ & $ 2.519$ & $ 3.876$ & $-0.075$ & $-0.576$ & $-0.706$ \\
${\Sigma^*}^0$         & $                $ & $-0.303$ & $ 0.000$ & $ 0.000$ & $-0.303$ & $ 0.000$ \\
${\Sigma^*}^-$         & $                $ & $-3.126$ & $-3.876$ & $ 0.075$ & $-0.031$ & $ 0.706$ \\
${\Xi^*}^0$            & $                $ & $ 0.149$ & $ 0.000$ & $ 0.034$ & $-0.265$ & $ 0.379$ \\
${\Xi^*}^-$            & $                $ & $-2.596$ & $-3.876$ & $ 0.110$ & $ 0.085$ & $ 1.085$ \\
$\Omega^-$             & $-2.02 \pm 0.05$ & $-2.042$ & $-3.876$ & $ 0.144$ & $ 0.226$ & $ 1.465$ \\
$\Delta^+p$            & $ 3.51 \pm 0.09$ & $ 3.481$ & $ 3.496$ & $ 1.887$ & $-1.266$ & $-0.637$ \\
$\Delta^0n$            & $                $ & $ 3.481$ & $ 3.496$ & $ 1.887$ & $-1.266$ & $-0.637$ \\
${\Sigma^*}^0\Lambda$  & $                $ & $-2.863$ & $-3.027$ & $-2.163$ & $ 1.776$ & $ 0.551$ \\
${\Sigma^*}^0\Sigma^0$ & $                $ & $ 1.924$ & $ 1.748$ & $ 2.393$ & $-1.556$ & $-0.660$ \\
${\Sigma^*}^+\Sigma^+$ & $                $ & $ 3.639$ & $ 3.496$ & $ 4.252$ & $-3.130$ & $-0.979$ \\
${\Sigma^*}^-\Sigma^-$ & $                $ & $ 0.210$ & $ 0.000$ & $ 0.534$ & $ 0.018$ & $-0.342$ \\
${\Xi^*}^0\Xi^0$       & $                $ & $ 3.464$ & $ 3.496$ & $ 4.252$ & $-3.304$ & $-0.979$ \\
${\Xi^*}^-\Xi^-$       & $                $ & $ 0.110$ & $ 0.000$ & $ 0.534$ & $-0.082$ & $-0.342$ \\
\end{tabular}
\end{ruledtabular}
\end{table*}

In Table \ref{t:numerical}, the third column (from left to right) displays the predicted magnetic moments within the combined expansion in $m_q$ and $1/N_c$. The remaining columns display the contributions to these predicted values arising from tree-level and loop graphs from Fig.~\ref{fig:mmloop1}, \ref{fig:mmloop2}(a-d), and Fig.~\ref{fig:mmloop2}(e). The predicted magnetic moments are in good agreement with the existing experimental ones. We are able to also provide some predictions of the unmeasured magnetic moments. They are in good agreement with some other predictions presented in the literature \cite{lb,geng2} and will not be reproduced here. We only mention that, for instance, $\mu_{{\Sigma^*}^0}=0$ at tree level and up to corrections of order $\mathcal{O}(m_q^{1/2})$, but a nonvanishing contribution is picked up due to terms of order $\mathcal{O}(m_q\ln m_q)$. We also note in passing that, for the transitions $\Lambda\Sigma^0$, ${\Sigma^*}^0\Lambda$, ${\Sigma^*}^-\Sigma^-$ and ${\Xi^*}^-\Xi^-$, although their predicted magnetic moments are in magnitude comparable to the ones reported in Ref.~\cite{lb}, they carry the opposite sign.

\section{Summary and conclusions}\label{sec:conclusion}

In this paper we evaluated the magnetic moments of baryons up to one-loop order within heavy baryon chiral perturbation theory in the large-$N_c$ limit, considering corrections of the types $m_q^{1/2}$ and $m_q\ln m_q$. As a starting point, we used the fact that in the large-$N_c$ limit both the baryon axial-vector couplings and the baryon magnetic moments share the same kinematical properties so they can be analyzed in terms of the same operators. Hence, in Sec.~\ref{sec:tree}, we constructed the $1/N_c$ expansion of the baryon magnetic moment operator $M^k$ based on the expansion deduced for the axial-vector current operator. At this level, the matrix elements of $M^k$ yield the tree-level values of magnetic moments. As a byproduct, the Coleman-Glashow relations could be straightforwardly derived. In Sec.~\ref{sec:oneloop} we turned to compute one-loop corrections to the tree-level values, analyzing separately the corresponding Feynman diagrams of the two kinds of loops as they involve rather different mathematical complication. We should stress that one of the most important assumptions was carrying out the study in the degeneracy limit $\Delta\equiv M_T-M_B\to 0$, where $M_T$ and $M_B$ are the SU(3) invariant masses of the decuplet and octet baryon multiplets, respectively. This assumption is not a withdrawal of the analysis due to the fact that in the large-$N_c$ limit, $\Delta$  is order $\mathcal{O}(1/N_c)$ so this limit constitutes a very good first approximation.

The final analytic expression could be cast into Eq.~(\ref{eq:fullexp}). This expression was crosschecked with other calculations and with experiment. Existing analytic results in conventional heavy baryon chiral perturbation theory comprise complete one-loop corrections only for octet baryons \cite{jen92,loyal,puglia} whereas for decuplet baryons and decuplet-octet transitions only corrections of the type $m_q^{1/2}$ are available \cite{geng2,tib}. Barring a few exceptions (global multiplicative factors and/or opposite signs), the comparison with existing analytic results has been a successful one. The advantage of our approach is that one only needs to construct a universal operator $M^k + \delta M^k$, where $\delta M^k$ stands for one-loop corrections, evaluate the matrix elements of this operator between SU(6) baryon states and compute the integrals over the loops. Here we have performed the analysis to relative order $\mathcal{O}(1/N_c^3)$. It is now clear that if we had involved ourselves in evaluating higher order contributions, we would have faced a much more complicated computation, perhaps not yet needed.

On the other hand, the comparison with the available experimental data has been performed through a least-squares fit to evaluate the unknown parameters in the theory (eight in total). This analysis was indeed illuminating. Like Ref.~\cite{jen92}, we also found evidence that the invariant couplings $F$, $D$, $\mathcal{C}$ and $\mathcal{H}$ [related to the parameters of the $1/N_c$ expansion of the axial current by Eq.~(\ref{eq:rel1})], neither at tree level nor one-loop corrected, produce physically admissible fits. We had no other choice but to let all eight parameters enter as free ones in the analysis. The best-fit parameters agree very well with expectations. These parameters are then used to provide numerical values of magnetic moments from the theoretical standpoint; all this information is displayed in Table \ref{t:numerical}.
The available experimental magnetic moments are fairly well reproduced by their theoretical counterparts. In a general fashion, our results can also be compared with other numerical evaluations \cite{geng2,lb} and the agreement is acceptable. It is evident that, in order to be definitive, it should be interesting to redo the analysis for $\Delta \neq 0$. This, however, requires a rather formidable effort which goes beyond the scope of the present paper.

Returning to the main discussion about the comparison of this approach with conventional heavy baryon chiral perturbation theory, it should be emphasized that these two formulations yield to identical results. Nonetheless, in a given context, one or the other might be more inviting for computational ease.

\acknowledgments

The author would like to express his gratitude to Aneesh V.\ Manohar for interesting discussions and for his valuable comments on the manuscript. This work has been partially supported by Consejo Nacional de Ciencia y Tecnolog{\'\i}a and Fondo de Apoyo a la Investigaci\'on (Universidad Aut\'onoma de San Luis Potos{\'\i}), Mexico.

\appendix

\begin{widetext}

\section{Reduction of baryon operators: structure of diagrams of Fig.~\ref{fig:mmloop1} \label{app:bo1}}

In this appendix we turn to explicitly present the computation of the product operator $\epsilon^{ijk} A^{ia}A^{jb} \Gamma^{ab}$ introduced in Eq.~(\ref{eq:corrloop1}). Here $\Gamma^{ab}$ contains two pieces which transforms as flavor $\mathbf{8}$ and flavor $\mathbf{10+\overline{10}}$, respectively. We have performed the calculation by keeping $N_c$ and $N_f$ arbitrary, although the physical values $N_f=3$ and $N_c=3$ are used in the numerical evaluations.

For the flavor $\mathbf{8}$ piece we explicitly have
\begin{equation}
\epsilon^{ijk} f^{abc} G^{ia}G^{jb} = -\frac12 (N_c+N_f)G^{kc} + \frac12 \mathcal{D}_2^{kc},
\end{equation}
\begin{equation}
\epsilon^{ijk} f^{abc} ( G^{ia}\mathcal{D}_2^{jb} + \mathcal{D}_2^{ia}G^{jb} ) = -N_f G^{kc} - \mathcal{O}_3^{kc},
\end{equation}
\begin{equation}
\epsilon^{ijk} f^{abc} \mathcal{D}_2^{ia}\mathcal{D}_2^{jb} = -\frac12 N_f \mathcal{D}_2^{kc},
\end{equation}
\begin{equation}
\epsilon^{ijk} f^{abc} ( G^{ia}\mathcal{D}_3^{jb} + \mathcal{D}_3^{ia}G^{jb} ) = -2(N_c+N_f) G^{kc} -(N_f-2) \mathcal{D}_2^{kc} - (N_c+N_f) \mathcal{O}_3^{kc},
\end{equation}
\begin{equation}
\epsilon^{ijk} f^{abc} ( G^{ia}\mathcal{O}_3^{jb} + \mathcal{O}_3^{ia}G^{jb} ) = \frac32 N_f \mathcal{D}_2^{kc} - \frac12 (N_c+N_f) \mathcal{D}_3^{kc} - \frac12 (N_c+N_f) \mathcal{O}_3^{kc} + \mathcal{D}_4^{kc},
\end{equation}
\begin{equation}
\epsilon^{ijk} f^{abc} (\mathcal{D}_2^{ia}\mathcal{D}_3^{jb} + \mathcal{D}_3^{ia}\mathcal{D}_2^{jb} ) = -N_f \mathcal{D}_3^{kc},
\end{equation}
\begin{equation}
\epsilon^{ijk} f^{abc} (\mathcal{D}_2^{ia}\mathcal{O}_3^{jb} + \mathcal{O}_3^{ia}\mathcal{D}_2^{jb} ) = -N_f \mathcal{O}_3^{kc} - \mathcal{O}_5^{kc}.
\end{equation}

For the flavor $\mathbf{10} + \overline{\mathbf{10}}$ contribution we have
\begin{equation}
\epsilon^{ijk} ( f^{aec}d^{be8} - f^{bec}d^{ae8} - f^{abe}d^{ec8} ) G^{ia}G^{jb} = \frac12 \{T^c,G^{k8}\} - \frac12 \{G^{kc},T^8\} + \frac{1}{N_f} [J^2,[T^8,G^{kc}]],
\end{equation}
\begin{eqnarray}
\epsilon^{ijk} ( f^{aec}d^{be8} - f^{bec}d^{ae8} - f^{abe}d^{ec8} )( G^{ia}\mathcal{D}_2^{jb} + \mathcal{D}_2^{ia}G^{jb} ) & = & -\{G^{kc},\{J^r,G^{r8}\}\} + \{G^{k8},\{J^r,G^{rc}\}\} \nonumber \\
&  & \mbox{} + \frac{N_c+N_f}{N_f} [J^2,[T^8,G^{kc}]],
\end{eqnarray}
\begin{equation}
\epsilon^{ijk} ( f^{aec}d^{be8} - f^{bec}d^{ae8} - f^{abe}d^{ec8} ) \mathcal{D}_2^{ia}\mathcal{D}_2^{jb} = 0,
\end{equation}
\begin{eqnarray}
&  & \epsilon^{ijk} ( f^{aec}d^{be8} - f^{bec}d^{ae8} - f^{abe}d^{ec8} ) ( G^{ia}\mathcal{D}_3^{jb} + \mathcal{D}_3^{ia}G^{jb} ) = 2\{T^c,G^{k8}\} - 2\{G^{kc},T^8\} + \frac{4}{N_f} [J^2,[T^8,G^{kc}]]  \nonumber \\
&  & \mbox{} + \{J^2,\{T^c,G^{k8}\}\} - \{J^2,\{G^{kc},T^8\}\} - \{\mathcal{D}_2^{kc},\{J^r,G^{r8}\}\} + \{\mathcal{D}_2^{k8},\{J^r,G^{rc}\}\}  + \frac{2}{N_f} \{J^2,[J^2,[T^8,G^{kc}]]\},
\end{eqnarray}
\begin{eqnarray}
&  & \epsilon^{ijk} ( f^{aec}d^{be8} - f^{bec}d^{ae8} - f^{abe}d^{ec8} ) ( G^{ia}\mathcal{O}_3^{jb} + \mathcal{O}_3^{ia}G^{jb} ) = \frac12 \{J^2,\{T^c,G^{k8}\}\} - \frac12 \{J^2,\{G^{kc},T^8\}\} \nonumber \\
&  & \mbox{} + \frac12 \{\mathcal{D}_2^{kc},\{J^r,G^{r8}\}\} - \frac12 \{\mathcal{D}_2^{k8},\{J^r,G^{rc}\}\} + \frac{1}{N_f} \{J^2,[J^2,[T^8,G^{kc}]]\},
\end{eqnarray}
\begin{equation}
\epsilon^{ijk} ( f^{aec}d^{be8} - f^{bec}d^{ae8} - f^{abe}d^{ec8} ) (\mathcal{D}_2^{ia}\mathcal{D}_3^{jb} + \mathcal{D}_3^{ia}\mathcal{D}_2^{jb} ) = 0,
\end{equation}
\begin{eqnarray}
&  & \epsilon^{ijk} ( f^{aec}d^{be8} - f^{bec}d^{ae8} - f^{abe}d^{ec8} ) (\mathcal{D}_2^{ia}\mathcal{O}_3^{jb} + \mathcal{O}_3^{ia}\mathcal{D}_2^{jb} ) = - \frac38 \{J^2,[G^{kc},\{J^r,G^{r8}\}]\} + \frac38  \{J^2,[G^{k8},\{J^r,G^{rc}\}]\} \nonumber \\
&  & \mbox{} + \frac38 \{J^k,[\{J^m,G^{mc}\},\{J^r,G^{r8}\}]\} + \frac{N_c+N_f}{N_f} \{J^2,[J^2,[T^8,G^{kc}]]\} - \frac38 \{[J^2,G^{kc}],\{J^r,G^{r8}\}\} \nonumber \\
&  & \mbox{} + \frac38 \{[J^2,G^{k8}],\{J^r,G^{rc}\}\} - \{J^2,\{G^{kc},\{J^r,G^{r8}\}\}\} + \{J^2,\{G^{k8},\{J^r,G^{rc}\}\}\}.
\end{eqnarray}

We notice that the product operator $A^{ia}A^{jb}$ is at most of order $\mathcal{O}(N_c^2)$, so no large-$N_c$ cancellations were expected in the reduction of this operator in terms of the operator basis.

\section{Reduction of baryon operators: structure of diagrams of Fig.~\ref{fig:mmloop2} \label{app:bo2}}

In this appendix we now present the reduction of the product operator $[A^{ia},[A^{ib},A^{kc}]] \Pi^{ab}$ introduced in Eq.~(\ref{eq:corrloop2}). Here $\Pi^{ab}$ decomposes into flavor singlet, flavor $\mathbf{8}$ and flavor $\mathbf{27}$ representations, according to Eq.~(\ref{eq:pisym}). To the order in $1/N_c$ implemented in this work, the results can be organized as follows:

Flavor singlet contribution:

\begin{equation}
[G^{ia}, [G^{ia}, G^{kc}]] = \frac{3N_f^2-4}{4N_f} G^{kc},
\end{equation}
\begin{equation}
[G^{ia},[\mathcal{D}_2^{ia},G^{kc}]] + [\mathcal{D}_2^{ia},[G^{ia},G^{kc}]] = \frac{(N_c+N_f)(N_f-2)}{N_f} G^{kc} + \frac{N_f+2}{2} \mathcal{D}_2^{kc},
\end{equation}
\begin{equation}
[G^{ia},[G^{ia},\mathcal{D}_2^{kc}]] = -(N_c+N_f) G^{kc} + \frac{7N_f^2+4N_f-4}{4N_f} \mathcal{D}_2^{kc},
\end{equation}
\begin{equation}
[G^{ia},[\mathcal{D}_2^{ia},\mathcal{D}_2^{kc}]] + [\mathcal{D}_2^{ia},[G^{ia},\mathcal{D}_2^{kc}]] = -2N_f G^{kc} + \frac{2(N_c+N_f)(N_f-1)}{N_f}\mathcal{D}_2^{kc} + \frac{N_f}{2} \mathcal{D}_3^{kc} - 2\mathcal{O}_3^{kc},
\end{equation}
\begin{equation}
[\mathcal{D}_2^{ia},[\mathcal{D}_2^{ia},G^{kc}]] = \frac{N_c(N_c+2N_f)(N_f-2)-2N_f^2}{2N_f}G^{kc} + \frac{N_f+2}{4} \mathcal{D}_3^{kc} + \frac{N_f+4}{2} \mathcal{O}_3^{kc},
\end{equation}
\begin{equation}
[G^{ia},[\mathcal{D}_3^{ia},G^{kc}]] + [\mathcal{D}_3^{ia},[G^{ia},G^{kc}]] = 2(N_f-2) G^{kc} + (N_c+N_f) \mathcal{D}_2^{kc} + \frac{N_f^2+2N_f-4}{2N_f} \mathcal{D}_3^{kc} + \frac{N_f^2+2N_f-8}{N_f} \mathcal{O}_3^{kc},
\end{equation}
\begin{equation}
[G^{ia},[G^{ia},\mathcal{D}_3^{kc}]] = -[N_c(N_c+2N_f)+4] G^{kc} - 4(N_c+N_f) \mathcal{D}_2^{kc}  + \frac{11N_f^2+12N_f-4}{4N_f} \mathcal{D}_3^{kc},
\end{equation}
\begin{equation}
[G^{ia},[\mathcal{O}_3^{ia},G^{kc}]] + [\mathcal{O}_3^{ia},[G^{ia},G^{kc}]] = - \frac32(N_c+N_f) \mathcal{D}_2^{kc} + \frac{N_f+1}{2} \mathcal{D}_3^{kc} + N_f \mathcal{O}_3^{kc},
\end{equation}
\begin{equation}
[G^{ia},[G^{ia},\mathcal{O}_3^{kc}]] = [-N_c(N_c+2N_f)+N_f] G^{kc} + (N_c+N_f) \mathcal{D}_2^{kc} + \frac{11N_f^2+12N_f-4}{4N_f} \mathcal{O}_3^{kc},
\end{equation}
\begin{equation}
[\mathcal{D}_2^{ia},[\mathcal{D}_2^{ia},\mathcal{D}_2^{kc}]]= \frac{N_c(N_c+2N_f)(N_f-2)-2N_f^2}{2N_f} \mathcal{D}_2^{kc} + \frac{N_f+2}{2}\mathcal{D}_4^{kc},
\end{equation}
\begin{eqnarray}
[\mathcal{D}_2^{ia},[\mathcal{D}_3^{ia},G^{kc}]] + [\mathcal{D}_3^{ia},[\mathcal{D}_2^{ia},G^{kc}]] & = & -4(N_c+N_f) G^{kc} - 2(N_f-2) \mathcal{D}_2^{kc} + \frac{(N_c+N_f)(3N_f-2)}{N_f} \mathcal{D}_3^{kc} \nonumber \\
&  & \mbox{} + \frac{2(N_c+N_f)(5N_f-4)}{N_f} \mathcal{O}_3^{kc} + (N_f-2) \mathcal{D}_4^{kc},
\end{eqnarray}
\begin{eqnarray}
[G^{ia},[\mathcal{D}_3^{ia},\mathcal{D}_2^{kc}]] + [\mathcal{D}_3^{ia},[G^{ia},\mathcal{D}_2^{kc}]] & = & -4(N_c+N_f) G^{kc} + [N_c(N_c+2N_f)+2N_f] \mathcal{D}_2^{kc} + (N_c+N_f) \mathcal{D}_3^{kc}  \nonumber \\
&  & \mbox{} - 2(N_c+N_f) \mathcal{O}_3^{kc} + \frac{N_f^2-4}{N_f} \mathcal{D}_4^{kc},
\end{eqnarray}
\begin{eqnarray}
[G^{ia},[\mathcal{D}_2^{ia},\mathcal{D}_3^{kc}]] + [\mathcal{D}_2^{ia},[G^{ia},\mathcal{D}_3^{kc}]] & = & -4(N_c+N_f) G^{kc} - 2(N_f-2) \mathcal{D}_2^{kc} + \frac{(N_c+N_f)(3N_f-2)}{N_f} \mathcal{D}_3^{kc} \nonumber \\
&  & \mbox{} - 2(N_c+N_f) \mathcal{O}_3^{kc} + (N_f-2) \mathcal{D}_4^{kc},
\end{eqnarray}
\begin{eqnarray}
&  & [\mathcal{D}_2^{ia},[\mathcal{O}_3^{ia},G^{kc}]] + [\mathcal{O}_3^{ia},[\mathcal{D}_2^{ia},G^{kc}]] = 3N_f \mathcal{D}_2^{kc} - (N_c+N_f) \mathcal{D}_3^{kc} - (N_c+N_f) \mathcal{O}_3^{kc} + 2 \mathcal{D}_4^{kc}, \nonumber \\
\end{eqnarray}
\begin{eqnarray}
[G^{ia},[\mathcal{O}_3^{ia},\mathcal{D}_2^{kc}]] + [\mathcal{O}_3^{ia},[G^{ia},\mathcal{D}_2^{kc}]] & = & -\frac32 [N_c(N_c+2N_f)-4N_f] \mathcal{D}_2^{kc} - \frac52(N_c+N_f) \mathcal{D}_3^{kc} - (N_c+N_f) \mathcal{O}_3^{kc} \nonumber \\
&  & \mbox{} + 3(N_f+2) \mathcal{D}_4^{kc},
\end{eqnarray}
\begin{equation}
[G^{ia},[\mathcal{D}_2^{ia},\mathcal{O}_3^{kc}]] + [\mathcal{D}_2^{ia},[G^{ia},\mathcal{O}_3^{kc}]] = 3N_f \mathcal{D}_2^{kc} - (N_c+N_f) \mathcal{D}_3^{kc} + \frac{2(N_c+N_f)(N_f-1)}{N_f} \mathcal{O}_3^{kc} + 2 \mathcal{D}_4^{kc}.
\end{equation}

Flavor $\mathbf{8}$ contribution

\begin{equation}
d^{ab8} [G^{ia},[G^{ib},G^{kc}]] = \frac{3N_f^2-16}{8N_f} d^{c8e} G^{ke} + \frac{N_f^2-4}{2N_f^2} \delta^{c8} J^k,
\end{equation}
\begin{eqnarray}
&  & d^{ab8} \left([G^{ia},[\mathcal{D}_2^{ib},G^{kc}]] + [\mathcal{D}_2^{ia},[G^{ib},G^{kc}]]\right) = \frac{(N_c+N_f)(N_f-4)}{2N_f} d^{c8e} G^{ke} + \frac{(N_c+N_f)(N_f-2)}{N_f^2} \delta^{c8} J^k \nonumber \\
&  & \mbox{} + \frac{N_f+2}{4} d^{c8e} \mathcal{D}_2^{ke} + \frac12 \{T^c,G^{k8}\} + \frac{N_f-4}{2N_f} \{G^{kc},T^8\} + \frac{N_f^2+2N_f-4}{4N_f} [J^2,[T^8,G^{kc}]],
\end{eqnarray}
\begin{eqnarray}
d^{ab8} [G^{ia}, [G^{ib},\mathcal{D}_2^{kc}]] & = & -\frac{N_c+N_f}{2} d^{c8e} G^{ke} + \frac{3N_f+4}{8} d^{c8e} \mathcal{D}_2^{ke} + \frac{N_f^2+N_f-4}{2N_f} \{T^c,G^{k8}\} - \frac12 \{G^{kc},T^8\} \nonumber \\
&  & \mbox{} + \frac{1}{N_f} [J^2,[T^8,G^{kc}]],
\end{eqnarray}
\begin{eqnarray}
&  & d^{ab8} \left([G^{ia},[\mathcal{D}_2^{ib},\mathcal{D}_2^{kc}]] + [\mathcal{D}_2^{ia},[G^{ib},\mathcal{D}_2^{kc}]] \right) = - N_f d^{c8e} G^{ke} + \frac{(N_c+N_f)(N_f-2)}{N_f} \{T^c,G^{k8}\} + \frac{N_c+N_f}{N_f} [J^2,[T^8,G^{kc}]] \nonumber \\
&  & \mbox{} + \frac{N_f}{4} d^{c8e} \mathcal{D}_3^{ke} - d^{c8e} \mathcal{O}_3^{ke} + \frac{N_f-2}{2N_f} \{J^k,\{T^c,T^8\}\} - \{G^{kc},\{J^r,G^{r8}\}\} + \{G^{k8},\{J^r,G^{rc}\}\},
\end{eqnarray}
\begin{eqnarray}
&  & d^{ab8} [\mathcal{D}_2^{ia},[\mathcal{D}_2^{ib},G^{kc}]] = - \frac{N_f}{2} d^{c8e} G^{ke} + \frac{(N_c+N_f)(N_f-4)}{2N_f} \{G^{kc},T^8\} + \frac{(N_c+N_f)(N_f-4)}{4N_f}[J^2,[T^8,G^{kc}]] \nonumber \\
&  & \mbox{} + \frac{N_f}{8} d^{c8e} \mathcal{D}_3^{ke} + \frac{N_f+2}{4} d^{c8e} \mathcal{O}_3^{ke} + \frac32 \{G^{kc},\{J^r,G^{r8}\}\} - \frac12 \{G^{k8},\{J^r,G^{rc}\}\},
\end{eqnarray}
\begin{eqnarray}
&  & d^{ab8} \left([G^{ia},[\mathcal{D}_3^{ib},G^{kc}]] + [\mathcal{D}_3^{ia},[G^{ib},G^{kc}]]\right) = (N_f-4) d^{c8e} G^{ke} + \frac{N_c(N_c+2N_f)+4N_f-8}{2N_f} \delta^{c8} J^k + \frac{N_c+N_f}{2} d^{c8e} \mathcal{D}_2^{ke} \nonumber \\
&  & \mbox{} + (N_c+N_f) [J^2,[T^8, G^{kc}]] + \frac{N_f^2+2N_f-8}{4N_f} d^{c8e} \mathcal{D}_3^{ke} + \frac{N_f^2+2N_f-20}{2N_f} d^{c8e} \mathcal{O}_3^{ke} + \frac14 \{J^k,\{T^c,T^8\}\} \nonumber \\
&  & \mbox {} - \{J^k,\{G^{rc},G^{r8}\}\} + \frac{N_f-6}{N_f} \{G^{kc},\{J^r,G^{r8}\}\} + \frac{N_f+2}{N_f} \{G^{k8},\{J^r,G^{rc}\}\} + \frac{N_f-4}{N_f^2} \delta^{c8} \{J^2,J^k\},
\end{eqnarray}
\begin{eqnarray}
&  & d^{ab8} [G^{ia},[G^{ib},\mathcal{D}_3^{kc}]] = -4 d^{c8e} G^{ke} - \frac{2[N_c(N_c+2N_f)-N_f+2]}{N_f} \delta^{c8} J^k - 2(N_c+N_f) d^{c8e} \mathcal{D}_2^{ke} - (N_c+N_f) \{G^{kc},T^8\} \nonumber \\
&  & \mbox{} + \frac12 (N_c+N_f) [J^2, [T^8, G^{kc}]] + \frac{3N_f+8}{8} d^{c8e} \mathcal{D}_3^{ke} - \frac{2}{N_f} d^{c8e} \mathcal{O}_3^{ke} - \{J^k,\{T^c,T^8\}\} + (N_f+2) \{J^k,\{G^{rc},G^{r8}\}\} \nonumber \\
&  & \mbox {} + \frac{2}{N_f} \{G^{kc},\{J^r,G^{r8}\}\} + \frac{N_f^2+2N_f-6}{N_f} \{G^{k8},\{J^r,G^{rc}\}\} + \frac{N_f+2}{N_f} \delta^{c8} \{J^2,J^k\},
\end{eqnarray}
\begin{eqnarray}
&  & d^{ab8} \left([G^{ia},[\mathcal{O}_3^{ib},G^{kc}]] + [\mathcal{O}_3^{ia},[G^{ib},G^{kc}]]\right) = - \frac{3N_c(N_c+2N_f)}{4N_f} \delta^{c8} J^k - \frac34 (N_c+N_f) d^{c8e} \mathcal{D}_2^{ke} \nonumber \\
&  & \mbox{} - \frac14(N_c+N_f) [J^2, [T^8, G^{kc}]] + \frac{N_f^2+N_f-4}{4N_f} d^{c8e} \mathcal{D}_3^{ke} + \frac{N_f^2-2}{2N_f} d^{c8e} \mathcal{O}_3^{ke} - \frac38 \{J^k,\{T^c,T^8\}\} \nonumber \\
&  & \mbox{} + \frac{N_f+4}{2N_f} \{J^k,\{G^{rc},G^{r8}\}\} + \frac{1}{N_f} \{G^{kc},\{J^r,G^{r8}\}\} - \frac{1}{N_f} \{G^{k8},\{J^r,G^{rc}\}\} + \frac{2N_f^2+N_f-4}{2N_f^2} \delta^{c8}\{J^2,J^k\}, \nonumber \\
\end{eqnarray}
\begin{eqnarray}
&  & d^{ab8} [G^{ia},[G^{ib},\mathcal{O}_3^{kc}]] = \frac{N_f}{2} d^{c8e}G^{ke} + \frac{N_c(N_c+2N_f)}{2N_f} \delta^{c8} J^k + \frac{N_c+N_f}{2} d^{c8e} \mathcal{D}_2^{ke} - (N_c+N_f)\{G^{kc},T^8\} \nonumber \\
&  & \mbox{} - \frac34(N_c+N_f) [J^2, [T^8, G^{kc}]] - \frac{1}{N_f} d^{c8e} \mathcal{D}_3^{ke} + \frac{3N_f^2+8N_f-8}{8N_f} d^{c8e} \mathcal{O}_3^{ke} + \frac14 \{J^k,\{T^c,T^8\}\} \nonumber \\
&  & \mbox{} - \frac{N_f^2+2N_f-4}{2N_f} \{J^k,\{G^{rc},G^{r8}\}\} + \frac{N_f^2+2N_f-1}{N_f} \{G^{kc},\{J^r,G^{r8}\}\} - \frac{N_f^2+2N_f-2}{2N_f} \{G^{k8},\{J^r,G^{rc}\}\} \nonumber \\
&  & \mbox{} - \frac{2}{N_f^2} \delta^{c8} \{J^2,J^k\},
\end{eqnarray}
\begin{equation}
d^{ab8} [\mathcal{D}_2^{ia},[\mathcal{D}_2^{ib},\mathcal{D}_2^{kc}]] = - \frac{N_f}{2} d^{c8e}\mathcal{D}_2^{ke} + \frac{(N_c+N_f)(N_f-4)}{4N_f} \{J^k,\{T^c,T^8\}\} + \frac{N_f}{4} d^{c8e} \mathcal{D}_4^{ke} + \{\mathcal{D}_2^{kc},\{J^r,G^{r8}\}\},
\end{equation}
\begin{eqnarray}
&  & d^{ab8} \left([\mathcal{D}_2^{ia},[\mathcal{D}_3^{ib},G^{kc}]] + [\mathcal{D}_3^{ia},[\mathcal{D}_2^{ib},G^{kc}]] \right) = - 2 (N_c+N_f) d^{c8e} G^{ke} - (N_f-2) d^{c8e} \mathcal{D}_2^{ke} - 2 \{G^{kc},T^8\} \nonumber \\
&  & \mbox{}  + 2 \{T^c,G^{k8}\} - \frac{N_f^2-2N_f-4}{N_f} [J^2,[T^8,G^{kc}]] + \frac{N_c+N_f}{2} d^{c8e} \mathcal{D}_3^{ke} + \frac{2(N_c+N_f)(N_f-1)}{N_f} d^{c8e} \mathcal{O}_3^{ke} \nonumber \\
&  & \mbox{} + \frac{3(N_c+N_f)(N_f-2)}{N_f} \{G^{kc},\{J^r,G^{r8}\}\} - \frac{(N_c+N_f)(N_f-2)}{N_f} \{G^{k8},\{J^r,G^{rc}\}\} + \frac{N_f-2}{2} d^{c8e} \mathcal{D}_4^{ke} \nonumber \\
&  & \mbox{} - \{J^2,\{T^c,G^{k8}\}\}  + \frac{5N_f-8}{N_f} \{J^2,\{G^{kc},T^8\}\} - \frac{2(N_f-2)}{N_f} \{\mathcal{D}_2^{k8},\{J^r,G^{rc}\}\} + \frac{5}{32} \{J^2,[G^{kc},\{J^r,G^{r8}\}]\} \nonumber \\
&  & \mbox{} - \frac{5}{32} \{J^2,[G^{k8},\{J^r,G^{rc}\}]\} + \frac{5}{32} \{[J^2,G^{kc}],\{J^r,G^{r8}\}\} - \frac{5}{32} \{[J^2,G^{k8}],\{J^r,G^{rc}\}\} \nonumber \\
&  & \mbox{} - \frac{5}{32} \{J^k,[\{J^m,G^{mc}\},\{J^r,G^{r8}\}]\} + \frac{N_f^2+2N_f-12}{2N_f} \{J^2,[J^2,[T^8,G^{kc}]]\},
\end{eqnarray}
\begin{eqnarray}
&  & d^{ab8} \left([G^{ia},[\mathcal{D}_3^{ib},\mathcal{D}_2^{kc}]] + [\mathcal{D}_3^{ia},[G^{ib},\mathcal{D}_2^{kc}]] \right) = - 2 (N_c+N_f) d^{c8e} G^{ke} - (N_f-2) d^{c8e} \mathcal{D}_2^{ke} - 2 \{G^{kc},T^8\} \nonumber \\
&  & \mbox{} + 2 (N_f-1) \{T^c,G^{k8}\} + \frac{4}{N_f} [J^2,[T^8,G^{kc}]] + \frac{N_c+N_f}{2} d^{c8e} \mathcal{D}_3^{ke} - (N_c+N_f) d^{c8e} \mathcal{O}_3^{ke} \nonumber \\
&  & \mbox{} + \frac{N_c+N_f}{2} \{J^k,\{T^c,T^8\}\} + \frac{N_f-2}{2} d^{c8e}\mathcal{D}_4^{ke} + \frac{N_f^2+3N_f-8}{N_f}\{J^2,\{T^c,G^{k8}\}\} - \{J^2,\{G^{kc},T^8\}\} \nonumber \\
&  & \mbox{} - (N_f+2) \{\mathcal{D}_2^{kc},\{J^r,G^{r8}\}\} + 2 \{\mathcal{D}_2^{k8},\{J^r,G^{rc}\}\} + \frac{2}{N_f}\{J^2,[J^2,[T^8,G^{kc}]]\},
\end{eqnarray}
\begin{eqnarray}
&  & d^{ab8} \left([G^{ia},[\mathcal{D}_2^{ib},\mathcal{D}_3^{kc}]] + [\mathcal{D}_2^{ia},[G^{ib},\mathcal{D}_3^{kc}]]
\right) = - 2 (N_c+N_f) d^{c8e} G^{ke} - (N_f-2) d^{c8e} \mathcal{D}_2^{ke} - 2 \{G^{kc},T^8\} + 2 \{T^c,G^{k8}\} \nonumber \\
&  & \mbox{} + 2(N_f-1) [J^2,[T^8,G^{kc}]] + \frac12 (N_c+N_f) d^{c8e} \mathcal{D}_3^{ke} - \frac{2(N_c+N_f)}{N_f} d^{c8e} \mathcal{O}_3^{ke} - \frac{(N_c+N_f)(N_f-2)}{N_f} \{G^{kc},\{J^r,G^{r8}\}\} \nonumber \\
&  & \mbox{}  \nonumber \\
&  & \mbox{} + \frac{3(N_c+N_f)(N_f-2)}{N_f} \{G^{k8},\{J^r,G^{rc}\}\} + \frac12 (N_f-2) d^{c8e} \mathcal{D}_4^{ke} + \{J^2,\{T^c,G^{k8}\}\} - \{J^2,\{G^{kc},T^8\}\} \nonumber \\
&  & \mbox{} - 2 \{\mathcal{D}_2^{kc},\{J^r,G^{r8}\}\} + \frac{4(N_f-1)}{N_f} \{\mathcal{D}_2^{k8},\{J^r,G^{rc}\}\} - \frac{15}{64} \{J^2,[G^{kc},\{J^r,G^{r8}\}]\} + \frac{15}{64} \{J^2,[G^{k8},\{J^r,G^{rc}\}]\} \nonumber \\
&  & \mbox{} - \frac{15}{64} \{[J^2,G^{kc}],\{J^r,G^{r8}\}\} + \frac{15}{64}\{[J^2,G^{k8}],\{J^r,G^{rc}\}\} + \frac{15}{64}\{J^k,[\{J^m,G^{mc}\},\{J^r,G^{r8}\}]\} \nonumber \\
&  & \mbox{} + \{J^2,[J^2,[T^8,G^{kc}]]\},
\end{eqnarray}
\begin{eqnarray}
&  & d^{ab8} \left([\mathcal{D}_2^{ia},[\mathcal{O}_3^{ib},G^{kc}]] + [\mathcal{O}_3^{ia},[\mathcal{D}_2^{ib},G^{kc}]]\right) = \frac32 N_f d^{c8e} \mathcal{D}_2^{ke} + \frac{N_f-2}{2} [J^2,[T^8,G^{kc}]] - \frac{N_c+N_f}{N_f} d^{c8e} \mathcal{D}_3^{ke} \nonumber \\
&  & \mbox{} - \frac{N_c+N_f}{N_f} d^{c8e} \mathcal{O}_3^{ke} - \frac{(N_c+N_f)(N_f-2)}{N_f} \{J^k,\{G^{rc},G^{r8}\}\} - \frac{(N_c+N_f)(N_f-2)}{2N_f} \{G^{kc},\{J^r,G^{r8}\}\} \nonumber \\
&  & \mbox{} + \frac{(N_c+N_f)(N_f-2)}{2N_f} \{G^{k8},\{J^r,G^{rc}\}\} + \frac{(N_c+N_f)(N_f-2)}{N_f^2} \delta^{c8} \{J^2,J^k\} + d^{c8e} \mathcal{D}_4^{ke} - \frac12 \{J^2,\{G^{kc},T^8\}\} \nonumber \\
&  & \mbox{} + \frac12 \{J^2,\{G^{k8},T^c\}\}  + \frac12 \{\mathcal{D}_2^{kc},\{J^r,G^{r8}\}\} - \frac12 \{\mathcal{D}_2^{k8},\{J^r,G^{rc}\}\} - \frac{13}{64} \{J^2,[G^{kc},\{J^r,G^{r8}\}]\} \nonumber \\
&  & \mbox{} + \frac{13}{64} \{J^2,[G^{k8},\{J^r,G^{rc}\}]\} - \frac{13}{64} \{[J^2,G^{kc}],\{J^r,G^{r8}\}\} + \frac{13}{64} \{[J^2,G^{k8}],\{J^r,G^{rc}\}\} \nonumber \\
&  & \mbox{} + \frac{13}{64} \{J^k,[\{J^m,G^{mc}\},\{J^r,G^{r8}\}]\} + \frac12 \{J^2,[J^2,[T^8,G^{kc}]]\},
\end{eqnarray}
\begin{eqnarray}
&  & d^{ab8} \left([G^{ia},[\mathcal{O}_3^{ib},\mathcal{D}_2^{kc}]] + [\mathcal{O}_3^{ia},[G^{ib},\mathcal{D}_2^{kc}]] \right) = 3 N_f d^{c8e} \mathcal{D}_2^{ke} - \frac54 (N_c+N_f) d^{c8e} \mathcal{D}_3^{ke} - \frac12 (N_c+N_f) d^{c8e}\mathcal{O}_3^{ke} \nonumber \\
&  & \mbox{} - \frac34 (N_c+N_f) \{J^k,\{T^c,T^8\}\} + \frac{N_f+5}{2} d^{c8e} \mathcal{D}_4^{ke} - \frac12 \{J^2,\{G^{kc},T^8\}\} + \frac{N_f^2+N_f-4}{2N_f} \{J^2,\{G^{k8},T^c\}\} \nonumber \\
&  & \mbox{} + \frac{N_f^2+6N_f+4}{2N_f} \{\mathcal{D}_2^{kc},\{J^r,G^{r8}\}\} - 2 \{\mathcal{D}_2^{k8},\{J^r,G^{rc}\}\} + \frac{1}{N_f} \{J^2,[J^2,[T^8,G^{kc}]]\},
\end{eqnarray}
\begin{eqnarray}
&  & d^{ab8} \left([G^{ia},[\mathcal{D}_2^{ib},\mathcal{O}_3^{kc}]] + [\mathcal{D}_2^{ia},[G^{ib},\mathcal{O}_3^{kc}]] \right) = \frac32 N_f d^{c8e} \mathcal{D}_2^{ke} - \frac12 (N_f-2) [J^2,[T^8,G^{kc}]] - \frac{N_c+N_f}{N_f} d^{c8e} \mathcal{D}_3^{ke} \nonumber \\
&  & \mbox{} + \frac{(N_c+N_f)(N_f-2)}{2N_f} d^{c8e} \mathcal{O}_3^{ke} - \frac{(N_c+N_f)(N_f-2)}{N_f} \{J^k,\{G^{rc},G^{r8}\}\} + \frac{(N_c+N_f)(N_f-2)}{2N_f} \{G^{kc},\{J^r,G^{r8}\}\} \nonumber \\
&  & \mbox{} - \frac{(N_c+N_f)(N_f-2)}{2N_f} \{G^{k8},\{J^r,G^{rc}\}\} + \{\mathcal{D}_2^{kc},\{J^r,G^{r8}\}\} + \frac{(N_c+N_f)(N_f-2)}{N_f^2} \delta^{c8} \{J^2,J^k\} + d^{c8e} \mathcal{D}_4^{ke} \nonumber \\
&  & \mbox{} + \frac{N_f-2}{N_f} \{J^2,\{G^{kc},T^8\}\} - \frac{2(N_f-1)}{N_f} \{\mathcal{D}_2^{k8},\{J^r,G^{rc}\}\} - \frac{45}{128} \{J^2,[G^{kc},\{J^r,G^{r8}\}]\} + \frac{45}{128} \{J^2,[G^{k8},\{J^r,G^{rc}\}]\} \nonumber \\
&  & \mbox{} - \frac{45}{128} \{[J^2,G^{kc}],\{J^r,G^{r8}\}\} + \frac{45}{128} \{[J^2,G^{k8}],\{J^r,G^{rc}\}\} + \frac{45}{128} \{J^k,[\{J^m,G^{mc}\},\{J^r,G^{r8}\}]\} \nonumber \\
&  & \mbox{} + \frac{N_f^2-4}{4N_f} \{J^2,[J^2,[T^8,G^{kc}]]\},
\end{eqnarray}

Flavor $\mathbf{27}$  contribution

\begin{equation}
[G^{i8}, [G^{i8}, G^{kc}]] = \frac14 \left(2d^{c8e} d^{8eg} + f^{c8e} f^{8eg} \right) G^{kg} + \frac{1}{N_f} \delta^{c8} G^{k8} + \frac{1}{2N_f} d^{c88} J^k,
\end{equation}
\begin{eqnarray}
[G^{i8},[\mathcal{D}_2^{i8},G^{kc}]] + [\mathcal{D}_2^{i8},[G^{i8},G^{kc}]] & = & \frac12 f^{c8e} f^{8eg} \mathcal{D}_2^{kg} + \frac{2}{N_f} \delta^{c8} \mathcal{D}_2^{k8} + d^{c8e} \{G^{ke},T^8\} + \frac{i}{2} f^{c8e}[G^{k8},\{J^r,G^{re}\}] \nonumber \\
&  & \mbox{} + \frac{i}{2} f^{c8e}[G^{ke},\{J^r,G^{r8}\}],
\end{eqnarray}
\begin{equation}
[G^{i8},[G^{i8},\mathcal{D}_2^{kc}]] = \frac54 f^{c8e} f^{8eg} \mathcal{D}_2^{kg} + \frac{1}{N_f} \delta^{88} \mathcal{D}_2^{kc} + \frac12 d^{88e} \{G^{ke},T^c\} + \frac{i}{2} f^{c8e}[G^{k8},\{J^r,G^{re}\}] - \frac{i}{2} f^{c8e}[G^{ke},\{J^r,G^{r8}\}],
\end{equation}
\begin{eqnarray}
[G^{i8},[\mathcal{D}_2^{i8},\mathcal{D}_2^{kc}]] + [\mathcal{D}_2^{i8},[G^{i8},\mathcal{D}_2^{kc}]] & = & - f^{c8e} f^{8eg} G^{kg} + \frac12 f^{c8e} f^{8eg} \mathcal{D}_3^{kg} + \{G^{k8},\{T^c,T^8\}\} \nonumber \\
&  & \mbox{} - \frac12 \epsilon^{kim} f^{c8e} \{T^e,\{J^i,G^{m8}\}\} + \frac12 \epsilon^{kim} f^{c8e} \{T^8,\{J^i,G^{me}\}\},
\end{eqnarray}
\begin{equation}
[\mathcal{D}_2^{i8},[\mathcal{D}_2^{i8},G^{kc}]] = - f^{c8e} f^{8eg} G^{kg} + \frac14 f^{c8e} f^{8eg} \mathcal{D}_3^{kg} + \frac12 f^{c8e} f^{8eg} \mathcal{O}_3^{kg} + \frac12 \{G^{kc},\{T^8,T^8\}\} - \frac12 \epsilon^{kim} f^{c8e} \{T^8,\{J^i,G^{me}\}\},
\end{equation}
\begin{eqnarray}
[G^{i8},[\mathcal{D}_3^{i8},G^{kc}]] + [\mathcal{D}_3^{i8},[G^{i8},G^{kc}]] & = & \frac12 f^{c8e} f^{8eg} \mathcal{D}_3^{kg} + \frac{2}{N_f} \delta^{c8} \mathcal{D}_3^{k8} +
d^{c8e} d^{8eg} \mathcal{O}_3^{kg}  + 3 d^{c8e} \{G^{ke},\{J^r,G^{r8}\}\} \nonumber \\
&  & \mbox{} - d^{c8e} \{G^{k8},\{J^r,G^{re}\}\},
\end{eqnarray}
\begin{eqnarray}
&  & [G^{i8},[G^{i8},\mathcal{D}_3^{kc}]] = - \frac32 f^{c8e} f^{8eg} G^{kg} + \frac12 d^{c8e} d^{8eg} \mathcal{D}_3^{kg} + \frac14 f^{c8e} f^{8eg} \mathcal{D}_3^{kg} - \frac{1}{N_f} \delta^{c8} \mathcal{D}_3^{k8} + \frac{2}{N_f} \delta^{88} \mathcal{D}_3^{kc} + \frac{1}{N_f} d^{c88} \{J^2,J^k\} \nonumber \\
&  & \mbox{} - 2 \{G^{kc},\{G^{r8},G^{r8}\}\} + 2 \{G^{k8},\{G^{rc},G^{r8}\}\} - 3 d^{c8e} \{J^k,\{G^{re},G^{r8}\}\}  + d^{88e} \{J^k,\{G^{rc},G^{re}\}\} \nonumber \\
&  & \mbox{} + d^{c8e} \{G^{ke},\{J^r,G^{r8}\}\} + d^{88e} \{G^{ke},\{J^r,G^{rc}\}\} - \frac12 \epsilon^{kim} f^{c8e} \{T^e,\{J^i,G^{m8}\}\},
\end{eqnarray}
\begin{eqnarray}
&  & [G^{i8},[\mathcal{O}_3^{i8},G^{kc}]] + [\mathcal{O}_3^{i8},[G^{i8},G^{kc}]] = \frac12 d^{c8e} d^{8eg} \mathcal{D}_3^{kg} + \frac12 d^{c8e} d^{8eg} \mathcal{O}_3^{kg} + \frac12 f^{c8e} f^{8eg} \mathcal{O}_3^{kg} + \frac{2}{N_f} \delta^{c8} \mathcal{O}_3^{k8} \nonumber \\
&  & \mbox{} + \frac{1}{N_f} d^{c88} \{J^2,J^k\} - d^{c8e} \{J^k,\{G^{re},G^{r8}\}\} - \frac12 d^{c8e} \{G^{ke},\{J^r,G^{r8}\}\} + \frac12 d^{c8e} \{G^{k8},\{J^r,G^{re}\}\},
\end{eqnarray}

\begin{eqnarray}
&  & [G^{i8},[G^{i8},\mathcal{O}_3^{kc}]] = \frac34 f^{c8e} f^{8eg} G^{kg} + \frac{1}{N_f} \delta^{c8} \mathcal{D}_3^{k8} + \frac12 d^{c8e} d^{8eg} \mathcal{O}_3^{kg} + \frac14 f^{c8e} f^{8eg} \mathcal{O}_3^{kg} + \frac{3}{N_f} \delta^{c8} \mathcal{O}_3^{k8} + \frac{2}{N_f} \delta^{88} \mathcal{O}_3^{kc} \nonumber \\
&  & \mbox{}  - \{G^{kc},\{G^{r8},G^{r8}\}\} - \{G^{k8},\{G^{rc},G^{r8}\}\} + \frac12 d^{c8e} \{J^k,\{G^{re},G^{r8}\}\} - \frac12 d^{88e} \{J^k,\{G^{rc},G^{re}\}\} + d^{88e} \{G^{kc},\{J^r,G^{re}\}\} \nonumber \\
&  & \mbox{}  - \frac12 d^{c8e} \{G^{ke},\{J^r,G^{r8}\}\} - \frac12 d^{88e} \{G^{ke},\{J^r,G^{rc}\}\} + d^{c8e} \{G^{k8},\{J^r,G^{re}\}\} + \frac34 \epsilon^{kim} f^{c8e} \{T^e,\{J^i,G^{m8}\}\},
\end{eqnarray}
\begin{eqnarray}
[\mathcal{D}_2^{i8},[\mathcal{D}_2^{i8},\mathcal{D}_2^{kc}]] = -f^{c8e} f^{8eg} \mathcal{D}_2^{kg} + \frac12 \{\mathcal{D}_2^{kc},\{T^8,T^8\}\} + \frac12 f^{c8e} f^{8eg} \mathcal{D}_4^{kg},
\end{eqnarray}
\begin{eqnarray}
&  & [\mathcal{D}_2^{i8},[\mathcal{D}_3^{i8},G^{kc}]] + [\mathcal{D}_3^{i8},[\mathcal{D}_2^{i8},G^{kc}]] = - if^{c8e}[G^{k8},\{J^r,G^{re}\}] -2 if^{c8e}[G^{ke},\{J^r,G^{r8}\}] + d^{c8e}\{J^2,\{G^{ke},T^8\}\} \nonumber \\
&  & \mbox{} - d^{c8e}\{\mathcal{D}_2^{k8},\{J^r,G^{re}\}\} - \{\{J^r,G^{rc}\},\{G^{k8},T^8\}\} + 3 \{\{J^r,G^{r8}\},\{G^{kc},T^8\}\} + 2 if^{c8e} \{J^2,[G^{ke},\{J^r,G^{r8}\}]\} \nonumber \\
&  & \mbox{} - if^{c8e} \{\{J^r,G^{r8}\},[J^2,G^{ke}]\},
\end{eqnarray}
\begin{eqnarray}
&  & [G^{i8},[\mathcal{D}_3^{i8},\mathcal{D}_2^{kc}]] + [\mathcal{D}_3^{i8},[G^{i8},\mathcal{D}_2^{kc}]] = -2i f^{c8e}[G^{ke},\{J^r,G^{r8}\}] + d^{88e}\{J^2,\{G^{ke},T^c\}\} - d^{88e} \{\mathcal{D}_2^{kc},\{J^r,G^{re}\}\} \nonumber \\
&  & \mbox{} + 2 \{\{J^r,G^{r8}\},\{G^{k8},T^c\}\} + if^{c8e}\{J^k,[\{J^i,G^{ie}\},\{J^r,G^{r8}\}]\} - if^{c8e} \{\{J^r,G^{re}\},[J^2,G^{k8}]\} \nonumber \\
&  & \mbox{} + if^{c8e} \{\{J^r,G^{r8}\},[J^2,G^{ke}]\},
\end{eqnarray}
\begin{eqnarray}
&  & [G^{i8},[\mathcal{D}_2^{i8},\mathcal{D}_3^{kc}]] + [\mathcal{D}_2^{i8},[G^{i8},\mathcal{D}_3^{kc}]] = 5 if^{c8e}[G^{k8},\{J^r,G^{re}\}] + d^{c8e}\{J^2,\{G^{ke},T^8\}\} - d^{c8e}\{\mathcal{D}_2^{k8},\{J^r,G^{re}\}\} \nonumber \\
&  & \mbox{} + 3 \{\{J^r,G^{rc}\},\{G^{k8},T^8\}\} - \{\{J^r,G^{r8}\},\{G^{kc},T^8\}\} + if^{c8e}\{J^k,[\{J^i,G^{ie}\},\{J^r,G^{r8}\}]\} \nonumber \\
&  & \mbox{} - if^{c8e} \{\{J^r,G^{re}\},[J^2,G^{k8}]\},
\end{eqnarray}
\begin{eqnarray}
&  & [\mathcal{D}_2^{i8},[\mathcal{O}_3^{i8},G^{kc}]] + [\mathcal{O}_3^{i8},[\mathcal{D}_2^{i8},G^{kc}]] = \frac32 f^{c8e} f^{8eg} \mathcal{D}_2^{kg}  + \frac{i}{2} f^{c8e}[G^{k8},\{J^r,G^{re}\}] + \frac12 d^{c8e} \{J^2,\{G^{ke},T^8\}\} \nonumber \\
&  & \mbox{} + \frac12 f^{c8e} f^{8eg} \mathcal{D}_4^{kg} + \frac{2}{N_f} \delta^{c8} \mathcal{D}_4^{k8} - 2 \{\mathcal{D}_2^{k8},\{G^{rc},G^{r8}\}\} + \frac12 d^{c8e} \{\mathcal{D}_2^{k8},\{J^r,G^{re}\}\} + \frac12 \{\{J^r,G^{rc}\},\{G^{k8},T^8\}\} \nonumber \\
&  & \mbox{} - \frac12 \{\{J^r,G^{r8}\},\{G^{kc},T^8\}\} + \frac{i}{2} f^{c8e} \{J^2,[G^{k8},\{J^r,G^{re}\}]\} - \frac{i}{2} f^{c8e} \{J^2,[G^{ke},\{J^r,G^{r8}\}]\} \nonumber \\
&  & \mbox{} - \frac{i}{2} f^{c8e} \{\{J^r,G^{r8}\},[J^2,G^{ke}]\},
\end{eqnarray}
\begin{eqnarray}
&  & [G^{i8},[\mathcal{O}_3^{i8},\mathcal{D}_2^{kc}]] + [\mathcal{O}_3^{i8},[G^{i8},\mathcal{D}_2^{kc}]] = 6 f^{c8e} f^{8eg} \mathcal{D}_2^{kg} + \frac12 d^{88e} \{J^2,\{G^{ke},T^c\}\} + \frac72 f^{c8e} f^{8eg} \mathcal{D}_4^{kg} + \frac{2}{N_f} \delta^{88} \mathcal{D}_4^{kc} \nonumber \\
&  & \mbox{} - 2 \{\mathcal{D}_2^{kc},\{G^{r8},G^{r8}\}\} + \frac12 d^{88e} \{\mathcal{D}_2^{kc},\{J^r,G^{re}\}\} + if^{c8e} \{J^2,[G^{k8},\{J^r,G^{re}\}]\} - if^{c8e} \{J^2,[G^{ke},\{J^r,G^{r8}\}]\} \nonumber \\
&  & \mbox{} - \frac32 if^{c8e}\{J^k,[\{J^i,G^{ie}\},\{J^r,G^{r8}\}]\} + \frac{i}{2} f^{c8e} \{\{J^r,G^{re}\},[J^2,G^{k8}]\} - \frac{i}{2} f^{c8e} \{\{J^r,G^{r8}\},[J^2,G^{ke}]\},
\end{eqnarray}
\begin{eqnarray}
&  & [G^{i8},[\mathcal{D}_2^{i8},\mathcal{O}_3^{kc}]] + [\mathcal{D}_2^{i8},[G^{i8},\mathcal{O}_3^{kc}]] = \frac32 f^{c8e} f^{8eg} \mathcal{D}_2^{kg} - \frac{i}{2} f^{c8e}[G^{k8},\{J^r,G^{re}\}] + \frac12 d^{c8e}\{J^2,\{G^{ke},T^8\}\} \nonumber \\
&  & \mbox{} + \frac12 f^{c8e} f^{8eg} \mathcal{D}_4^{kg} + \frac{2}{N_f} \delta^{c8} \mathcal{D}_4^{k8} - 2 \{\mathcal{D}_2^{k8},\{G^{rc},G^{r8}\}\} + \frac12 d^{c8e} \{\mathcal{D}_2^{k8},\{J^r,G^{re}\}\} - \frac12 \{\{J^r,G^{rc}\},\{G^{k8},T^8\}\} \nonumber \\
&  & \mbox{} + \frac12 \{\{J^r,G^{r8}\},\{G^{kc},T^8\}\} + \frac{i}{2} f^{c8e} \{J^2,[G^{k8},\{J^r,G^{re}\}]\} + \frac{i}{2} f^{c8e} \{J^2,[G^{ke},\{J^r,G^{r8}\}]\} \nonumber \\
&  & \mbox{} - \frac{i}{2} f^{c8e}\{J^k,[\{J^i,G^{ie}\},\{J^r,G^{r8}\}]\} + \frac{i}{2} f^{c8e} \{\{J^r,G^{re}\},[J^2,G^{k8}]\},
\end{eqnarray}

In this case, the operator structure $[A^{ia},[A^{ib},A^{kc}]]$ does contain manifest large-$N_c$ cancellations in such a way that it is at most of order $\mathcal{O}(N_c)$, and is consistent with expectations.

\end{widetext}

\end{document}